\documentclass[twocolumn]{aastex62}
\usepackage[normalem]{ulem}
\usepackage{booktabs}
\usepackage{cancel}
\usepackage{CJK}
\usepackage{natbib}
\usepackage{morefloats}
\newcommand{\sersic}{S\'{e}rsic}
\newcommand{\galfit}{\texttt{GALFIT}}

\makeatletter

\newcommand{\Rmnum}[1]{\expandafter\@slowromancap\romannumeral #1@}
\makeatother
\defcitealias{2008AJ+Fisher}{FD08}
\defcitealias{2009MNRAS+Gadotti}{G09}

\begin{document}
\begin{CJK*}{UTF8}{gkai}
\title{The Carnegie-Irvine Galaxy Survey. \Rmnum{9}. Classification of Bulge Types and Statistical Properties of Pseudo Bulges}

\author[0000-0003-1015-5367]{Hua Gao (高桦)}
\affiliation{Department of Astronomy, School of Physics, Peking University, Beijing 100871, China}
\affiliation{Kavli Institute for Astronomy and Astrophysics, Peking University, Beijing 100871, China}

\author[0000-0001-6947-5846]{Luis C. Ho}
\affiliation{Kavli Institute for Astronomy and Astrophysics, Peking University, Beijing 100871, China}
\affiliation{Department of Astronomy, School of Physics, Peking University, Beijing 100871, China}

\author[0000-0002-3026-0562]{Aaron J. Barth}
\affiliation{Department of Physics and Astronomy, University of California at Irvine, 4129 Frederick Reines Hall, Irvine, CA 92697-4575, USA}

\author[0000-0001-5017-7021]{Zhao-Yu Li}
\affiliation{Department of Astronomy, Shanghai Jiao Tong University, Shanghai 200240, China}

\correspondingauthor{Hua Gao}
\email{hgao.astro@gmail.com}

\begin{abstract}
  We study the statistical properties of 320 bulges of disk galaxies in the
  Carnegie-Irvine Galaxy Survey, using robust structural parameters of galaxies
  derived from image fitting. We apply the Kormendy relation to classify
  classical and pseudo bulges and characterize bulge dichotomy with respect to
  bulge structural properties and physical properties of host galaxies. We
  confirm previous findings that pseudo bulges on average have smaller \sersic{}
  indices, smaller bulge-to-total ratios, and fainter surface brightnesses when
  compared with classical bulges. Our sizable sample statistically shows that
  pseudo bulges are more intrinsically flattened than classical bulges. Pseudo
  bulges are most frequent (incidence $\ga 80\%$) in late-type spirals (later
  than Sc). Our measurements support the picture in which pseudo bulges arose
  from star formation induced by inflowing gas, while classical bulges were born
  out of violent processes such as mergers and coalescence of clumps. We reveal
  differences with the literature that warrant attention: (1) the bimodal
  distribution of \sersic{} indices presented by previous studies is not
  reproduced in our study; (2) classical and pseudo bulges have similar relative
  bulge sizes; and (3) the pseudo bulge fraction is considerably smaller in
  early-type disks compared with previous studies based on one-dimensional
  surface brightness profile fitting. We attribute the above differences to our
  improved image quality, more robust bulge-to-disk decomposition technique, and
  different classification criteria applied.  Moreover, we find that barred
  galaxies do not host more pseudo bulges or more prominent pseudo bulges than
  unbarred galaxies. Various implications of these findings are discussed.
\end{abstract} 


\section{Introduction}
\label{sec:intro}

The fundamental plane of elliptical galaxies was first established in the 1980s
\citep{1987ApJ+Djorgovski,1987ApJ+Dressler} and has long proved to be a
useful tool to estimate distances and to study the formation and evolution of
these systems \citep{1987ApJ+Dressler,1987nngp+Faber, 1988ASPC+Djorgovski,
1988ApJ+Lynden-Bell,2013MNRAS+DOnofrio,2017ApJ+DOnofrio,2016MNRAS+Zhang}. The
fundamental plane is a three-parameter correlation of the galaxy half-light 
(effective) radius $r_{e}$, central velocity dispersion $\sigma_{0}$, and 
average effective surface brightness $\langle\mu_{e}\rangle$:
\begin{equation}
  \label{eq:FP_gen}
  \log r_{e}=a\log \sigma_{0}+b\langle\mu_{e}\rangle+c.
\end{equation}
At the most basic level, the physical root of the relation is the virial theorem
\citep{1987ApJ+Djorgovski,1988ASPC+Djorgovski, 1989ARA&A+Kormendy},
\begin{equation}
  \label{eq:vir_theo}
  \frac{GM}{r_{g}}\sim \langle v^{2}\rangle,
\end{equation}
where $M$ is the total mass, $r_{g}$ is the gravitational radius, and
$\langle v^{2}\rangle$ is the mean square of stellar velocity
\citep{2008gady+Binney}. To translate the physical parameters in the virial
theorem into observables, one has to make assumptions of the density, kinematic
structure, and dynamical mass-to-light ratio ($M/L$) of the galaxies. Rewriting
Equation~\ref{eq:vir_theo} to follow the formulation of
Equation~\ref{eq:FP_gen}, we have
\begin{equation}
  \label{eq:vir_pla}
  \log r_{e}=2\log \sigma_{0}+0.4\langle\mu_{e}\rangle+\log \left[k\left(\frac{M}{L}\right)^{-1}\right]+C,
\end{equation}
where $k$ accounts for the projection effect and intrinsic density and kinematic
structure of the galaxies. The observed coefficients $a$ and $b$ in
Equation~\ref{eq:FP_gen} generally deviate from 2 and 0.4, respectively, which
is commonly termed the tilt of the fundamental plane with respect to the virial
plane. The heterogeneity of density and kinematic structures and dynamical
mass-to-light ratios of ellipticals may contribute to the tilt, although which
factor plays a dominant role is still debated. Ellipticals have long been
approximated as homologous stellar systems, so the contribution of heterogeneity
of their structures to the tilt was often neglected. Assuming that ellipticals
have homologous density and kinematics, some authors found $M/L\sim M^{0.2}$
\citep{1987IAUS+Djorgovski,1987nngp+Faber,1998AJ+Pahre,2005ApJ+Treu}. Factors
that may lead to variations of dynamical mass-to-light ratio include metallicity
\citep[e.g.,][]{1987ApJ+Djorgovski}, initial mass function
\citep[e.g.,][]{2001MNRAS+Kroupa,1993ApJ+Renzini,2012Natur+Cappellari}, and dark
matter \citep[e.g.,][]{1996MNRAS+Ciotti,2006MNRAS+Cappellari,
  2013MNRAS+Cappellari1}. In addition to the aforementioned possibilities, the
heterogeneity of density and kinematics of ellipticals may also tilt the
fundamental plane. Ellipticals are now known to be not strictly homologous in
terms of light profile \citep{1991A&A+Nieto,1995AJ+Lauer,2005AJ+Lauer,
  2007ApJ+Lauer,1996IAUS+Kormendy,2009ApJS+Kormendy,2002A&A+Bertin} and
kinematic structure \citep{1977ApJ+Illingworth,1983ApJ+Davies,1987MitAG+Bender,
  1988A&A+Bender,2011MNRAS+Emsellem}. \citet{1997A&A+Busarello} stressed the
importance of kinematic heterogeneity in causing the tilt, and that structural
heterogeneity and stellar population effects play a minor role. On the contrary,
\citet{2013MNRAS+DOnofrio} concluded that the tilt is mainly affected by
structural heterogeneity. However, \citet{1996MNRAS+Ciotti} and
\citet{2006MNRAS+Cappellari,2013MNRAS+Cappellari1} disagree. Apart from the
tilt, studying the residuals of the fundamental plane provides insights into and
constrains the regularity/stochasticity of the formation and evolution of
elliptical galaxies \citep{1993ApJ+Renzini,2003MNRAS+Borriello,
  2013MNRAS+DOnofrio,2017ApJ+DOnofrio}.

Even though ellipticals are not a strictly homologous population, the tightness
of the fundamental plane implies their highly regulated formation and evolution
pathways, especially when compared with other stellar systems. For example,
star-forming disks on the other end of the Hubble sequence differ from the
ellipticals in every aforementioned aspect that could cause the tilt and amplify
the scatter of the scaling relations. Hence, the fundamental plane relation is
useful to distinguish stellar systems of distinct physical nature. One notably
successful example in application of the fundamental plane projections and their
variants was the discovery of the dichotomy between elliptical and spheroidal
galaxies \citep{1985ApJ+Kormendy,1987nngp+Kormendy,1992ApJ+Bender,
  1994ESOC+Binggeli}. These authors showed that spheroidal galaxies form a
sequence distinct from that of ellipticals in the core parameter correlations
(i.e., central surface brightness and core radius---a projection of the core
fundamental plane; see \citealp{1985ApJ+Lauer,1985ApJ+Kormendy,
  1987nngp+Kormendy,1994ESOC+Binggeli}) and in the $\kappa$ formulation of the
fundamental plane \citep{1992ApJ+Bender}. Surprisingly, in these parameter
spaces spheroidals occupy a locus that overlaps with the disks of spiral
galaxies. The dichotomy between slow-rotator and fast-rotator ellipticals
\citep[e.g.,][]{1977ApJ+Illingworth,1983ApJ+Davies,2011MNRAS+Emsellem} is also
another manifestation of the application of the core parameter correlations, but
it is much less obvious than the elliptical vs. spheroidal dichotomy
\citep{1997AJ+Faber}.

Since the seminal work of \citet{1976ApJ+van_den_Bergh1},
\citet{1981A&A+Combes}, \citet{1982ApJ+Gallagher}, \citet{1982ApJ+Kormendy2,
  1982SAAS+Kormendy,1993IAUS+Kormendy}, \citet{1982ApJ+Kormendy1,
  1983ApJ+Kormendy}, \citet{1984A&A+Pfenniger,1985A&A+Pfenniger},
\citet{1990A&A+Combes}, and \citet{1990ApJ+Pfenniger}, two classes of bulges are
recognized in disk (S0 and spiral) galaxies. Classical bulges have similar
observational properties as ellipticals \citep{1977egsp.conf+Faber,
  1977ARA&A+Gott,1999fgb+Renzini} and are therefore considered to form out of
rapid, violent processes such as mergers and coalescence of clumps
\citep{1977egsp+Tmoore,2016ASSL+Bournaud}. They follow the scaling relations
defined by ellipticals \citep{1985ApJ+Kormendy,1987nngp+Kormendy,
  1992ApJ+Bender}. On the other hand, pseudo bulges are more likely central
miniature disks embedded in the large-scale disks. They have younger, more
composite stellar populations, more flattened stellar light distribution, and
more rotation-dominated kinematics compared with classical bulges
\citep[e.g.,][]{1997ARA&A+Wyse,2004ARA&A+Kormendy,2016ASSL+Fisher,
  2016ASSL+Kormendy}. In contrast with the violent origin of classical bulges,
pseudo bulges are deemed to form out of redistributed disk material by gradual,
secular processes facilitated by non-axisymmetry in the galaxy potential
\citep{1981A&A+Combes,1981seng.proc+Kormendy,1982SAAS+Kormendy,1990ApJ+Pfenniger,
  1993RPPh+Sellwood}. Pseudo bulges are found to be low-surface brightness
outliers in the \citet{1977ApJ+Kormendy2} relation---the correlation between
surface brightness and effective radius of ellipticals, which constitutes the
photometric projection of the fundamental plane (e.g., \citealp{1999ApJ+Carollo,
  2009MNRAS+Gadotti,2010ApJ+Fisher,2010MNRAS+Laurikainen}; but see
\citealp{2017ApJS+Kim,2019ApJ+Kim,2019ApJ+Zhao} for the case of active
galaxies). The dichotomy between classical and pseudo bulges is the main subject
of this paper.

In addition to using the Kormendy relation, there are alternative ways to
distinguish classical bulges from pseudo bulges. For example, the
\citet{1968adga+Sersic} index $n$ is widely used to separate the two categories
of bulges. Based on the bimodal distribution of $n$ found by
\citet{2008AJ+Fisher,2016ASSL+Fisher}, these authors proposed that pseudo bulges
be defined by $n<2$. Bulge type is statistically linked with Hubble type (pseudo
bulges occur most frequently in spirals of type Sbc and later; \citealp[][and
references therein]{2004ARA&A+Kormendy}), and thus bulge-to-total ratio ($B/T$),
but the correspondence is imperfect because the correlation between Hubble type
and $B/T$ has large scatter \citep[e.g.,][]{1985ApJS+Kent,1986ApJ+Simien,
  1996A&A+de_Jong3,2004A&A+Grosbol,2007MNRAS+Laurikainen,2010MNRAS+Laurikainen,
  2009ApJ+Weinzirl,2017A&A+Mendez-Abreu,2019ApJS+Gao}. While a low $B/T$ does
not guarantee a pseudo bulge, pseudo bulges are generally found in galaxies with
$B/T\la 0.35$; if $B/T\ga 0.5$, the bulge is very likely classical
\citep{2013ARA&A+Kormendy,2016ASSL+Kormendy}. Bulge size was used in conjunction
with $B/T$ to classify bulges in \citet{2006MNRAS+Allen}. Central morphological
features, such as nuclear bars/rings/spirals that are indicative of
kinematically cold bulges, boxy/peanut bulges that are actually thickened bars,
or apparent flattenings similar to those of disks, are useful diagnostics to
recognize pseudo bulges \citep[e.g.,][]{2004ARA&A+Kormendy,2008AJ+Fisher,
  2010ApJ+Fisher,2009ApJ+Fisher,2013ARA&A+Kormendy,2016ASSL+Kormendy}. Stellar
kinematics provide perhaps the cleanest diagnostic, but they are not readily
available for large samples. Pseudo bulges are characterized by their larger
degree of rotation relative to random motion \citep{1982ApJ+Kormendy1,
  1982ApJ+Kormendy2}, stand out as low-$\sigma$ outliers in the Faber-Jackson
\citeyearpar{1976ApJ+Faber2} relation \citep{1983ApJ+Kormendy,
  2004ARA&A+Kormendy}, and have flat central $\sigma$ profiles
\citep{2016ASSL+Fisher}. Some extreme cases even show ``$\sigma$ drops'' at the
center. Vigorous star formation and young stellar populations, if not induced by
mergers, are signposts of on-going growth of pseudo bulges
\citep{2004ARA&A+Kormendy,2009ApJ+Fisher,2016ASSL+Fisher}. For complete reviews
of the observational criteria to classify classical and pseudo bulges, see
\citet{2004ARA&A+Kormendy}, supplemental material of \citet{2013ARA&A+Kormendy},
\citet{2016ASSL+Fisher}, and \citet{2016ASSL+Kormendy}. Unfortunately, these
many criteria do not necessarily produce consistent results. Robust
classifications can be achieved by applying as many criteria as possible. To
resolve the ambiguity, \citet{2017A&A+Neumann} performed a comprehensive
comparison among the commonly adopted criteria: \sersic{} index, concentration
index, the Kormendy relation, and the inner slope of radial velocity dispersion
profile. They found that the Kormendy relation is the best single criterion, as
it is able to recover 39 out of 40 ``safe'' classifications, where safe
classifications were defined as consistent classifications resulted from at
least three out of the four criteria. 

In order to characterize bulge dichotomy in the local Universe, we will
consistently apply the Kormendy relation to classify classical and pseudo
bulges, as it has a strong physical basis and provides the best agreement with
classifications based on multiple criteria. We do not consider the full
fundamental plane because stellar central velocity dispersions are not available
for a significant fraction of the spiral galaxies in our sample. We will use
the robust structural parameters derived in \citet{2019ApJS+Gao}. The rest of
this paper will be devoted to introduction of the Carnegie-Irvine Galaxy Survey
(CGS; \citealp{2011ApJS+Ho}),  measuring the Kormendy relation of CGS
ellipticals and bulges, classifying classical and pseudo bulges, studying their
statistical properties, and discussing implications on their respective
formation and evolution paths.

\section{Sample and Data}
\label{sec:sample-data}

The CGS sample is defined by $B_{T}\leq12.9$ mag and $\delta<0\arcdeg$, without
any reference to morphology, size, or environment. Details of the observations
and data reduction are given in \citet{2011ApJS+Ho} and \citet{2011ApJS+Li}, 
and will not be repeated here. In this study, we mainly use products from the
$R$-band images. The majority of the images are of high quality, in terms of
field-of-view ($8\farcm9\times8\farcm9$), median seeing ($1\farcs01$), and
median surface brightness depth (26.4\,mag~arcsec$^{-2}$).

In this paper, we make use of all the successfully decomposed disk galaxies
presented in \citet{2019ApJS+Gao}. The CGS disk galaxy sample for bulge-to-disk
decomposition is restricted to the subset of galaxies with morphological type
index $-3\leq T\leq9.5$ and inclination angle $i\leq70\arcdeg$. After
complementing the sample with a handful of misclassified ellipticals and
removing  objects whose decomposition was unsuccessful, the final
sample consists of 320 S0s and spirals. Basic properties of the sample are
discussed in detail in \citet{2019ApJS+Gao} and will not be repeated here. Below
we briefly describe the decomposition strategy; we refer readers to Section~3 of
\citet{2019ApJS+Gao} for full details of the decomposition. We performed
multi-component decomposition of the $R$-band images to derive accurate bulge
structural parameters, following closely the technique described in
\citet{2017ApJ+Gao} and \citet{2018ApJ+Gao}. In addition to bulges and disks, we
successfully modeled nuclei, bars, disk breaks, nuclear/inner lenses, and inner
rings. Our modeling strategy treats nuclear rings and nuclear bars as part of
the bulge component, while other features such as spiral arms, outer lenses, and
outer rings were omitted from the fits because they are not crucial for accurate
bulge measurements according to the experiments in \citet{2017ApJ+Gao}. The
error budget of the bulge parameters includes the uncertainties from sky level
measurements and model assumptions.

In addition to the disk galaxy sample, we perform single-\sersic{} fits to 83
ellipticals to measure their structural parameters in the $R$ band
(Appendix~\ref{sec:str-par-ell}). The CGS elliptical sample is drawn from
\citet{2013ApJ+Huang1}. We remove misclassified ellipticals that have since been
classified as S0s in \citet{2018ApJ+Gao}. Although \citet{2013ApJ+Huang1}
promoted the three-component nature of ellipticals, single-\sersic{} fits
suffice for our purpose of deriving global structural parameters for reference
comparison with the bulges on the fundamental plane correlations. The
uncertainties of the structural parameters stem from the uncertainties in sky
subtraction. We estimate the uncertainties in a manner consistent with the bulge
measurements. Namely, we measure the sky-induced uncertainties as variations of
the best-fit parameters when perturbing the sky levels around
$\pm1\,\sigma_{\mathrm{sky}}$ of the measured sky levels, where uncertainties of
the sky level $\sigma_{\mathrm{sky}}$ are measured as the root-mean-square of
the residuals determined from randomly placed boxes in the sky-dominated region
of the sky-subtracted data image (see Appendix~B of \citealp{2017ApJ+Gao} for
details). Considering that ellipticals are more extended than disk galaxies and
a single \sersic{} function cannot accurately describe their light profiles
across a large dynamical range in radius, we strive to avoid introducing
model-induced uncertainties to sky level measurements, which in return may
induce errors in \sersic{} index measurements. Therefore, instead of
simultaneously fitting the sky with the galaxy, as was done for the disk
galaxies, we measure the sky level using the direct approach described in
\citeauthor{2017ApJ+Gao}~(\citeyear{2017ApJ+Gao}; their Appendix~B.1) and
subtract it before performing the fits.

\section{Results}

\subsection{The Kormendy Relation of Ellipticals}
\label{sec:best-rel-ell}

We derive the Kormendy relation of the ellipticals by minimizing the $\chi^{2}$,
defined as,
\begin{equation}
\label{eq:chi_sq1}
\chi^{2}=\sum_{i=1}^{N}\frac{\left(\langle\mu_{e,i}\rangle-\alpha\log r_{e,i}-
\beta\right)^{2}}{\xi_{\langle\mu_{e,i}\rangle}^{2}+\alpha^{2}\xi_{\log r_{e,i}}^{2}},
\end{equation}
where $\alpha$ and $\beta$ are the coefficients of the Kormendy relation
$\langle\mu_{e}\rangle=\alpha\log r_{e}+\beta$ and $\xi$ denotes the
uncertainties. The best-fit relation is
\begin{equation}
\label{eq:kr-ell}
\langle\mu_{e}\rangle=\left(2.38\pm0.07\right)\log r_{e}+\left(17.86\pm 0.04\right),
\end{equation}
with a scatter of 0.42\,dex in $\langle\mu_{e}\rangle$. We correct
$\langle\mu_{e}\rangle$ for Galactic extinction, following \citet{2011ApJS+Ho}
and \citet{2011ApJS+Li}.

\begin{figure}
  \epsscale{1.15}
  \plotone{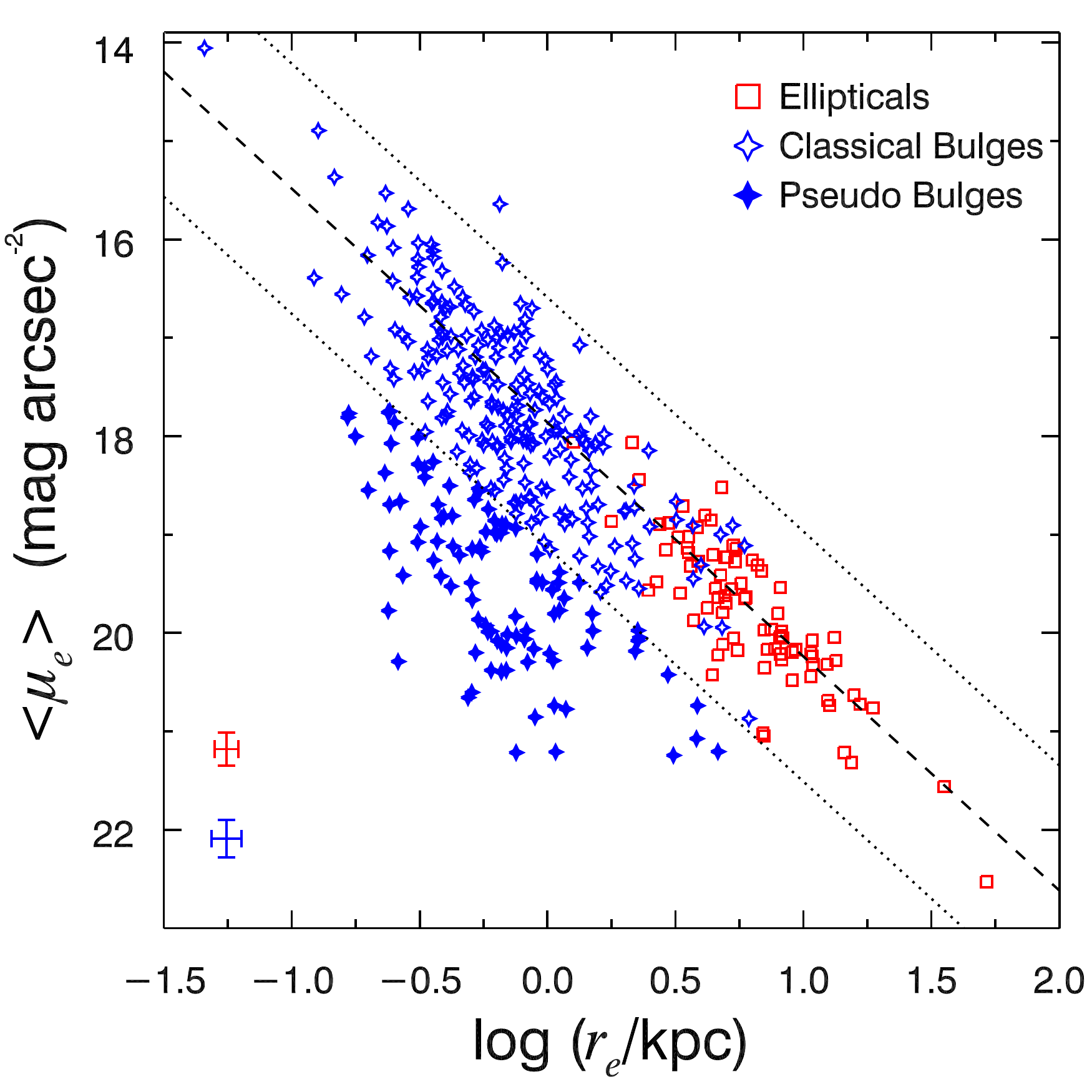}
  \caption{The Kormendy relation of CGS ellipticals and bulges. The best-fit
    relation and its $3\,\sigma$ boundary of the ellipticals are indicated by the
    short-dashed lines and dotted lines, respectively. Typical uncertainties are
    illustrated in the bottom-left corner. Blue filled stars represent pseudo
    bulges that are $3\,\sigma$ outliers from the best-fit relation of the
    ellipticals (red squares). The other bulges (blue open stars) are designated
    as classical bulges. \label{fig:KR}}
\end{figure}

The best-fit relation is shown in Figure~\ref{fig:KR}, with the bulges of the
disk galaxies overlaid. It is immediately apparent that some of the bulges
overlap with and continuously extend the high-$\langle\mu_{e}\rangle$, low-$r_e$
end of the Kormendy relation (the open blue stars), while others (the filled
stars) scatter to low values of $\langle\mu_{e}\rangle$ at fixed $r_e$.  We
classify as pseudo bulges those that scatter more than $3\,\sigma$ below the
best-fit Kormendy relation of ellipticals and the rest as classical bulges.
Namely, pseudo bulges satisfy
\begin{equation}
\label{eq:pb-def}
\langle\mu_{e}\rangle>2.38\log r_{e}+19.12,
\end{equation}
where $\langle\mu_{e}\rangle$ and $r_{e}$ are measured in the $R$ band and
$r_{e}$ is in units of kpc. This relation can be translated to other photometric
bands using standard colors of ellipticals from \citet{1995+Fukugita}. This
yields 101 pseudo bulges.  Interestingly, only one of the pseudo bulges resides
in an S0 galaxy---NGC~4802. Its bulge is dusty but overall blue, signaling
recent star formation.

\subsection{Classical Bulges Are Not Necessarily Prominent}
\label{sec:small-but-dense}

\begin{figure*}
  \epsscale{1.18}
  \plotone{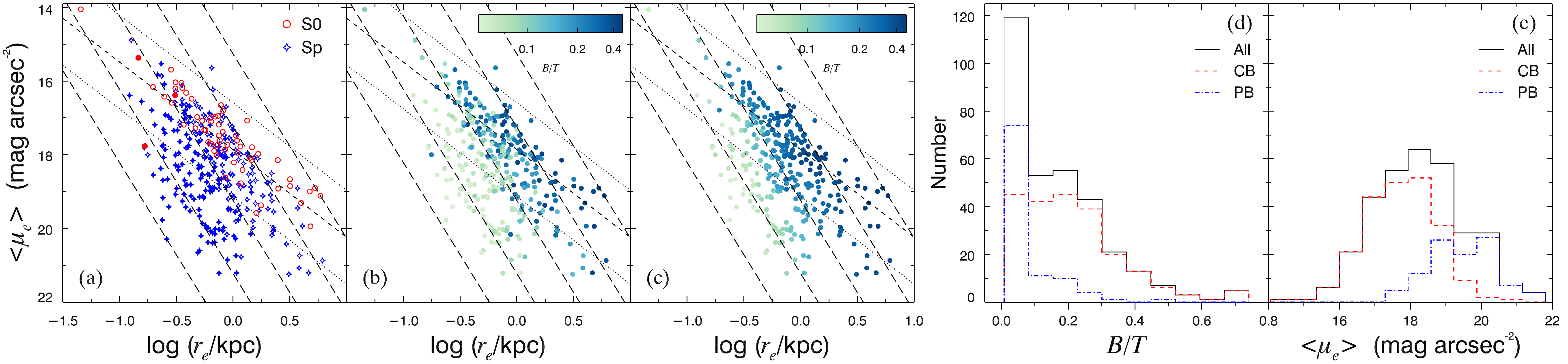}
  \caption{Bulges in the Kormendy relation: (a) distinguished by Hubble types,
    (b) color-coded according to their $B/T$, and (c) LOESS-smoothed version of
    panel~(b). The filled symbols in panel~(a) represent weak bulges ($B/T<0.1$)
    in S0s (red) and spirals (blue). The best-fit relation and its $3\,\sigma$
    boundary of the ellipticals are indicated by the short-dashed lines and
    dotted lines, respectively. The long-dashed lines approximately represent
    trajectories of constant bulge luminosities (ignoring cosmological effects).
    Panel~(d) displays the distribution of $B/T$ for the classical (red) and
    pseudo (blue) bulges; the overlap is significant. The two bulge types are
    better distinguished in surface brightness (e). \label{fig:KR_frac}}
\end{figure*}

Figure~\ref{fig:KR_frac} shows the distribution of $B/T$ in the Kormendy
relation (panel~b). To mitigate the effects of object overlay, we present a
version smoothed by locally weighted regression (LOESS;
\citealp{1988J.AM.STAT.ASSOC+Cleveland})\footnote{We adopt the implementation of
  the two-dimensional LOESS method by \citet{2013MNRAS+Cappellari2}, who provide
  the code in \url{http://purl.org/cappellari/idl}.} to highlight the general
trend (panel~c).  We find that bulges with constant $B/T$ approximately follow
lines of constant bulge luminosity, although, as expected, considerable
curvature is observed, considering that the host galaxies of pseudo bulges are
generally less massive and less luminous. We find that, along the direction of
the Kormendy relation, bulges form a sequence of increasing $B/T$, from small
$r_{e}$ to large $r_{e}$.  This again reinforces the point that both classical
and pseudo bulges can have small and large $B/T$, even if on average classical
bulges have larger $B/T$.

\begin{figure}
  \epsscale{1.17}
  \plotone{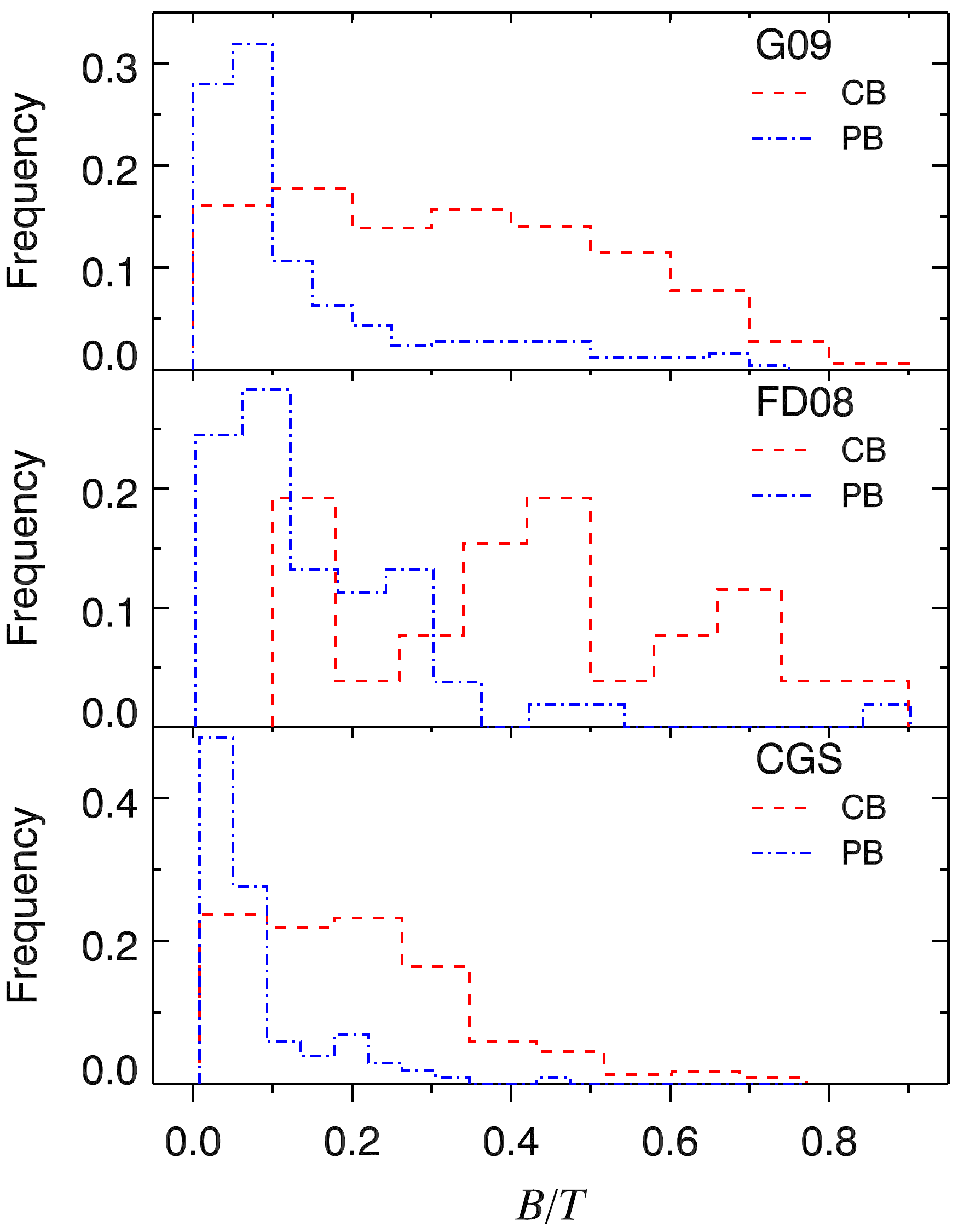}
  \caption{Distributions of $B/T$ for classical (red) and pseudo (blue) bulges
    in (top) \citetalias{2009MNRAS+Gadotti}, (middle)
    \citetalias{2008AJ+Fisher}, and (bottom) CGS. \label{fig:frac_comp}}
\end{figure}

\begin{figure}
  \epsscale{1.18}
  \plotone{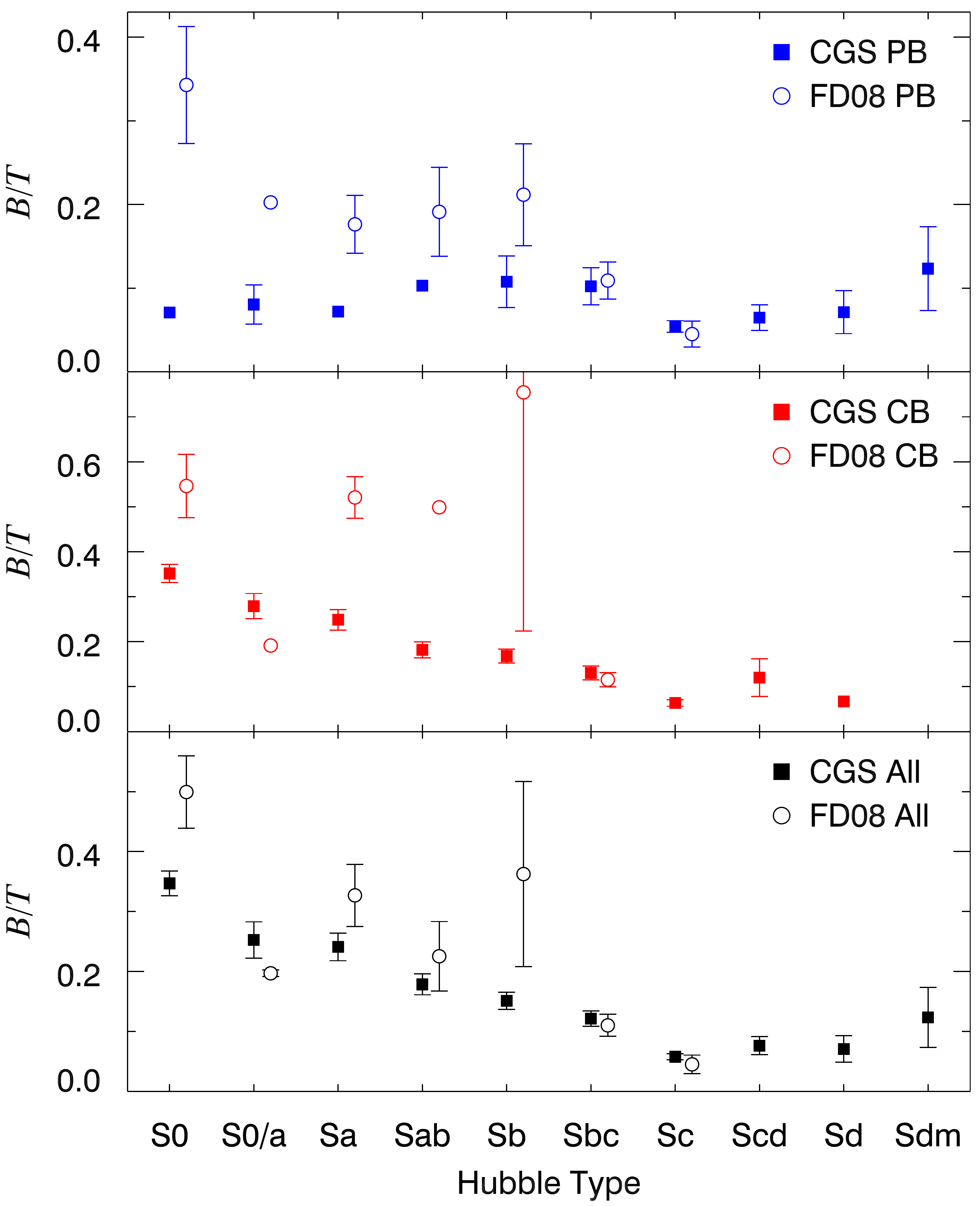}
  \caption{Distributions of $B/T$ as a function of Hubble type in this study
    (CGS) with those from \citetalias{2008AJ+Fisher} for (top) pseudo bulges,
    (middle) classical bulges, and (bottom) both bulge types combined. Symbols
    and error bars represent mean and error of the mean in each bin. Symbols are
    horizontally offset for clarity. \label{fig:frac_T_comp}}
\end{figure}

Pseudo bulges are generally weak, with 78\% of them having $B/T<0.1$
(Figure~\ref{fig:KR_frac}d). Meanwhile, classical bulges are not always strong;
26\% of them have $B/T<0.1$. The distributions of $B/T$ for classical and pseudo
bulges significantly overlap. The mean $B/T$ of classical bulges is 0.21, not
dramatically larger than that of pseudo bulges ($\langle B/T \rangle =
0.08$). \citet[][hereafter \citetalias{2008AJ+Fisher}]{2008AJ+Fisher} and
\citet[][hereafter \citetalias{2009MNRAS+Gadotti}]{2009MNRAS+Gadotti} also see
such an overlap, but in general their values of $B/T$ , both of classical and
pseudo bulges, are significantly larger than ours (Figure~\ref{fig:frac_comp}).
This systematic difference repeatedly shows up when comparing bulge properties
derived from two-dimensional (2D) image fitting and one-dimensional (1D) surface
brightness profile fitting (see Section~5 of \citealp{2019ApJS+Gao}). In
Figure~\ref{fig:frac_T_comp}, we illustrate this effect by comparing our $B/T$
values with those of \citetalias{2008AJ+Fisher} at fixed Hubble
type. \citetalias{2008AJ+Fisher} find systematically larger $B/T$ than we do in
disks of early to intermediate type, while their values are generally consistent
with ours for Hubble types later than Sb. The difference is most dramatic in
early-type disks.\footnote{The S0/a bin is an exception probably because it
  contains only two galaxies.} The trend holds true for both classical and
pseudo bulges.

We speculate that the systematic discrepancies between our results and those of
\citetalias{2008AJ+Fisher} are a direct consequence of the different fitting
techniques applied to perform the bulge decomposition and the different input
models used. We believe that our 2D fits provide more robust bulge parameters
because our multi-component models properly separate the bulge from surrounding
and overlapping substructures such as bars and lenses. The 1D approach of
\citetalias{2008AJ+Fisher} explicitly masks such substructures during the
fitting. As shown by \citet{2017ApJ+Gao}, this introduces large uncertainties in
bulge parameters, especially in early-type disks where such substructures are
most prevalent.  We find better agreement with \citetalias{2009MNRAS+Gadotti}'s
results, presumably because his 2D decomposition technique is similar to ours,
as is his criteria for classifying bulges. Nevertheless, the average bulge
strength found by Gadotti ($\langle B/T \rangle=0.33$ for classical bulges;
$\langle B/T \rangle=0.15$ for pseudo bulges) is also larger than ours
(Figure~\ref{fig:frac_comp}). Although Gadotti's 2D method is certainly more
trustworthy than 1D methods, and it does systematically treat bars, our 2D
method (\galfit{}; \citealp{2002AJ+Peng, 2010AJ+Peng}) can handle a vastly more
intricate array of internal substructures
\citep{2017ApJ+Gao,2018ApJ+Gao,2019ApJS+Gao}, which ultimately leads to more
robust bulge parameters. \citet{2016ASSL+Kormendy}'s statement that almost all
pseudo bulges have $B/T\la 0.35$ and $B/T\ga 0.5$ guarantees that the bulge is
classical still applies to our measurements, but it is not representative
because $B/T\la 0.2$ already includes most pseudo bulges and few bulges of
either type have $B/T\ga 0.5$.  We echo previous works
\citep[e.g.,][]{2017A&A+Neumann} that stress that $B/T$ is not a good parameter
to distinguish between classical and pseudo bulges. The two bulge types simply
overlap too heavily in $B/T$. The implication of such overlap at low $B/T$ will
be discussed in Section~\ref{sec:violent-vs.-secular}.

Although classical bulges are not necessarily prominent, they do mostly have
high surface brightnesses (Figure~\ref{fig:KR_frac}e), a direct outcome of their
classification based on the Kormendy relation. It is noteworthy that the
distribution of $\langle\mu_{e}\rangle$ shows the strongest separation between
classical and pseudo bulges among all the structural parameters of bulges. A
demarcation line at $\sim$18.5\,mag~arcsec$^{-2}$ can isolate most of the
classical bulges from the pseudo bulges. If any single bulge structural
parameter should be used to classify classical and pseudo bulges,
$\langle\mu_{e}\rangle$ would be the best choice to produce consistent
classifications based on the Kormendy relation.  Adopting a boundary of
$\langle\mu_{e}\rangle = 18.5\,\mathrm{mag~arcsec^{-2}}$ (in the $R$ band) would
correctly classify 87\% of the pseudo bulges and 77\% of the classical bulges.

S0 galaxies exhibit a wide range of bulge prominence, from traditionally
bulge-dominated systems to those with bulges as weak as those in spirals of type
Sc and later \citep{2018ApJ+Gao}. However, in stark contrast to late-type
spirals, which host mostly pseudo bulges , the bulges of S0s essentially {\it all}\
form a uniform sequence with the ellipticals on the Kormendy relation. In other
words, almost all bulges in CGS S0s are classical. Based on this observation,
\citet{2018ApJ+Gao} argue that, unless historically popular processes such as
ram-pressure stripping can modify bulge structures substantially, most S0s
cannot simply be descendants of faded late-type spirals (later than Sc). The
present study further reinforces the conclusions of \citet{2018ApJ+Gao}. S0
bulges, independent of their $B/T$, occupy almost exclusively the upper envelope
of the $\langle\mu_{e}\rangle - r_e$ plane. The weakest bulges found in S0s
($B/T \lesssim 0.1$) have surface brightnesses at least 2~magnitudes brighter
than the majority of the comparably weak bulges in late-type spirals
(Figure~\ref{fig:KR_frac}a).


\subsection{Bimodality or Continuity in \sersic{} Indices}
\label{sec:pse-n2}

\begin{figure*}
  \epsscale{1.15}
  \plotone{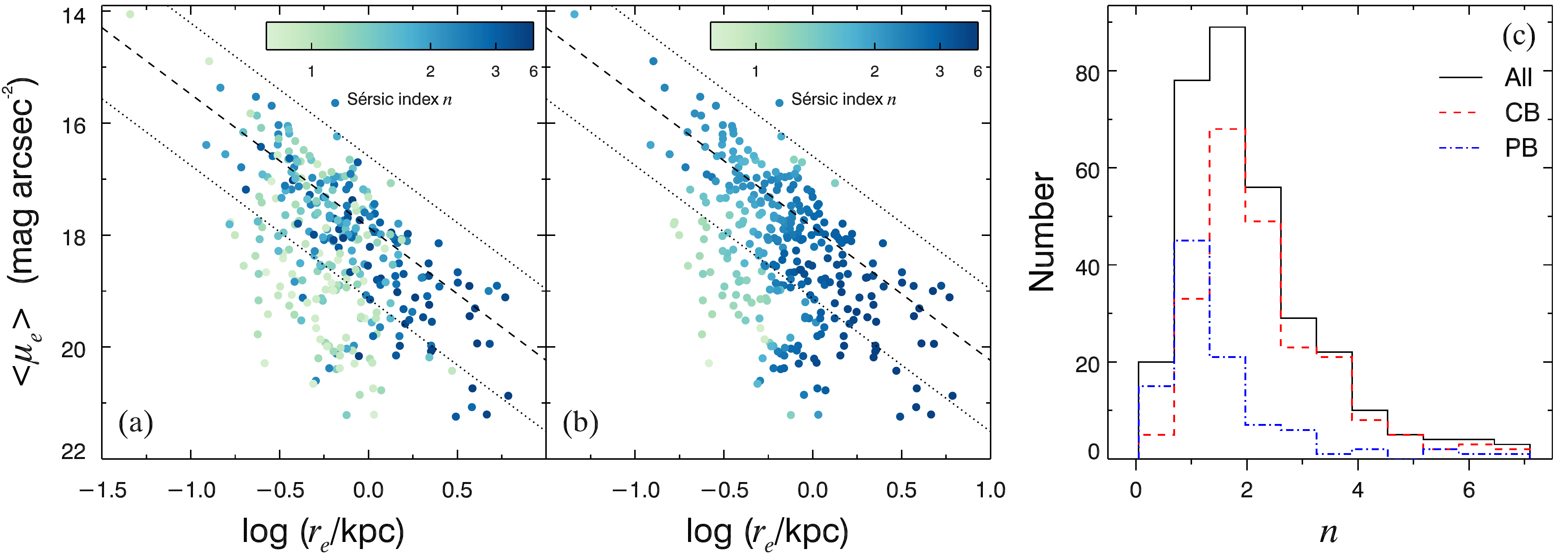}
  \caption{(a) Bulges in the Kormendy relation color-coded according to their
    \sersic{} indices. Panel (b) is the same as panel (a), but the data have 
    been LOESS-smoothed. The best-fit relation and its $3\,\sigma$ boundary of 
    the ellipticals are indicated by short-dashed lines and dotted lines,
    respectively. Panel (c) shows the distribution of \sersic{} indices for the
    classical (red) and pseudo (blue) bulges. The overlap is significant, and 
    the bimodality presented in \citetalias{2008AJ+Fisher} is not observed here.
    \label{fig:KR_nser}}
\end{figure*}

Figure~\ref{fig:KR_nser} shows the distribution of CGS bulges on the Kormendy
relation, with symbols color-coded according to their \sersic{} indices. We show
both the original data (panel~a) and the LOESS-smoothed version (panel~b) to
facilitate visualization of the general trend. Along the direction of the
Kormendy relation, bulges in general form a sequence of increasing \sersic{}
indices, from small to large $r_{e}$. This hints that both classical and pseudo
bulge can have small and large \sersic{} indices, even if on average classical
bulges have larger values. As with \citetalias{2009MNRAS+Gadotti}, we find that
a minority of the pseudo bulges have \sersic{} indices larger than 2.
Meanwhile, a substantial fraction of classical bulges have $n<2$ (panel~c).
Although on average pseudo bulges have smaller \sersic{} indices than classical
bulges, we do {\it not}\ observe the bimodal distribution of \sersic{} indices
reported in \citetalias{2008AJ+Fisher}, which was the empirical basis of their
\sersic{} $n$-based classification criterion. \citetalias{2008AJ+Fisher}
identified pseudo bulges mainly based on their nuclear morphologies (presence of
nuclear bars, rings, and spirals). However, the recent analysis of a small
sample of double-barred galaxies by \citet{2019MNRAS+de_Lorenzo-Caceres} showed
that most of the underlying bulges are classical. Moreover,
\citet{2017MNRAS+Tabor} found some pressure-supported bulges in S0s with
$n \approx 1$. In the same vein, \citet{2018MNRAS+Mendez-Abreu} discuss the lack
of clear correspondence between bulge \sersic{} index and their
kinematics. \citet{2018MNRAS+Costantin} tested various observational diagnostics
to separate classical and pseudo bulges and also concluded that \sersic{} index
is disfavored. Although \sersic{} indices do carry physically significant
information \citep[e.g., mergers result in higher \sersic{}
indices:][]{2001A&A+Aguerri,2006A&A+Eliche-Moral, 2016ASSL+Brooks}, their large
measurement error \citep{2008MNRAS+Gadotti} and sensitivity to spatial
resolution \citep{2003ApJ+Balcells} and nuclear fine structures
\citep{2017ApJ+Gao} may hamper their application in classifying bulges. Combined
with the discrepant classification results mentioned above, we suggest that
$n=2$ is not an appropriate demarcation line for separating classical and pseudo
bulges. As with \citet{2017A&A+Neumann}, we find that $n=1.5$ is a better
criterion; based on the statistics of CGS, this revised criterion would
correctly classify 65\% of the pseudo bulges and 74\% of the classical bulges.
Note that even with this modified threshold \sersic{} $n$ is still a less
effective classifier of bulge type than $\langle\mu_{e}\rangle$
(Section~\ref{sec:small-but-dense}).

\begin{figure}
  \epsscale{1.15}
  \plotone{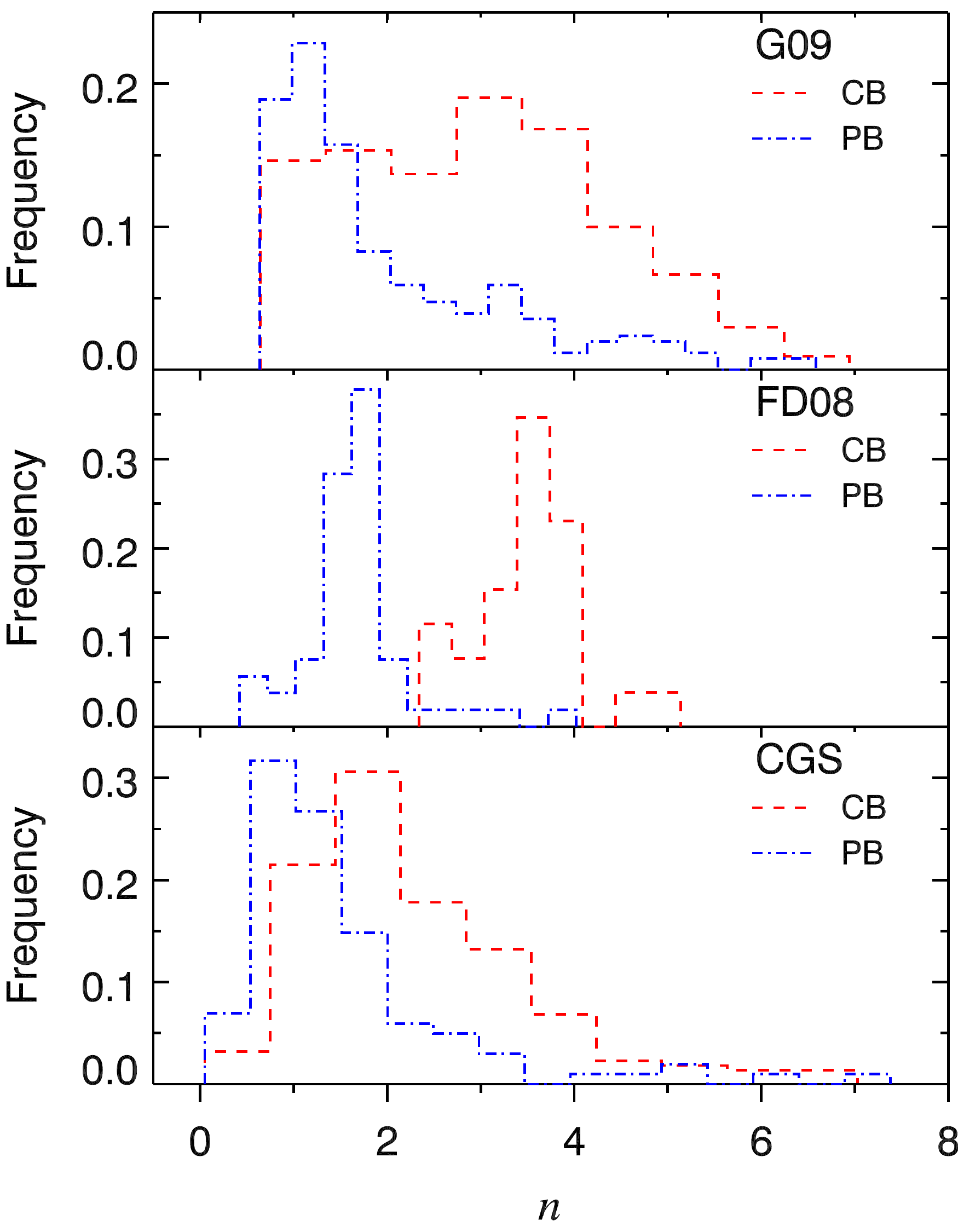}
  \caption{Distributions of \sersic{} $n$ for classical (red) and pseudo (blue)
    bulges in (top) \citetalias{2009MNRAS+Gadotti}, (middle)
    \citetalias{2008AJ+Fisher}, and (bottom) CGS. \label{fig:nser_comp}}
\end{figure}

\begin{figure}
  \epsscale{1.18}
  \plotone{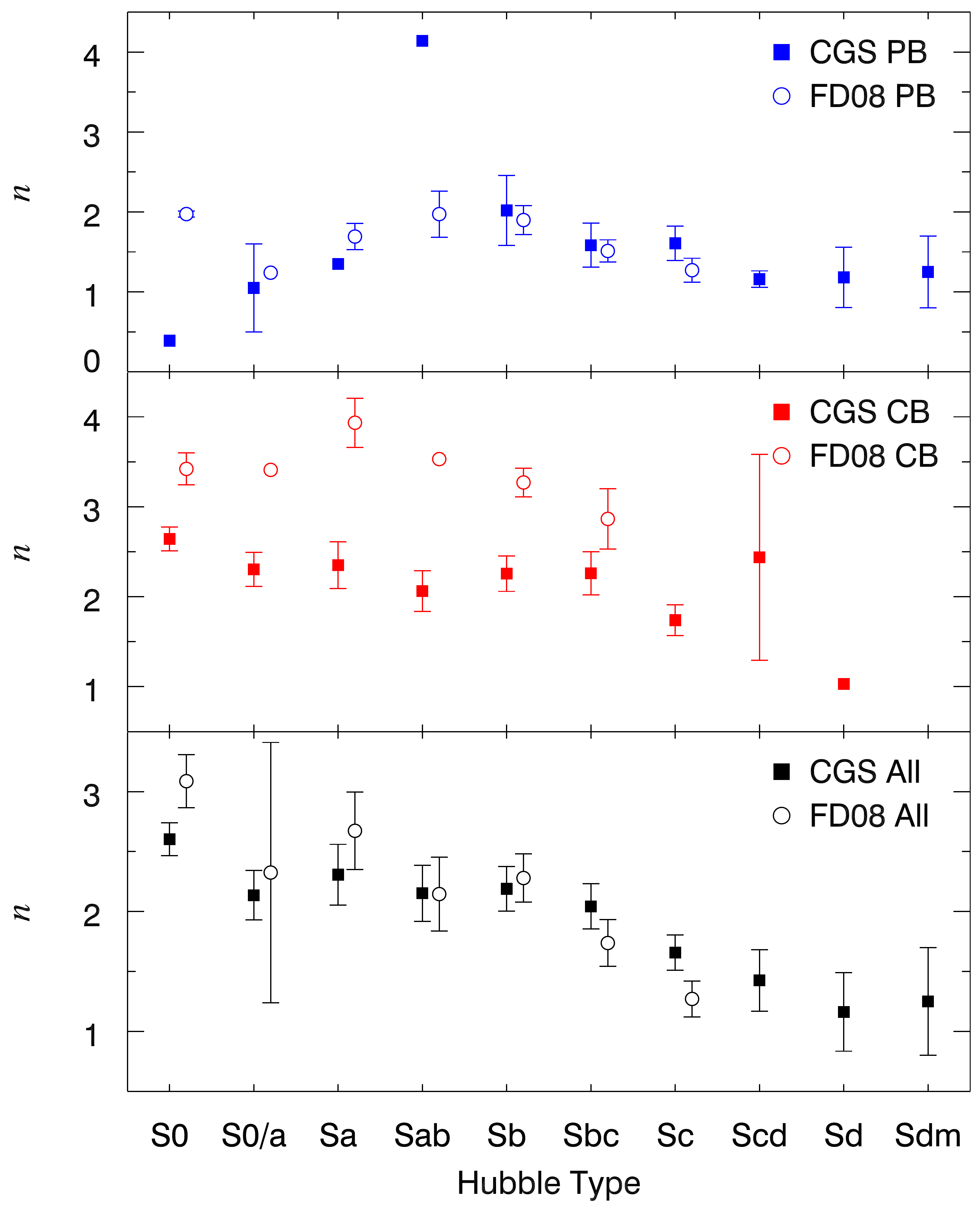}
  \caption{Distributions of bulge \sersic{} $n$ as a function of Hubble type in
    this study (CGS) with those from \citetalias{2008AJ+Fisher} for (top) pseudo
    bulges, (middle) classical bulges, and (bottom) both bulge types
    combined. Symbols and error bars represent the mean and error of the mean in
    each bin. Symbols are horizontally offset for clarity.    
    \label{fig:nser_T_comp}}
\end{figure}

We also compare the distribution of bulge \sersic{} $n$ from CGS with those from
\citetalias{2008AJ+Fisher} and \citetalias{2009MNRAS+Gadotti}
(Figure~\ref{fig:nser_comp}). It is evident that both our measurements and those
of \citetalias{2009MNRAS+Gadotti} show significant overlap in $n$ for classical
and pseudo bulges, while in \citetalias{2008AJ+Fisher} the two bulge types are
well separated at $n\approx 2$. In contrast to the comparison of $B/T$ at fixed
Hubble type (Figure~\ref{fig:frac_T_comp}), Figure~\ref{fig:nser_T_comp} shows
that FD08's measurements show no noticeable systematic difference in bulge
\sersic{} $n$ at fixed Hubble type for all bulges. This is also true for pseudo
bulges. However, our values of $n$ are systematically larger than those in
\citetalias{2008AJ+Fisher} for classical bulges in galaxies of all Hubble types.


\subsection{Classical and Pseudo Bulges Share Similar Relative Bulge Size}
\label{sec:bul-size}

\begin{figure}
  \epsscale{1.17}
  \plotone{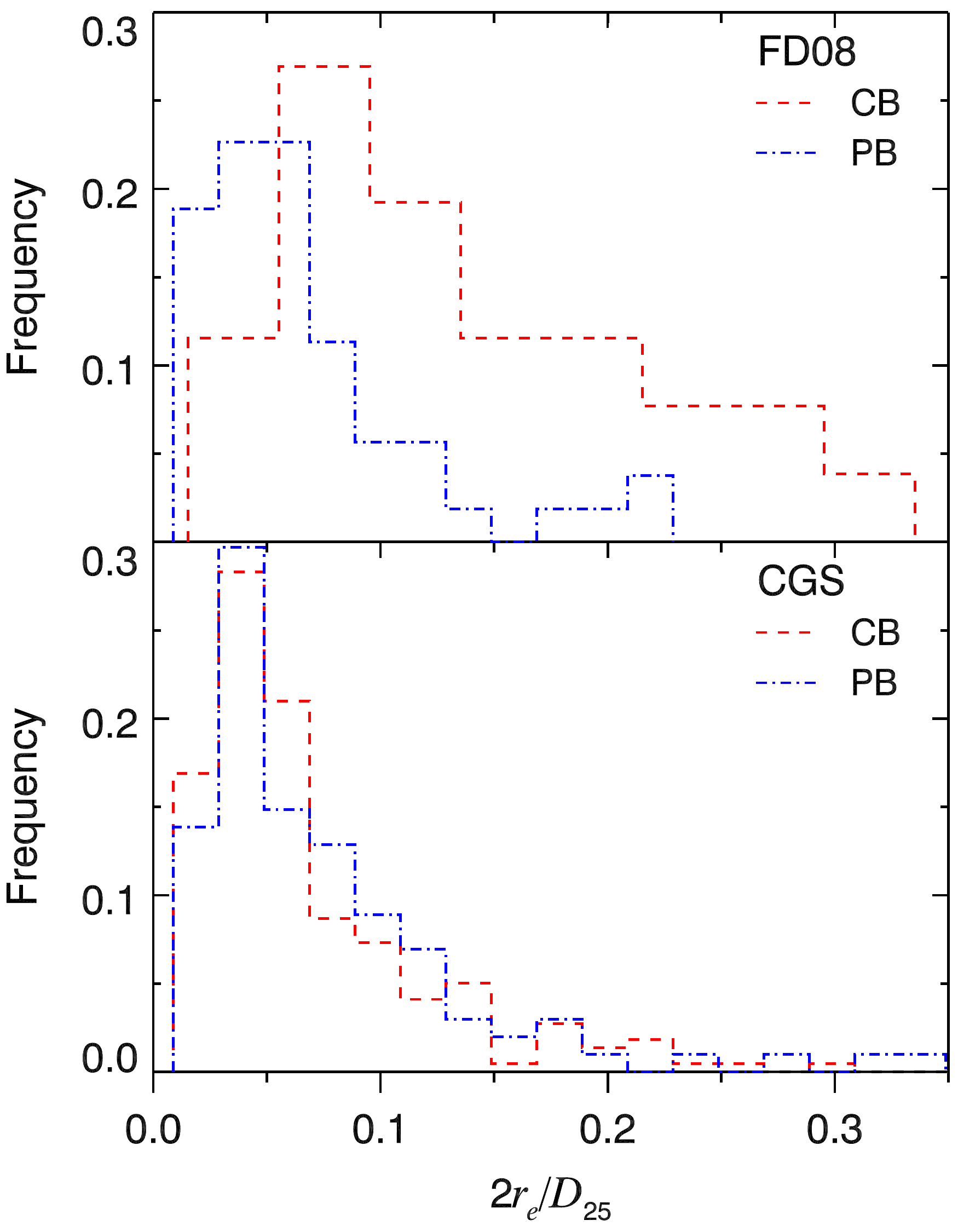}
  \caption{Distributions of relative size of classical (red) and pseudo (blue)
    bulges in (top) \citetalias{2008AJ+Fisher} and (bottom) CGS. The bulge
    effective radius $r_e$ is normalized to the $D_{25}$ isophotal diameter of
    the galaxy.\label{fig:re_comp}}
\end{figure}

\begin{figure*}
  \epsscale{1.17}
  \plotone{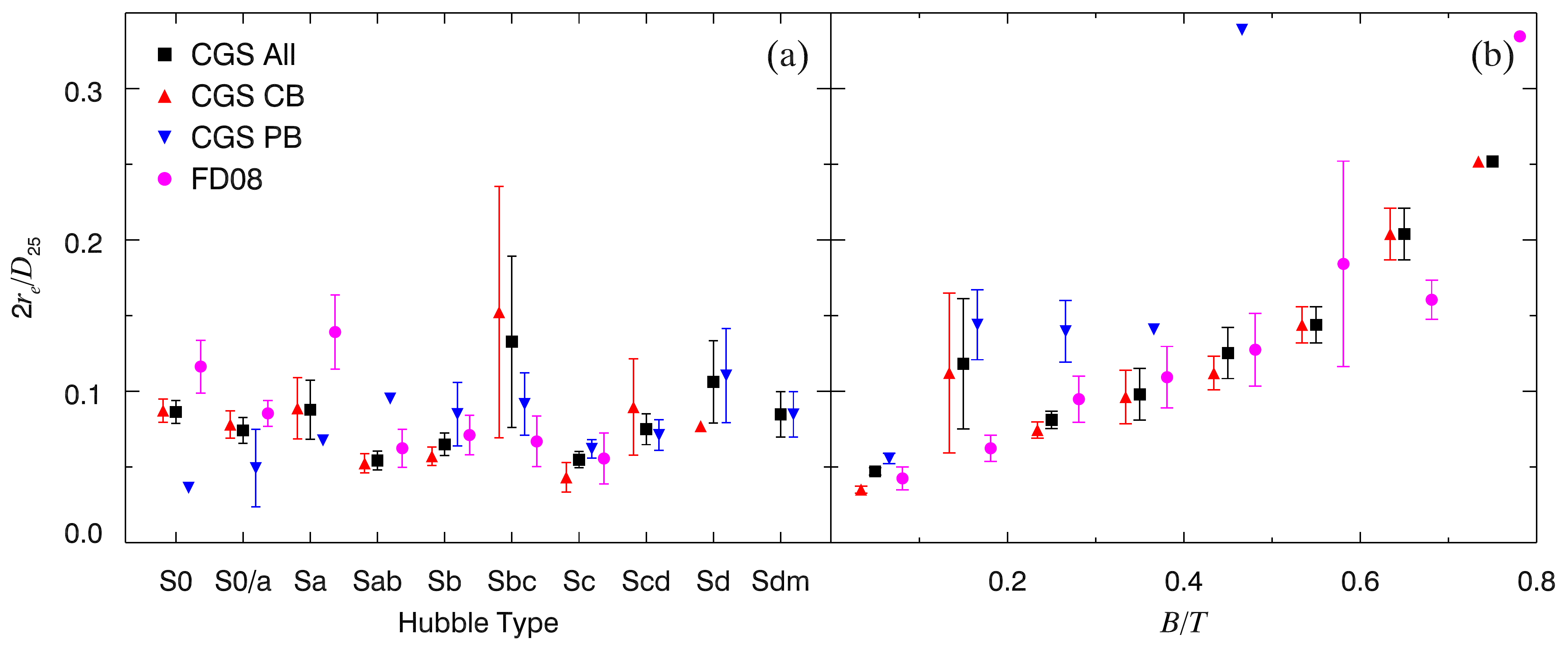}
  \caption{Relative size of the bulge as a function of (a) Hubble type and (b)
    $B/T$ for all CGS bulges (black), CGS classical bulges (red), CGS pseudo
    bulges (blue), and bulges in \citetalias{2008AJ+Fisher} (magenta). The bulge
    effective radius $r_e$ is normalized to the $D_{25}$ isophotal diameter of
    the galaxy. Symbols and error bars represent mean and error of the mean in
    each bin, respectively. Symbols are horizontally offset for clarity.
    \label{fig:re_T_frac}}
\end{figure*}

Our decomposition demonstrates that classical and pseudo bulges have similar
relative bulge size ($2r_{e}/D_{25}$), contrary to the results of
\citetalias{2008AJ+Fisher}, who found that classical bulges have significantly
larger relative sizes than pseudo bulges (Figure~\ref{fig:re_comp}). We use
$D_{25}$ from \citet{2011ApJS+Ho} for CGS galaxies and retrieve $D_{25}$ of the
sample in \citetalias{2008AJ+Fisher} from
HyperLeda\footnote{\url{http://leda.univ-lyon1.fr/}} \citep{2003A&A+Paturel}. To
investigate the origin of the discrepant results, we plot $2r_{e}/D_{25}$ as a
function of Hubble type and $B/T$ in Figure~\ref{fig:re_T_frac}. We find that at
fixed Hubble type or $B/T$, bulges in CGS and \citetalias{2008AJ+Fisher} have
similar mean relative bulge size, which increases as $B/T$ increases; except for
early-type disks (S0 and Sa), \citetalias{2008AJ+Fisher}'s measurements are
systematically larger than ours. It is also noteworthy that in the case of CGS,
at a fixed $B/T$, pseudo bulges are more diffuse (larger $2r_{e}/D_{25}$) than
classical bulges. It is evident in Figure~\ref{fig:frac_comp} that the
separation of $B/T$ of classical and pseudo bulges in CGS is much less
significant than in \citetalias{2008AJ+Fisher}.  Therefore, we speculate that
the discrepant bulge relative size distributions of classical and pseudo bulges
between CGS and \citetalias{2008AJ+Fisher} is mainly due to the discrepant $B/T$
measurements for the two populations of bulges.

\subsection{Bulge Dichotomy in Their Intrinsic Shapes}
\label{sec:bulge-shape}

\begin{figure}
  \epsscale{1.17}
  \plotone{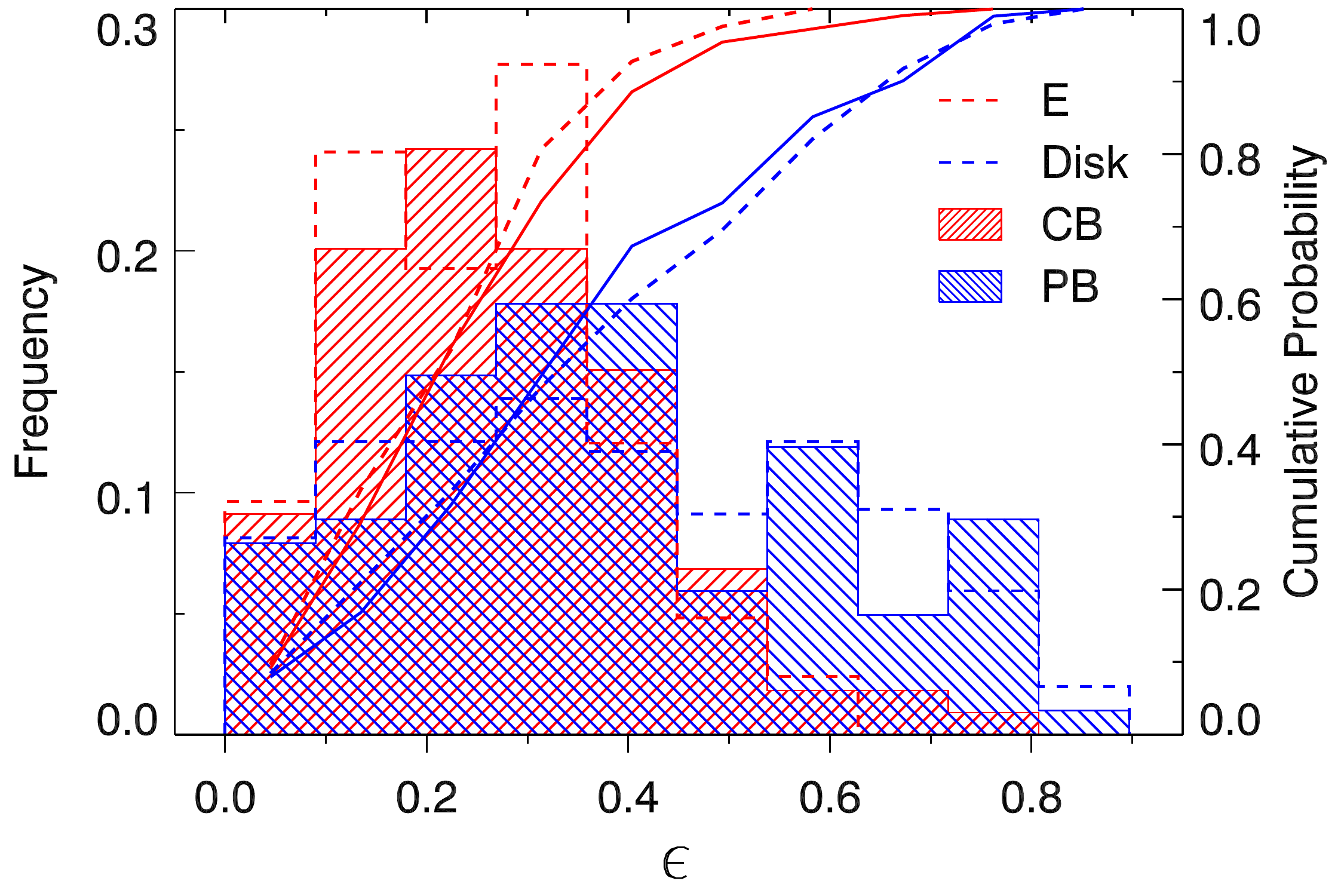}
  \caption{Distributions of apparent ellipticities of classical (red) and pseudo
    (blue) bulges, CGS ellipticals (red dashed), and CGS disks galaxies (blue
    dashed). The cumulative probability of each subgroup is overlaid with the 
    same line style. \label{fig:ell_comp}}
\end{figure}

Intrinsic shapes of bulges have been extensively investigated by
\citet{2008A&A+Mendez-Abreu,2010A&A+Mendez-Abreu}, \citet{2017A&A+Costantin,
  2018MNRAS+Costantin}, and \citet{2019MNRAS+de_Lorenzo-Caceres}. Without
constraints from stellar kinematics, however, it is extremely difficult to
measure the intrinsic shapes of individual bulges. Instead of looking into the
intrinsic shapes, we turn to examining the statistics of their intrinsic
flattenings.  Figure~\ref{fig:ell_comp} presents the apparent ellipticity
($\epsilon$) distributions of classical and pseudo bulges. Classical bulges are
on average apparently rounder than pseudo bulges. The two distributions are
statistically different. A Kolmogorov-Smirnov test rejects the null hypothesis
that the two samples are drawn from the same population at a probability of
$P_{\rm null}\approx 10^{-4}$. The distribution of the apparent
ellipticities of classical bulges is skewed markedly toward small ellipticities,
while pseudo bulges have a broader distribution, reminiscent of the difference
between apparent ellipticities of disks and spheroids \citep{1970ApJ+Sandage}.
Overlaying the apparent ellipticity distributions of CGS ellipticals and disk
galaxies from \citet{2011ApJS+Ho} reveals that classical bulges have similar
apparent ellipticities with ellipticals while the apparent ellipticities of
pseudo bulges closely resemble those of disk galaxies. This is formally
confirmed from Kolmogorov-Smirnov tests of the four populations:
$P_{\rm null}=0.51$ for classical bulges vs. ellipticals;
$P_{\rm null}=0.27$ for pseudo bulges vs. disk galaxies.
Classical bulges are intrinsically as round as ellipticals, and pseudo bulges
are intrinsically as flattened as disks. This is consistent with the picture
that classical bulges were formed in a manner similar to that of ellipticals,
while pseudo bulges originate from the gradual accumulation of disk material,
having preserved enough memory of their disky origin to be recognized. However,
as the apparent ellipticities of the two populations overlap with each other due
to projection effects, we do not suggest using ellipticity as a criterion to
separate classical and pseudo bulges.

\subsection{Dichotomies in Classical and Pseudo Bulge Hosts}
\label{sec:dich-hosts}

\begin{figure*}
  \epsscale{1.17}
  \plotone{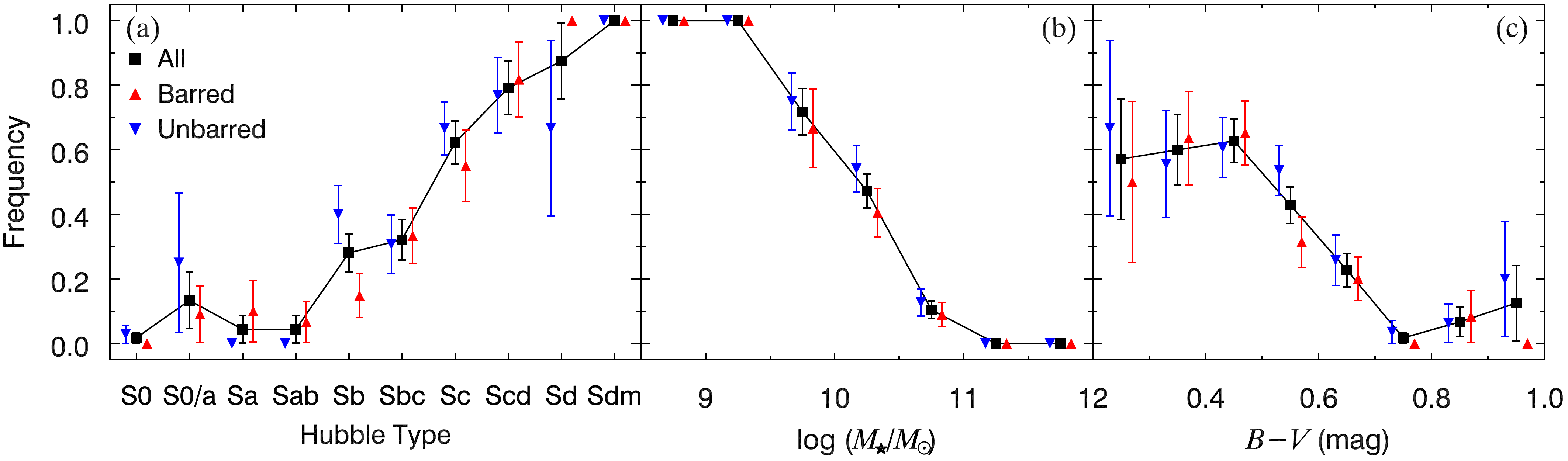}
  \caption{Frequency of pseudo bulges as a function of the (a) Hubble type, (b)
    stellar mass, and (c) optical color of the host galaxy. Black symbols with
    connected lines represent all galaxies, and red and blue symbols are barred
    and unbarred galaxies, respectively. Error bars are calculated assuming a
    binomial distribution. Symbols are horizontally offset for
      clarity. \label{fig:PB_T_mass_col}}
\end{figure*}

Section~\ref{sec:bulge-shape} shows that pseudo bulges are basically miniature
disks. As disks are vulnerable to violent processes such as mergers, pseudo
bulges are most likely to be found in late-type spirals, whose evolutionary
history has been driven more by secular evolution than major mergers.
Figure~\ref{fig:PB_T_mass_col} examines the dependence of pseudo bulge incidence
on Hubble type, stellar mass $M_{\star}$, and \bv{} color. As expected,
pseudo bulges are more frequent in galaxies with later Hubble types, lower
stellar masses, and bluer optical colors. We also contrast the pseudo bulge
incidence in barred and unbarred galaxies, but find no systematic differences
between the two subsamples. This is counterintuitive, because bars are often
thought to be the main driver of secular evolution. We will elaborate on this
point in the next section.

In broad agreement with the literature, we also find that pseudo bulges prefer
to reside in late-type spirals. Spirals with Hubble types later than Sbc are
indeed the most active manufacturing sites of pseudo bulges. At least $\sim60\%$
of them host pseudo bulges. However, our pseudo bulge incidence in most Hubble
types is significantly smaller than previously found. \citet{2004ARA&A+Kormendy}
summarized the statistics of pseudo bulges as a function of Hubble type from the
literature. Based on morphologies and \sersic{} indices of 75 galaxies with
\textit{Hubble Space Telescope (HST)} images analyzed by \citet{1997AJ+Carollo},
they found that 69\%, 50\%, 22\%, 11\%, and 0\% of S0--Sa, Sab, Sb, Sbc, and Sc
and later type galaxies host classical bulges, respectively.  The general trend
of increasing incidence of pseudo bulges toward late Hubble types is consistent
with our measurements. However, their pseudo bulge fraction is higher than ours
(Table~\ref{tab:frac_PB}). The same holds true for the work of
\citet{2004ARA&A+Kormendy}, who, re-analyzing the sample of
\citet{2002AJ+Carollo} based on \textit{HST} $V$-band and $H$-band images,
concluded that 50\%, 40\%, 56\%, 94\%, and 100\% of S0+Sa, Sab, Sb, Sbc, and
Sc--Sm galaxies host exponential bulges. Exponential bulges are considered a
close proxy for pseudo bulges.  Taken at face value, their pseudo bulge fraction
is again systematically higher than ours. We do not know the exact reason of the
quantitative discrepancy, as differences in bulge-to-disk decomposition
techniques, wavelength effects, spatial resolution of the data, sample size (CGS
is significantly larger than the above-cited samples), and, more directly, the
classification criteria affect the outcome.

\begin{deluxetable*}{cCCCCCCCCCC}
  \tablecaption{Percentage of Pseudo Bulge Hosts in Bins of Hubble Type and
    Stellar Mass\label{tab:frac_PB}}
  
  \tablehead{\colhead{Hubble Type} & \colhead{S0} & \colhead{S0/a}
    &\colhead{Sa} & \colhead{Sab} & \colhead{Sb} & \colhead{Sbc} & \colhead{Sc}
    & \colhead{Scd} & \colhead{Sd} & \colhead{Sdm}}

  \startdata
  All & 2\pm 2 & 13\pm 9 & 4\pm 4 & 4\pm 4 & 28\pm 6 & 32\pm 6 &
  62\pm 7 & 79\pm 8 & 88\pm 12 & 100\pm 0 \\
  Barred & 0\pm 0 & 9\pm 9 & 10\pm 10 & 7\pm 6 & 15\pm 7 & 33\pm 9 &
  55\pm 11 & 82\pm 12 & 100\pm 0 & 100\pm 0 \\
  Unbarred & 3\pm 3 & 25\pm 22 & 0\pm 0 & 0\pm 0 & 40\pm 9 & 31\pm 9 & 67\pm 8 &
  77\pm 12 & 67\pm 27 & 100\pm 0 \\
  \hline\hline \\[-0.38cm]
  $\log M_{\star}/M_{\sun}$ & \multicolumn2c{8.5--9.5} &
  \multicolumn2c{9.5--10.0} & \multicolumn2c{10.0--10.5} &
  \multicolumn2c{10.5--11.0} & \multicolumn2c{11.0--12.0} \\[0.08cm]
  \hline All & \multicolumn2c{$100\pm 0$} & \multicolumn2c{$72\pm 7$} &
  \multicolumn2c{$47\pm 5$} & \multicolumn2c{$10\pm 3$} &
  \multicolumn2c{$0\pm 0$} \\
  Barred & \multicolumn2c{$100\pm 0$} & \multicolumn2c{$67\pm 12$} &
  \multicolumn2c{$41\pm 8$} & \multicolumn2c{$9\pm 4$} & \multicolumn2c{$0\pm
    0$} \\
  Unbarred & \multicolumn2c{$100\pm 0$} & \multicolumn2c{$75\pm 9$} &
  \multicolumn2c{$54\pm 7$} & \multicolumn2c{$13\pm 4$} &
  \multicolumn2c{$0\pm 0$}
  \enddata
\end{deluxetable*}

\subsection{Pseudo Bulges and Their Relation to Bars}
\label{sec:pse-pre-host-bar}

\begin{figure}
  \epsscale{1.15}
  \plotone{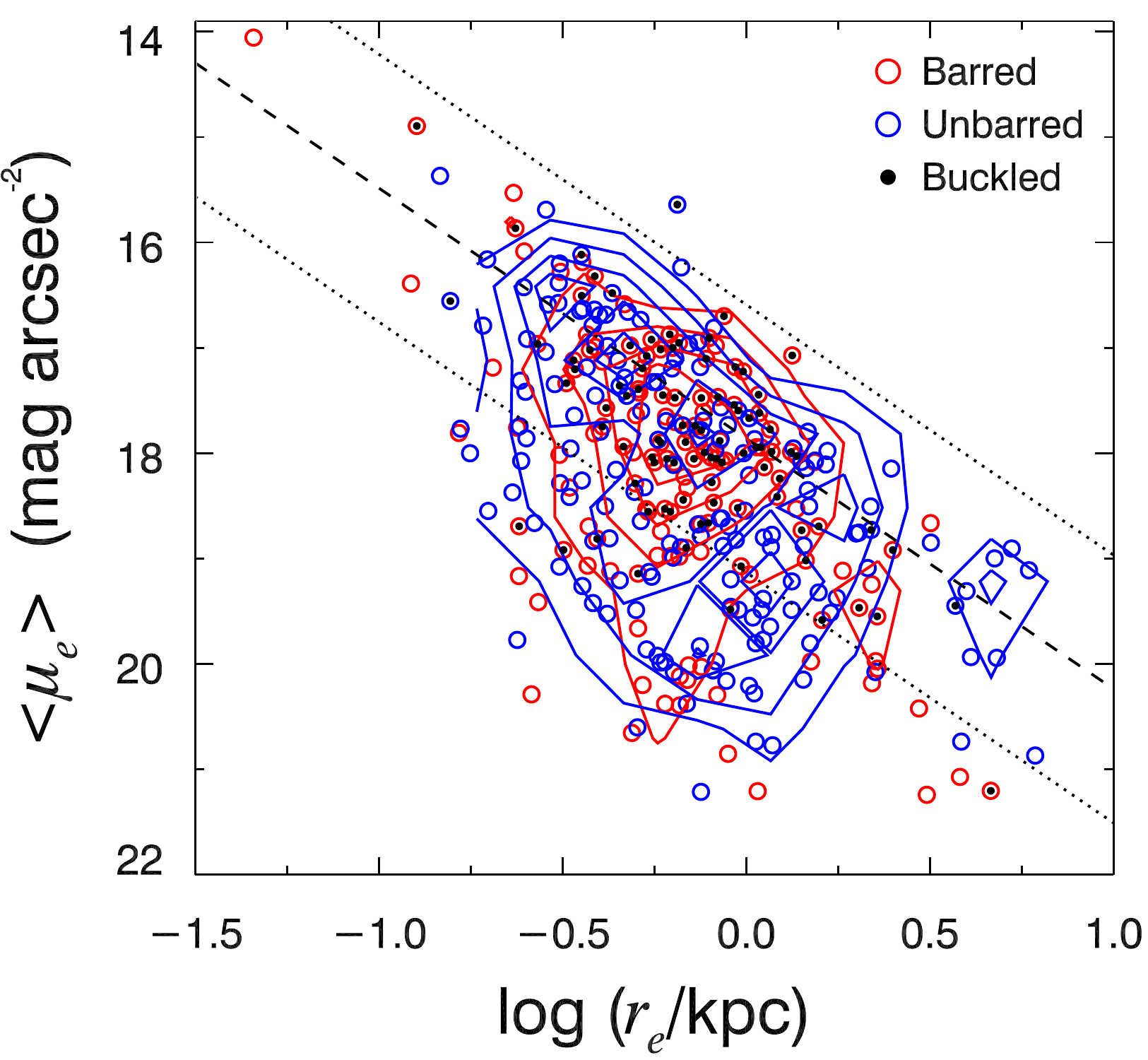}
  \caption{Bulges on the Kormendy relation for barred (red) and unbarred (blue)
    galaxies. The buckled bars recognized by \citet{2017ApJ+Li} are marked by
    solid black dots; due to different strategy of identifying barred galaxies,
    some of the buckled bars are designated unbarred according to the
    classification of this paper. The best-fit relation and its $3\,\sigma$
    boundary of the ellipticals are indicated by short-dashed lines and dotted
    lines, respectively. The barred and unbarred galaxies do not show any
    preference to host classical/pseudo bulges. Most of the boxy/peanut bulges
    follow the Kormendy relation of the ellipticals. \label{fig:KR_bar}}
\end{figure}

In Section~\ref{sec:dich-hosts} we already noted that barred galaxies do not
host more pseudo bulges than unbarred galaxies at fixed Hubble type, stellar
mass, or optical color. This is unexpected, if bars drive secular evolution, as
is often surmised \citep[e.g.,][]{2004ARA&A+Kormendy,2012MNRAS+Wang,2017ApJ+Lin,
  2019MNRAS+Chown}. Figure~\ref{fig:KR_bar} investigates this issue in the
context of the Kormendy relation.  Again, we find no indication that the
presence of a bar has any effect on skewing objects below the Kormendy relation
of the ellipticals (i.e., toward the locus of pseudo bulges). This seems to be
in conflict with the notion that bars help drive gas inflow and to build pseudo
bulges. However, one should bear in mind that bars can also contribute to bulge
growth through dissipationless processes, most notably via the buckling
instability. The difference in formation physics between bar buckling
instability and bar-driven gas inflow motivated \citet{2005MNRAS+Athanassoula}
to advocate that boxy/peanut bulges should be regarded as a separate class from
other pseudo bulges. The bulge-like structure produced by bar buckling
instability is not necessarily weak (small $B/T$) or disky (small $n$). This is
illustrated in Figure~\ref{fig:KR_bar}, where the bulges of buckled barred
galaxies occupy the locus of classical bulges with large $n$ ($\ga2$) and $B/T$
($\ga0.2$). The classifications of the buckled bars come from
\citet{2017ApJ+Li}, who visually identified boxy/peanut bulges and barlenses,
both of which are regarded as the same phenomena viewed from different
inclination angles, based on variations of isophotal shapes in the bar region.
We conclude that the Kormendy relation mostly only selects pseudo bulges that
are formed from dissipative secular processes.  We do not know whether this is
because of the inherent limitations of the Kormendy relation---given its
substantial intrinsic scatter---or because bulges with buckled bar features are
not genuine pseudo bulges. We observed a similar phenomenon before
\citep{2018ApJ+Gao}. The bulges of some S0 galaxies bear the appearance of
pseudo bulges but are not recognized as such on the Kormendy relation.  Disky
features do not reliably signify the overall photometric structure or the star
formation activities of the bulge. Given that buckled bars occur most frequently
in early-type disks, where mergers were once operative, they may not dominate
the overall bulge structure.

\begin{figure}
  \epsscale{1.19}
  \plotone{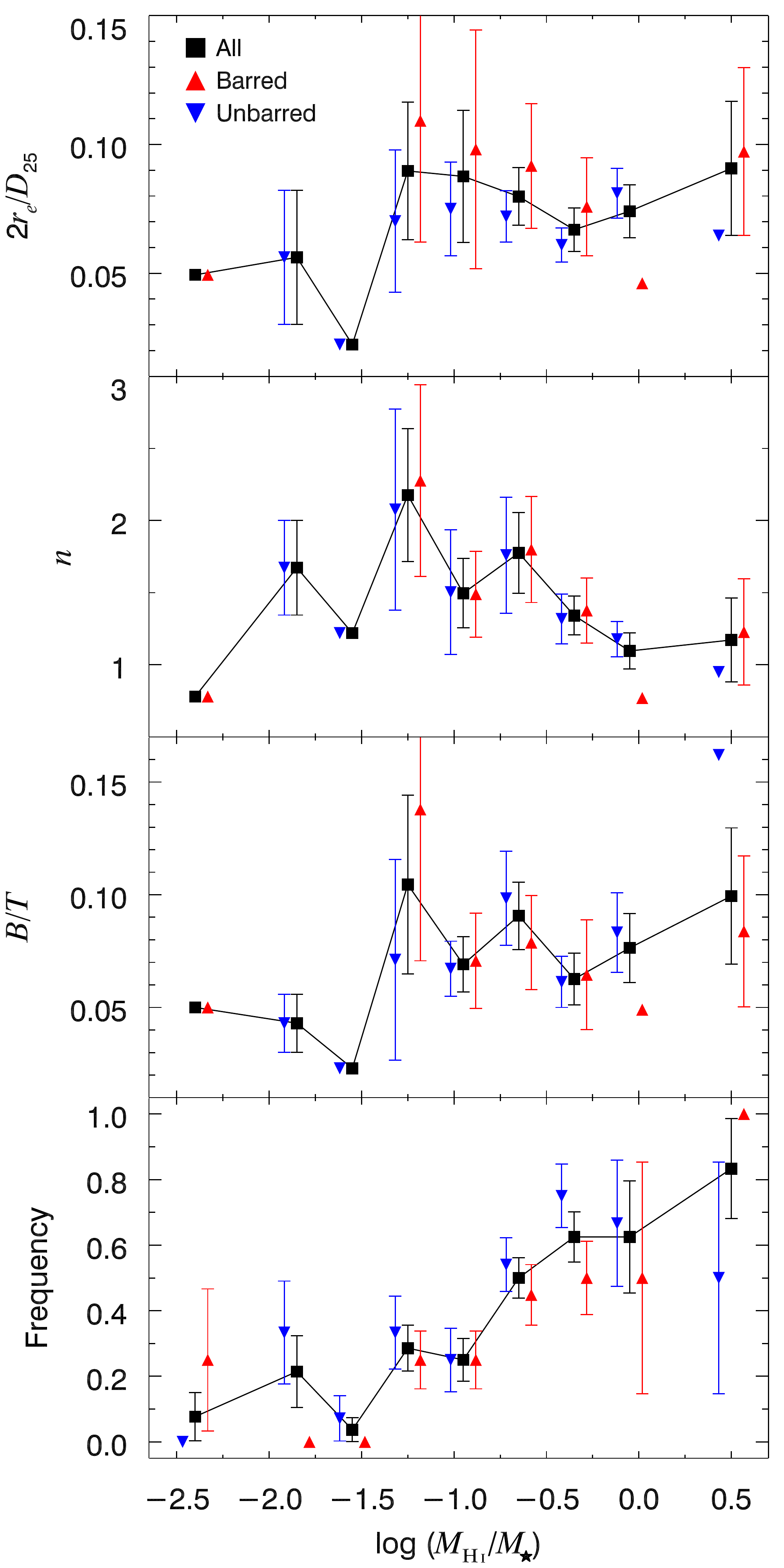}
  \caption{Correlation of pseudo bulge properties with the gas fraction of their
    hosts. From top to bottom, the panels show normalized pseudo bulge sizes,
    \sersic{} indices, bulge-to-total ratios, and detection frequency as a
    function of \ion{H}{1} gas fraction, for all galaxies (black), barred
    galaxies (red), and unbarred galaxies (blue). There are no systematic
    differences between barred and unbarred galaxies. Errors of pseudo bulge
    frequency are calculated assuming a binomial distribution. Symbols and error
    bars in the other panels represent mean and error of the mean in each gas
    fraction bin, respectively. Symbols are horizontally offset for clarity.
    \label{fig:gen_PB_gas}}
\end{figure}

Bars play a critical role in facilitating gas inflow, the raw material for star
formation in galaxy centers and hence for pseudo bulge growth.  It is
instructive to control for the amount of available gas and then compare pseudo
bulge properties of barred and unbarred galaxies.  While we expect the pseudo
bulge fraction to increase toward more gas-rich galaxies, a natural consequence
of the increase of specific gas content toward galaxies with higher specific
star formation rate and later Hubble type \citep[][and references
therein]{1994ARA&A+Roberts}, more available gas does not seem to be related to
increased pseudo bulge fraction for barred galaxies
(Figure~\ref{fig:gen_PB_gas}). At fixed gas fraction, the pseudo bulge fraction
of barred galaxies is not significantly different than that of unbarred
galaxies. If there is any systematic trend at all, it goes in the opposite
direction. Bulge prominence ($B/T$) and normalized bulge size ($2r_{e}/D_{25}$)
of pseudo bulges do not show any systematic trends with gas content, while
\sersic{} $n$ seems to increase toward gas-poor galaxies, but the results are
inconclusive because of the small number statistics in the three most gas-poor
bins. Again, at fixed gas fraction, the presence of a bar has no relation to
$B/T$, \sersic{} $n$, or $2r_{e}/D_{25}$. Due to the small number statistics, we
cannot afford to further control for galaxy stellar mass, but comparison of the
stellar mass distribution of barred and unbarred galaxies rules out stellar mass
as an important factor.  If barred galaxies were to host more and stronger
pseudo bulges, then, to account for the similar pseudo bulge properties of
barred and unbarred galaxies we would expect the stellar masses of barred
galaxies to deviate systematically from those of unbarred galaxies.  Comparison
of the stellar masses of barred and unbarred galaxies at fixed gas fraction does
not reveal any systematic differences.

To summarize: pseudo bulges identified by the Kormendy relation rarely include
bulges with boxy/peanut features (or barlenses when viewed face-on). Therefore,
the pseudo bulges identified in this paper largely refer to those formed out of
material accumulated from central gas inflows, not from the bar buckling
instability. We find no evidence that at a fixed gas fraction bars help the
formation or growth of pseudo bulges. Implications of these results will be
addressed in Section~\ref{sec:driv-engine-secul}.

\section{Discussion}
\label{sec:discussion}

\subsection{Bulge Definition: Comparison with Previous Studies}
\label{sec:comp-with-prev}

Measurements of bulge parameter depend on which part of the galaxy is regarded
as the bulge. As illustrated in \citet{2017ApJ+Gao} and \citet{2018ApJ+Gao,
  2019ApJS+Gao}, we consistently adopt the convention that bulges refer to extra
light above the inner extrapolation of disks/bars, after excluding any possible
nuclei. Note that disks here do not necessarily refer to a single exponential
component but instead to the analytic function or combination of functions that
best describes the disk surface brightness. Therefore, the bulge component
includes the contribution from nuclear bars/rings and boxy/peanut bulges
(thickened part of bar), if present. We do not model these substructures as
individual, separate components in addition to the bulge, although some are
actually part of the bar (e.g., boxy/peanut bulges). If we model all distinct
disky features in the bulge region as individual components, we risk isolating
every pseudo component from the rest of the photometric bulge. This betrays our
goal to characterize bulge dichotomies in the local Universe. We acknowledge
that other studies with different scientific goals may adopt different bulge
definitions from ours, and therefore their bulge parameters may not be
comparable to ours. For example, \citet{2016ApJ+Lasker} isolated everything else
to extract the pure classical bulges to study black hole--bulge relations.  The
study of \citet{2019MNRAS+de_Lorenzo-Caceres} modeled nuclear bars separately
because the latter were the objects of interest. In the same vein, we adopt the
definition that best suits our specific goal, which is to characterize the
structural properties of the overall photometric bulge, no matter how many
subcomponents it might have. Our definition is similar to that of
\citetalias{2009MNRAS+Gadotti}, only that our disk models are more detailed than
his, causing the bulge parameters to be different (Figure~\ref{fig:frac_comp}).

\citetalias{2008AJ+Fisher} employed a different method than we to measure bulge
parameters.  Due to the limitations of their 1D technique, non-axisymmetric
components (e.g., bars and lenses) cannot be modeled simultaneously with the
bulge and disk, and \citetalias{2008AJ+Fisher} excluded them from the fit.
Their bulge definition also differs subtly from ours, as they masked nuclear
rings/bars from their fits.  This might affect bulge structural parameters but
should not alter $B/T$ much \citep{2017ApJ+Gao,2019MNRAS+de_Lorenzo-Caceres}. To
better understand the underlying reasons for the discrepancy between our results
and those of \citetalias{2008AJ+Fisher}, we closely examined the 14 galaxies in
common between the two samples. Fortunately, none of the galaxies contains
complications from nuclear disky features, affording a relatively
straightforward comparison between \citetalias{2008AJ+Fisher}'s technique and
ours.

\begin{figure*}
  \epsscale{1.17}
  \plotone{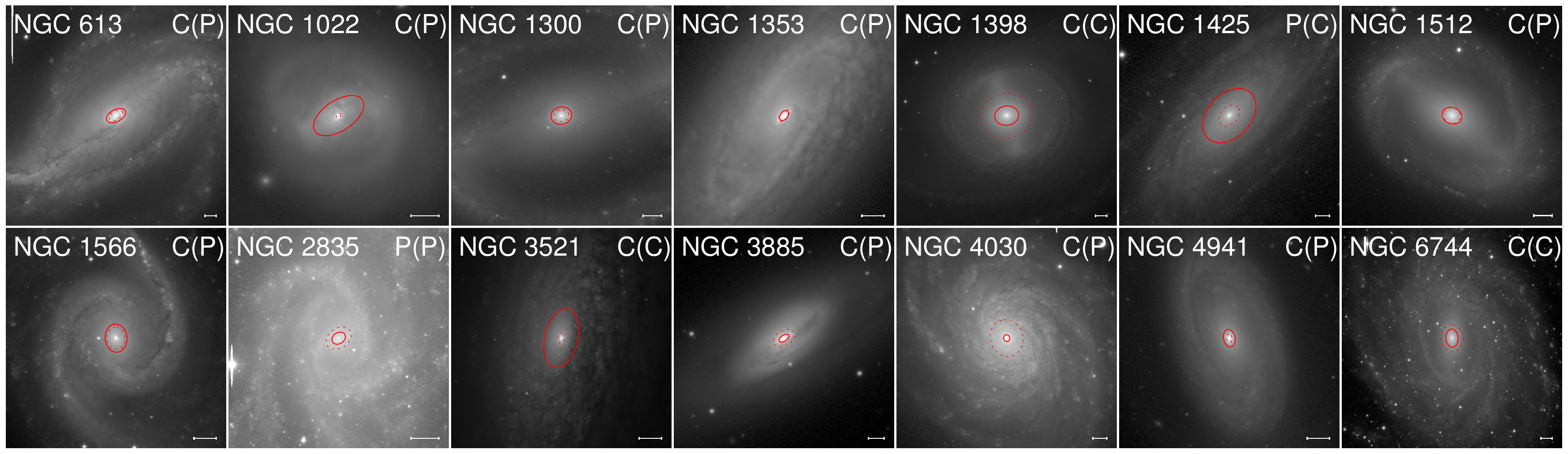}
  \caption{$R$-band images of 14 CGS galaxies in common with the
      \citetalias{2008AJ+Fisher} sample. The bulge region is overlaid with a red
      ellipse (solid = CGS; dashed = FD08), whose semi-major axis represents the
      bulge effective radius, with axis ratio and orientation following the
      best-fit model. Bulge types are indicated in the upper-right corner of
      each image (C = classical; P = pseudo); the classication from
      \citetalias{2008AJ+Fisher} is given in parentheses. The scale bar on the
      lower-right corner represents 1\,kpc.
    \label{fig:img_bul}}
\end{figure*}

Figure~\ref{fig:img_bul} illustrates the bulge-dominated region for the
overlapping objects. Because \citetalias{2008AJ+Fisher} did not provide the
position angles of their bulges, we assume that their values are the same as
ours. We find that except for NGC~1022~and NGC~3521 where
\citetalias{2008AJ+Fisher}’s bulges are treated as nuclei in our fits, both
studies ascribe the same physical entity to the bulge. The discrepancies in the
resulting bulge parameters mostly reflect differences in data quality, in the
techniques applied, and in details of the model construction. Our 2D approach
generally allows for more direct and robust model construction to account for
the multicomponent nature of nearby disk galaxies.

\begin{figure*}
  \epsscale{1.17}
  \plotone{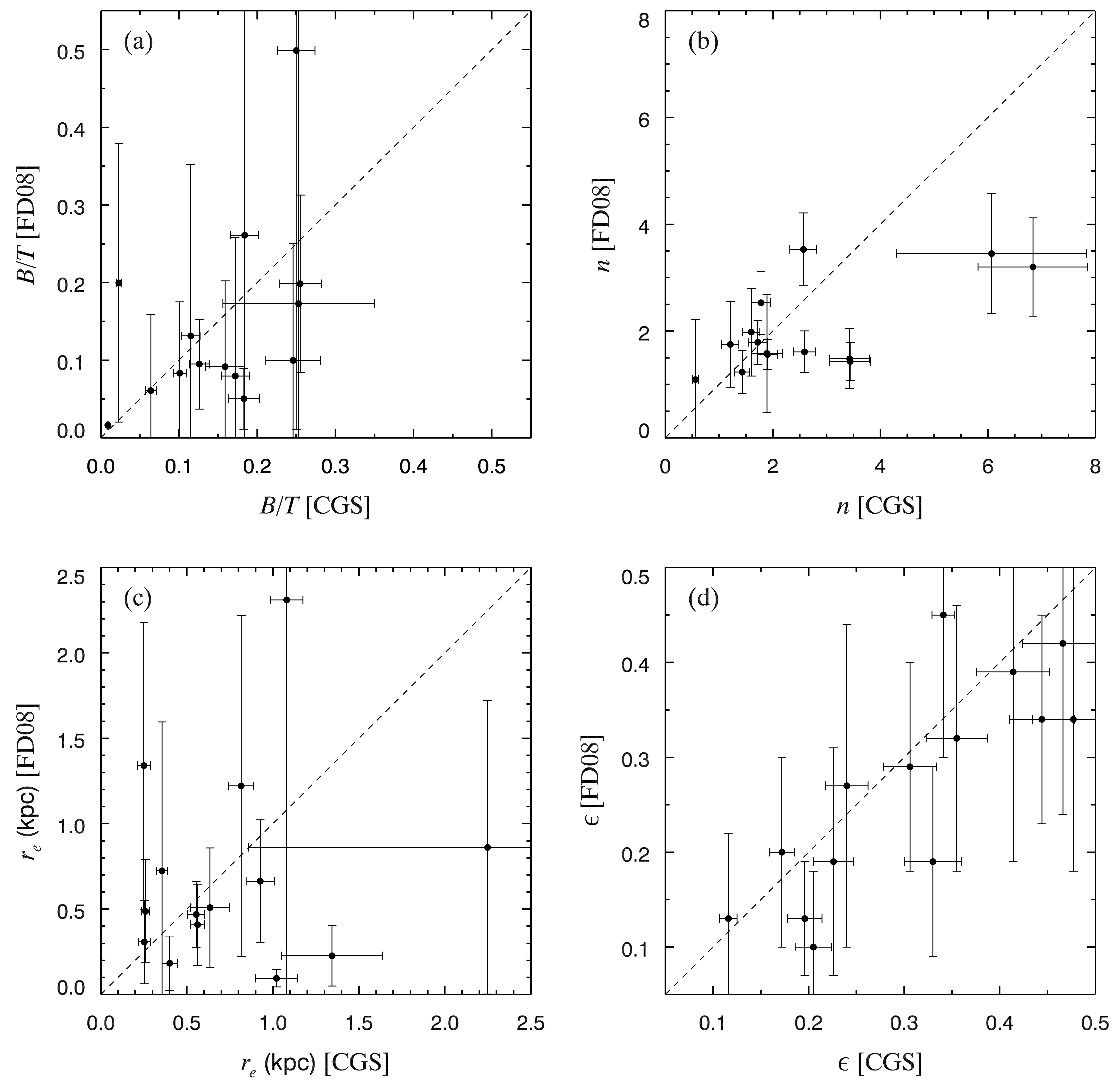}
  \caption{Comparison of CGS ($R$ band) bulge parameters with those obtained by
    \citetalias{2008AJ+Fisher} ($V$ band) for the 14 galaxies in common between
    the two studies, for (a) $B/T$, (b) $n$, (c) $r_{e}$, and (d) $\epsilon$.
    Dashed lines are one-to-one relations. \label{fig:comp_struct}}
\end{figure*}

Figure~\ref{fig:comp_struct} compares the bulge parameters from this study with
those from \citetalias{2008AJ+Fisher} for the comparison sample. Taking into
consideration the large uncertainties of the parameters in
\citetalias{2008AJ+Fisher}, their measurements, apart from a few outliers, are
broadly consistent with ours, and there are no significant systematic trends.
Unfortunately, the small size of the comparison sample precludes us from drawing
any firm conclusions as to the origin of the systematic discrepancies in $B/T$
and relative bulge size of early-type disks.  Among the 14 galaxies in common,
there are only two Sa galaxies, one Sa/0 galaxy, and no S0s, where the
differences in $B/T$ and relative bulge size are most significant. As mentioned
above, various reasons could be responsible for the differences. Our smaller
$B/T$ in early-type disks, where classical bulges are most abundant, in
conjunction with our classification criterion that includes less prominent
classical bulges (small $n$), results in a less clear separation between
classical and pseudo bulges in the $B/T$ distribution, when compared with
\citetalias{2008AJ+Fisher} (Figure~\ref{fig:frac_comp}). This leads to both
bulge types having similar relative sizes (Figure~\ref{fig:re_comp}), as argued
in Section~\ref{sec:bul-size}. We did not find substantial systematic
differences in bulge \sersic{} $n$, neither for the full sample at fixed Hubble
type (Figure~\ref{fig:nser_T_comp}) nor for the subset of galaxies in common
between CGS and \citetalias{2008AJ+Fisher} (Figure~\ref{fig:comp_struct}b).
Therefore, the significantly different distributions of $n$
(Figure~\ref{fig:nser_comp}) are probably due to the different classification
criteria used by us and \citetalias{2008AJ+Fisher}.

We also note that the consistency of the classifications is poor. Only four of
the 14 bulges have consistent classifications. If we adopted $n=2$ to separate
classical and pseudo bulges in CGS, our classifications would agree much better
with those of \citetalias{2008AJ+Fisher} (10/14). This suggests that adopting
different classification criteria is the major source of the inconsistency. In
contrast to our study that relies on scaling relations,
\citetalias{2008AJ+Fisher} used nuclear morphologies to distinguish bulge types.
While such an approach is physically meaningful and has the merit of being
immune from potential bias in bulge structural parameters, it is not
quantitative, relying on subjective, visual morphological classification.
Moreover, it may not work well in composite bulges whose disky features are not
fundamental to the overall bulge nature \citep[e.g.,][]{2015MNRAS+Erwin,
  2018ApJ+Gao}.

On the other hand, \citet{2013seg+Kormendy,2016ASSL+Kormendy} and
\citet{2016ASSL+Fisher} warned against using the Kormendy relation to identify
pseudo bulges, because high-density pseudo bulges would risk being
misclassified. Given that scaling relations are demonstrated useful in
distinguishing objects of different nature, an immediate question is why pseudo
bulges should follow the scaling relation of ellipticals. A plausible
explanation is that these are composite bulges, which contain contributions from
secular evolution and therefore can be recognized using \sersic{} index or
nuclear morphology. Unfortunately, this issue cannot be resolved in the context
of a binary classification scheme, namely classifying bulges as either classical
or pseudo bulges. In such an exercise, without independent measurements of the
relative importance of violent to secular processes in bulge growth, it is
difficult to tell which criterion is superior. However, it is possible to assess
the robustness of such classifications. For example, as mentioned in
Section~\ref{sec:intro}, \citet{2017A&A+Neumann} have done so by comparing
various bulge type indicators and found that using the Kormendy relation alone
can recover classifications based on multiple criteria to a high success
rate. Without access to other measurements that are useful to classify bulge
types (e.g., stellar kinematics and stellar populations), we are unable to
follow such a strategy. But we add to their advocacy by showing that pseudo
bulges selected by the Kormendy relation exhibit properties that consistently
imply a disky origin and that \sersic{} index is not an appropriate indicator of
bulge types.

Using the classifications from \citetalias{2008AJ+Fisher} for the comparison
sample, we find that 30\% of the pseudo bulges have $n>2$ in CGS, and 25\% of
the classical bulges have $n<2$ in CGS; in FD08, all the pseudo bulges have
$n<2$, and all the classical bulges have $n>2$. This suggests that differences
in \sersic{} $n$ measurements tend to weaken the separation of $n$ of classical
and pseudo bulges in CGS. The next section will demonstrate that the weakening
of the bimodality is not due to the relative large number of classical bulges in
the CGS sample.

To summarize: the small number of overlapping galaxies precludes us from
investigating the cause of the discrepancies between our study and that of FD08.
We speculate that differences in data quality, fitting techniques, model
construction, and classification criteria all play a role.


\subsection{Conflicts with Previous Results}
\label{sec:conflicts-with-known}

Some of our population statistics differ from those of
previous studies (Sections~\ref{sec:small-but-dense},
\ref{sec:pse-n2}, \ref{sec:bul-size}, and \ref{sec:dich-hosts}). First,
the basis of the widely adopted criterion for bulge classification---the bimodal
distribution of \sersic{} indices---cannot be reproduced in our measurements.
Does the discrepancy stem merely from the different methods used to classify
bulge type? After all, \citetalias{2008AJ+Fisher} identified pseudo bulges based
on their nuclear morphologies in \textit{HST} images, while we define pseudo
bulges as low-surface brightness outliers in the Kormendy relation.

\begin{figure}
  \epsscale{1.17}
  \plotone{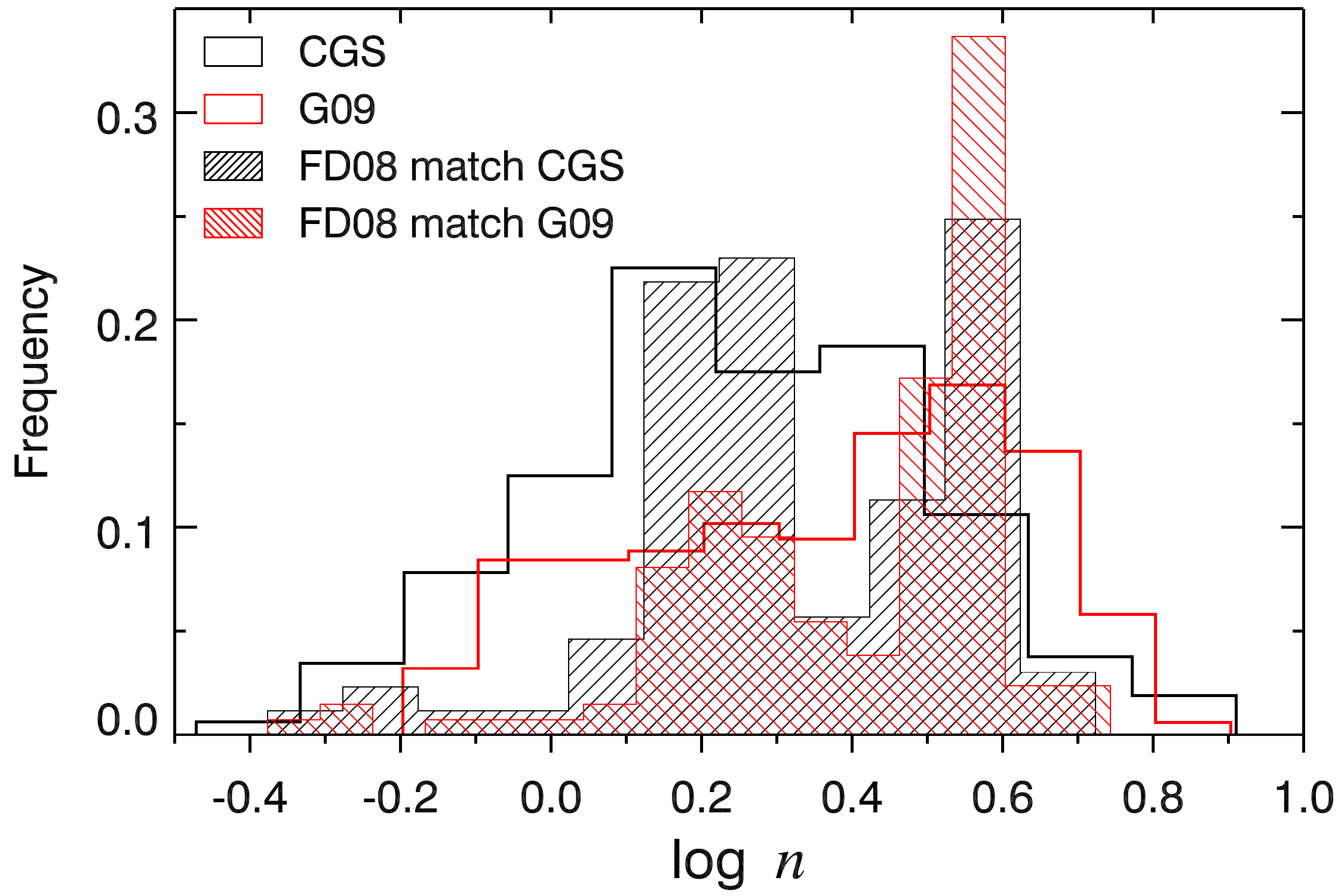}
  \caption{Comparison of bulge \sersic{} index distributions for the sample
      of CGS (unfilled black histogram), \citetalias{2009MNRAS+Gadotti}
      (unfilled red histogram), and \citetalias{2008AJ+Fisher} (filled black and
      red histogram), after matching the relative number of classical to pseudo
      bulges. Bulges with $\log n<-0.5$ in CGS are
      omitted. \label{fig:nser_comp_fake}}
\end{figure}

We do not think that this can be the whole story. If we suppose that classical
and pseudo bulges have distinct distributions in \sersic{} index and can be well
separated at $n\approx 2$, we would have 125 classical bulges and 195 pseudo
bulges in CGS, similar to but slightly different from the relative numbers in
\citetalias{2008AJ+Fisher} (26 vs. 53). On the contrary, the
\citetalias{2009MNRAS+Gadotti} sample is dominated by $n\ge 2$ bulges (421
vs. 267). The different relative number of classical to pseudo bulges could
erase the bimodality in the global distribution of bulge \sersic{} index. We
explore the effects of varying the relative fraction of classical to pseudo
bulges on the global distribution of bulge \sersic{} index in the
\citetalias{2008AJ+Fisher} sample, by matching the statistics of CGS and
\citetalias{2009MNRAS+Gadotti}. Following the style presented in Figure~9 of
\citetalias{2008AJ+Fisher}, we plot the distribution of $\log n$ in
Figure~\ref{fig:nser_comp_fake}. Although we observe a gentle dip at
$\log n\approx 0.3$ for CGS (unfilled black histogram), it is still not
comparable with the significant deficit of bulges in the
\citetalias{2008AJ+Fisher} sample at $\log n\approx 0.4$ (filled black
histogram), after matching the relative number of classical to pseudo
bulges. The same holds for the comparison of the \citetalias{2009MNRAS+Gadotti}
sample (unfilled red histogram) with the \citetalias{2008AJ+Fisher} sample
(filled red histogram). There is always a strong dip at $\log n\approx 0.4$ for
the \citetalias{2008AJ+Fisher} sample, no matter whether it is dominated by
classical or pseudo bulges. In concert with the comparisons for the objects in
common (Figure~\ref{fig:comp_struct}) and at fixed Hubble type
(Figure~\ref{fig:nser_T_comp}), we suspect that the small sample size and
selection effects of the \citetalias{2008AJ+Fisher} sample may be responsible
for their deficit of bulges with $n\approx 2-3$.

Instead, among all the bulge structural parameters considered in CGS, the
clearest separation between classical and pseudo
  bulges is seen in $\langle\mu_{e}\rangle$. This, of course, is a natural
outcome of using the Kormendy relation to classify bulges. A
clean separation in $\langle\mu_{e}\rangle$ is ensured
because pseudo bulges happen to have a similar range of $r_{e}$ as classical
bulges. Thus, $\langle\mu_{e}\rangle$ is an effective single criterion for
classifying bulge types consistent with classifications based on the Kormendy
relation.

Second, we reiterate that 1D fitting systematically overestimates bulge
luminosities (see also Section~5 of \citealp{2019ApJS+Gao}). Early 1D studies
often adopted a \citet{1948AnAp+de_Vaucouleurs} law ($n=4$) for the bulge
component, which was later shown to an overestimate the bulge contribution
\citep[e.g.,][]{1985ApJS+Kent,1986ApJS+Kodaira,1986ApJ+Simien}.
\citetalias{2008AJ+Fisher} presented significantly improved 1D fitting results
by adopting better model assumptions and carefully masking morphological
features that deviate from their simplified model. Still, comparison of our
results with those from \citetalias{2008AJ+Fisher} shows that this effect, most
severe in early-type disk galaxies, together with the different classification
criteria applied, leads to significant differences between their $B/T$
distributions and ours, both for classical and pseudo bulges. The same holds for
bulge sizes. Whereas we find that classical and pseudo bulges have similar
relative sizes, \citetalias{2008AJ+Fisher} concluded that the former have larger
relative sizes than the latter (Section~\ref{sec:bul-size}). 1D fitting comes
with several ambiguities and inherent weaknesses that cannot be overcome (see
Section~1 of \citealp{2017ApJ+Gao} for an overview). The most serious limitation
of the 1D technique is that it does not utilize the full 2D spatial information
in the image and therefore does not account for the full range of complexity
intrinsic to most galaxy disks. Sub-structures such as bars and lenses cannot be
adequately modeled in 1D fitting and therefore cannot be isolated properly from
the bulge, often causing the bulge to be overestimated
\citep[e.g.,][]{1985ApJS+Kent,1986ApJS+Kodaira,1986ApJ+Simien}. Although
\citetalias{2008AJ+Fisher} strived to mask these features in their fits,
\citet{2017ApJ+Gao} showed that such a compromised strategy amplifies
uncertainties in the resulting bulge parameters due to further loss of
information. Our results agree better with those of
\citetalias{2009MNRAS+Gadotti}, whose methodology, like ours, can account for
bars.  Indeed, we employ the latest version of \galfit{} to treat a variety of
other sub-structures as well \citep{2017ApJ+Gao}. Therefore, we conclude that
our measurements, based on better technique and more realistic models, are more
accurate than published 1D fitting results, and the differences mostly reflect
improvement of measurements.

Finally, we derive a lower pseudo bulge fraction in early-type disks than in
studies that use nuclear morphologies and/or \sersic{} indices to classify
bulges (Figure~\ref{fig:PB_T_mass_col}; Section~\ref{sec:dich-hosts}). This
mismatch is probably largely related to differences in classification criteria.
We have already noted inconsistencies between classifications based on nuclear
morphologies with those based on the Kormendy relation (Figure~3 of
\citealp{2018ApJ+Gao}). S0 bulges bearing disky features such as nuclear bars
and rings cannot be distinguished from classical bulges in terms of their
location on the Kormendy relation. We attributed this apparent inconsistency to
the fact that the bulges bearing disky features in early-type disks are most
likely composite bulges, the bulk of whose mass was assembled early and fast,
and secular processes subsequently superposed disky features on top of a
preexisting classical bulge. The disky features also help to lower the \sersic{}
index of the bulge (see the case of NGC~1326 in Section~4.7 of
\citealp{2017ApJ+Gao}). In our estimation, bulge classifications based on the
Kormendy relation are more secured than those that rely on nuclear morphology,
which neither comprehensively nor reliably signifies the overall photometric
structure and formation history of the bulge (see also
\citealp{2009ApJ+Fisher}).

\subsection{Implications for Bulge Formation and Evolution}
\label{sec:impl-bulge-form}

\subsubsection{Violent vs. Secular Processes}
\label{sec:violent-vs.-secular}

We summarize the key results from this study that are useful to diagnose 
the nature of bulges:

\begin{itemize}
\item Most pseudo bulges have $B/T\la 0.1$ and low surface brightness
  ($\langle\mu_{e, R}\rangle\ga 18.5\,\mathrm{mag~arcsec^{-2}}$), while
  classical bulges have a broad distribution of $B/T$ and exhibit higher surface
  brightness ($\langle\mu_{e, R}\rangle < 18.5\,\mathrm{mag~arcsec^{-2}}$).

\item Pseudo bulges have low \sersic{} indices ($n\la2$), while classical
  bulges have a broad distribution of $n$.

\item Classical bulges have an ellipticity distribution consistent with that of
  elliptical galaxies, while pseudo bulges have a broad distribution of
  ellipticity reminiscent of disks.

\item Pseudo bulges preferentially reside in late-type spirals that are less
  massive and more gas-rich.
\end{itemize}

The above-listed properties consistently support the prevailing hypothesis that
pseudo bulges are disky components masquerading as bulges. Pseudo bulges were
formed from stars born out of gas accumulated in the central regions of the
galaxy \citep{2004ARA&A+Kormendy,2014RvMP+Sellwood,2016MNRAS+Tonini,
  2019MNRAS+Izquierdo-Villalba}. Since their formation resembles the assembly of
galaxy disks, pseudo bulges manifest themselves as diffuse (low
$\langle\mu_{e}\rangle$), approximately exponential ($n \approx 1$), and
intrinsically flattened (broad distribution of $\epsilon$) structures.  Secular
processes being inefficient and gradual, only relatively modest masses
accumulate in the center (small $B/T$).  The products of secular processes are
most frequently found in late-type spirals, as these systems have more fuel to
feed central star formation and generally have experienced a more placid merger
history.  However, it is unclear which mechanisms are mainly responsible for
driving gas inflow. Nonaxisymmetries such as bars, lenses/ovals, and spiral arms
contribute to secular evolution, among them bars often thought to be the most
effective.  Intriguingly, the analysis in Section~\ref{sec:pse-pre-host-bar}
shows no preference for barred galaxies to host pseudo bulges.  We will revisit
this issue in Section~\ref{sec:driv-engine-secul}.

On the other hand, the broad distribution of $B/T$ and $n$ of classical bulges
implicates a variable formation efficiency. Ellipticals and classical bulges can
be made viably through major mergers \citep{2009ApJ+Hopkins,2009MNRAS+Hopkins,
  2010ApJ+Hopkins,2016ASSL+Brooks,2016MNRAS+Tonini,2017MNRAS+Rodriguez-Gomez}.
Violent relaxation of preexisting stars and central starbursts resulting from
the rapid inflow of gas create bulges of high \sersic{} indices and
pressure-supported stellar orbits. The efficiency of bulge formation depends on
the gas fraction of the merger, and the degree to which the merger remnant
reacquires a stellar disk \citep{2009ApJ+Hopkins,2009MNRAS+Hopkins,
  2010ApJ+Hopkins}. Minor mergers can also contribute to the formation or growth
of classical bulges, albeit in a less prominent \citep{2001A&A+Aguerri,
  2006A&A+Eliche-Moral,2010ApJ+Hopkins} and more complicated manner. Minor
mergers can either contribute directly to the bulge or trigger disk
instabilities and indirectly contribute to bulge growth in a secular manner. The
density of the satellite is a determining factor: a dense satellite more likely
survives to the center and contributes to the growth of a classical bulge, while a
diffuse satellite can be tidally disrupted halfway \citep{2001A&A+Aguerri,
  2006A&A+Eliche-Moral}. More massive satellites, of course, lead to more
dramatic effects.

Moreover, coalescence of giant clumps in high-redshift disks can form classical
bulges \citep{1999ApJ+Noguchi,2007ApJ+Bournaud,2009ApJ+Bournaud,
  2008ApJ+Elmegreen,2016ASSL+Bournaud}. The disks of gas-rich, vigorously
star-forming, high-redshift galaxies fragment into massive clumps of stars and
gas and exhibit irregular morphologies as a result of stellar feedback and
gravitationally instability. The clumps in what are commonly refer to as clumpy
or chain galaxies then migrate inward through dynamical friction and merge into
a bulge. Although the physical processes are driven by disk instability, they
are nothing like the secular evolution in $z\approx 0$ disks, as the migration
and coalescence of clumps operate on much shorter timescales (several hundreds
of Myr; see \citealp{1999ApJ+Noguchi} and \citealp{2008ApJ+Genzel}). Clump
coalescence, along with violent relaxation, creates highly concentrated,
dispersion-supported, high-$\left[\alpha/\mathrm{Fe}\right]$ ratio spheroids
that resemble ellipticals, which, by definition, are classical bulges
\citep{2004A&A+Immeli, 2008ApJ+Elmegreen}. The accretion timescale, mass, and
mass density of the galaxies can produce a broad spectrum of bulge prominence,
down to values as small as $B/T\approx 0.1$ \citep{1999ApJ+Noguchi,
  2007ApJ+Bournaud}. This seems to be consistent with the observed broad
distribution of $B/T$ for classical bulges (Figure~\ref{fig:frac_comp}).
However, not all investigators agree that bulges originate from clumps.
\citet{2017MNRAS+Oklopcic} used cosmological hydrodynamic simulations with a
realistic stellar feedback treatment to show that clumps are short-lived and do
not systematically migrate inward.

The broad outlines for the formation of classical and pseudo bulges seem more or
less in order. The secular processes that build pseudo bulges at $z\approx 0$
are inefficient and produce mostly weak bulges. The resulting disky component
retains enough memory of its origin that it can be distinguished from classical
bulges in various respects. Classical bulges form with variable efficiency
through rapid processes involving violent relaxation and gaseous dissipation at
an early epoch.  Their structural parameters follow the scaling relations (e.g.,
Kormendy relation and fundamental plane) of ellipticals, and they are spheroids
instead of flattened structures like pseudo bulges.  The formation picture,
however, still has some loose ends.  Some studies maintain that mergers can make
pseudo bulges \citep{2015MNRAS+Wang,2018A&A+Eliche-Moral,2018MNRAS+Sauvaget},
although the pseudo bulges may still be the products of secular processes of the
rebuilt disks after the gas-rich mergers settle down. \citet{2012MNRAS+Inoue}
suggest that clump coalescence can lead to pseudo bulges, even if, overall, the
ultimate fate of clumps is still controversial \citep{2017MNRAS+Oklopcic}. A
quantitative comparison between our measurements with theoretical predictions
would be highly desirable, but this lies beyond the scope of this paper.
Previous studies have revealed possible inconsistencies. According to
\citet{2009ApJ+Weinzirl}, semi-analytic $\Lambda$CDM models predict too many
prominent bulges compared with observations.  At the same time, other authors
consider the pseudo bulges to be overabundant compared to cosmological
predictions \citep{2008ASPC+Kormendy, 2010ApJ+Kormendy2}. It is also puzzling to
see no signs of a classical bulge in massive disk galaxies such as NGC~4565
(\citealp{2010ApJ+Kormendy1}).  The lower pseudo bulge fractions in early-type
disks reported in this study perhaps helps alleviate the tension.

\subsubsection{The Driving Engines of Secular Evolution}
\label{sec:driv-engine-secul}

In Section~\ref{sec:pse-pre-host-bar}, we found that bars do not bear any
obvious relation to pseudo bulges, not even in gas-rich galaxies.  This is
counterintuitive, as bars, the strongest nonaxisymmetric structure in the disk,
are thought to drive efficient gas inflow. Perhaps it indicates that
nonaxisymmetries other than bars, such as lenses/ovals and spiral arms, also
participate in driving gas inflow.  As these additional nonaxisymmetries are
ubiquitous features in disk galaxies, the presence or absence of pseudo bulges
does not have to rely on bars alone.

Still, because other nonaxisymmetries are weaker than bars, we need to
understand why barred galaxies have pseudo bulges of comparable strength and
frequency as unbarred galaxies (Figure~\ref{fig:gen_PB_gas}). We caution that
the observed $B/T$ of pseudo bulges represents the accumulated growth of their
prolonged history, and is not necessarily related to the present-day observable
properties of the galaxy, such as its current gas fraction and presence of a
bar. Today's gas-poor galaxies may once have been gas-rich.  At any rate, pseudo
bulges are generically weak ($B/T\la0.1$), and hence the host does not need to
be extraordinarily gas-rich to provide the fuel. This may explain why the $B/T$
of pseudo bulges does not depend on current gas fraction
(Figure~\ref{fig:gen_PB_gas}). Another outstanding issue pertains to whether
bars are long-lived \citep{2013MNRAS+Athanassoula} or not
\citep{2002A&A+Bournaud}. Central mass concentrations such as black holes and
compact bulges weaken and even dissolve bars \citep{1996ASPC+Combes,
  2004ApJ+Shen,2005MNRAS+Bournaud}, but they can be regenerated from gas
accretion \citep{1996ASPC+Combes,2002A&A+Bournaud,2002A&A+Block,
  2005MNRAS+Bournaud}. Therefore, present-day unbarred galaxies may once have
been barred, during which time their pseudo bulges were built.  If so, we expect
that at fixed $B/T$ and gas fraction, the pseudo bulges of barred galaxies
should have younger stellar populations. This hypothesis should be tested in the
future. The evolutionary state of the bar may also matter.  In gas-rich systems
where bars have the best chance to make a difference between barred and unbarred
galaxies, bars may have developed later and are weaker when mature
\citep{2013MNRAS+Athanassoula}.  Mature bars suffer from buckling instability
and form boxy/peanut bulges, which reduce gas inflow to the very center
\citep{2016MNRAS+Fragkoudi}. Bar-driven secular evolution itself may be
self-regulated: the very growth of a pseudo bulge may weaken the bar and thereby
curtail its own growth rate. This may weaken any systematic differences between
barred and unbarred systems.  Finally, external perturbations, such as
ram-pressure striping, flybys, and minor mergers may also promote gas inflows
\citep[e.g.,][]{1996Natur+Moore,1998ApJ+Moore,1999MNRAS+Moore,1999ApJ+Bekki,
  2011MNRAS+Bekki,2014MNRAS+Kaviraj,2017Natur+Poggianti}. Such external effects
would be incredibly difficult to disentangle from internal secular processes.

To summarize: despite the lack of clear observational evidence that bars 
directly relate to the pseudo bulge phenomenon, the above considerations 
prevent us from drawing any firm conclusions about their actual role.

\section{Summary}
\label{sec:summary}

We use robustly measured structural parameters of a large, homogeneous sample of
320 lenticular and spiral galaxies to investigate the Kormendy
($\langle\mu_{e}\rangle$ vs. $r_e$) relation of their bulge components. The
bulge parameters are derived from detailed 2D decomposition of high-quality
$R$-band images drawn from CGS. We critically discuss the empirical
classification of bulge types, reexamine the statistical properties of classical
and pseudo bulges, and consider their physical implications.

Our principal results are as follows.
\begin{itemize}
\item Despite the ambiguities in classification of bulge types, we find that
  pseudo bulges selected as low-$\langle\mu_{e}\rangle$ outliers of the Kormendy
  relation established by elliptical galaxies generally show physical properties
  consistent with their presumed disky origin.

\item The distribution of bulge \sersic{} indices is not bimodal. We recommend
  abandoning the common practice of adopting $n \approx 2$ to distinguish
  between classical and pseudo bulges.
  
\item In line with \citet{2009MNRAS+Gadotti} and \citet{2017A&A+Neumann}, we
  recommend using the Kormendy relation as a promising alternative to using
  \sersic{} index to classify bulges.

\item Pseudo bulges have low \sersic{} indices, most having $n\la 2$, with the
  distribution peaking at $n\approx 1$, consistent with disk-like profiles. 
  Most pseudo bulges are weak, comprising only $B/T \la 0.1$ of the total light.

\item Classical bulges display a broad distribution in $B/T$ and $n$, and are 
  on average more prominent than pseudo bulges, despite their significant
  overlap in $B/T$.

\item Pseudo bulges are intrinsically more flattened structures compared with
  classical bulges. The ellipticities of pseudo bulges are consistent with those
  of disks.

\item Pseudo bulges reside most frequently in less massive, gas-rich late-type
  spirals.

\item Contrary to naive expectations, barred galaxies do not host more pseudo
  bulges or more prominent pseudo bulges than unbarred galaxies.
\end{itemize}

The statistical properties of our sample reveal a number of quantitative
differences compared to previous studies that can be understood in terms of our
improved method of 2D image decomposition and different classification criteria
applied. In qualitative agreement with previous studies, we suggest that pseudo
bulges formed through internal secular processes in the large-scale disk, where
nonaxisymmetries drive gas inflow and build up central overdensities through
star formation. Their disky origin is still encoded in their low surface
brightnesses, near-exponential radial profiles, and flattened
geometry. Classical bulges formed via violent processes akin to those in
elliptical galaxies.

\acknowledgments

We thank the referee for constructive comments. This work was supported by the
National Science Foundation of China (11721303, 11991052) and the National Key
R\&D Program of China (2016YFA0400702). H.G. thanks Jing~Wang and Hassen~Yesuf
for useful discussions.

\appendix

\section{$R$-band Structural Parameters of CGS Ellipticals}
\label{sec:str-par-ell}

\begin{figure*}
  \epsscale{1.1}
  \plotone{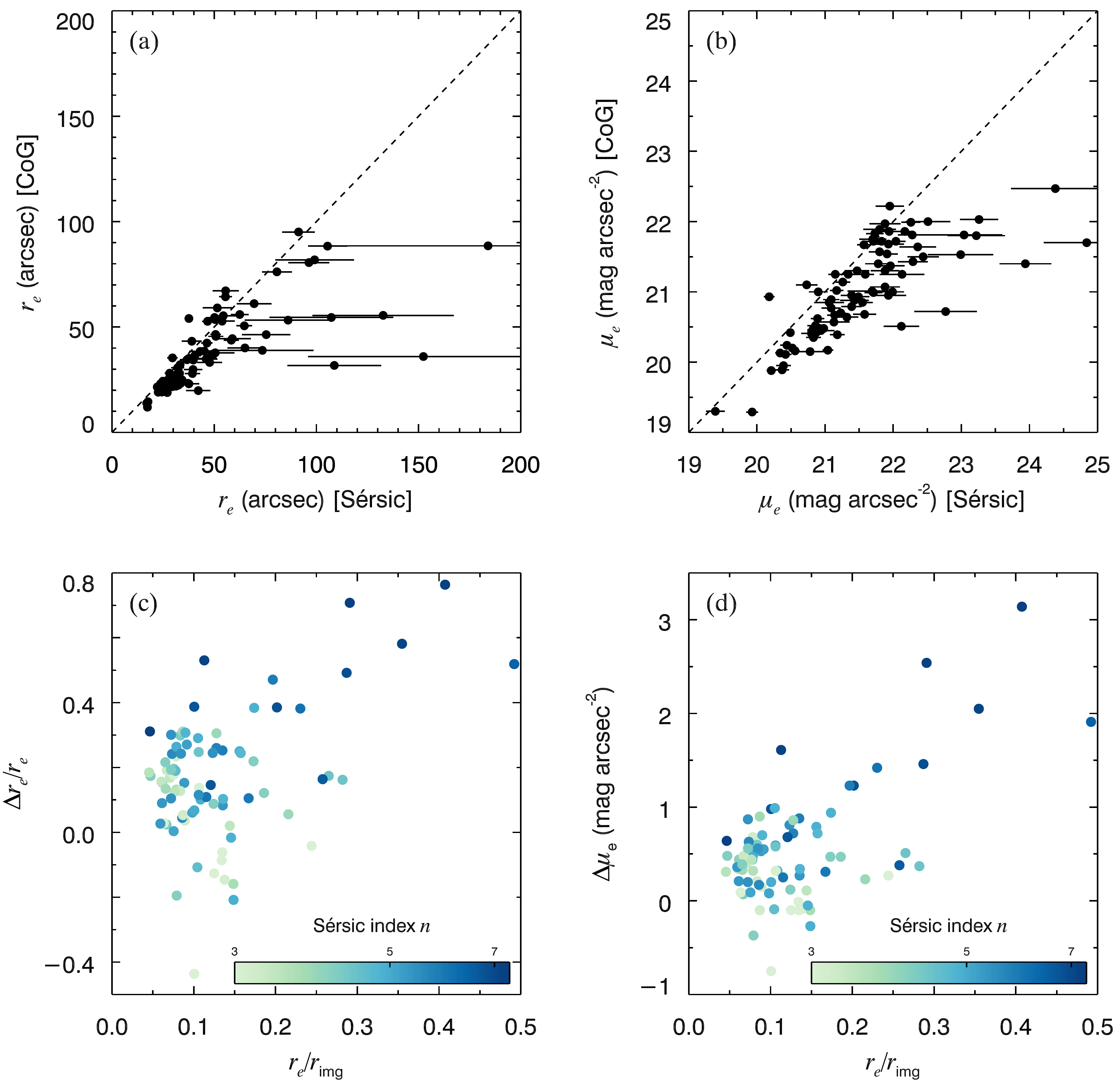}
  \caption{Comparison of \sersic{} structural parameters (a) $r_{e}$ and (b)
    $\mu_{e}$ with non-parametric ones based on the curve-of-growth (CoG) 
    analysis of the CGS ellipticals.  Dependence of the differences in (c)
    $r_{e}$ and (d) $\mu_{e}$ on galaxy size normalized by the image
    size and on \sersic{} index. \label{fig:comp_E_struct}}
\end{figure*}

The CGS ellipticals provide the fiducial scaling relations for comparison with
bulges. We performed single-\sersic{} fits to derive their global structural
parameters ($n$, $\mu_{e}$, and $r_{e}$) in the $R$ band. Although Li~et~al.
(2011) give access to non-parametric measurements of the structural parameters,
we consider our parametric analysis necessary because we wish to avoid possible
systematic biases due to measurement technique.  Comparison of the two sets of
measurements (Figure~\ref{fig:comp_E_struct}, top panels) reveals that that
\sersic{} fits indeed yield values of $\mu_{e}$ and $r_{e}$ that are
systematically larger from the non-parametric ones.  This is expected, as
\sersic{} fits integrate the flux to infinity, whereas non-parametric methods
measure the flux only within the radius where the galaxy light fades into the
noise. The tests of \citet{2001MNRAS+Trujillo} show that the differences between
the two kinds of measurements depend on the \sersic{} index and the size of the
galaxy relative to the aperture applied to measure its flux: the larger the
galaxy size and \sersic{} index, the more significant the differences between
the parametric and non-parametric measurements, as more light would be missed by
the non-parametric methods. The bottom panels of Figure~\ref{fig:comp_E_struct}
confirm this trend. Galaxies with larger normalized $r_{e}$ and larger $n$ show
larger offsets in the two structural parameters.

\begin{figure}
  \epsscale{1.}
  \plotone{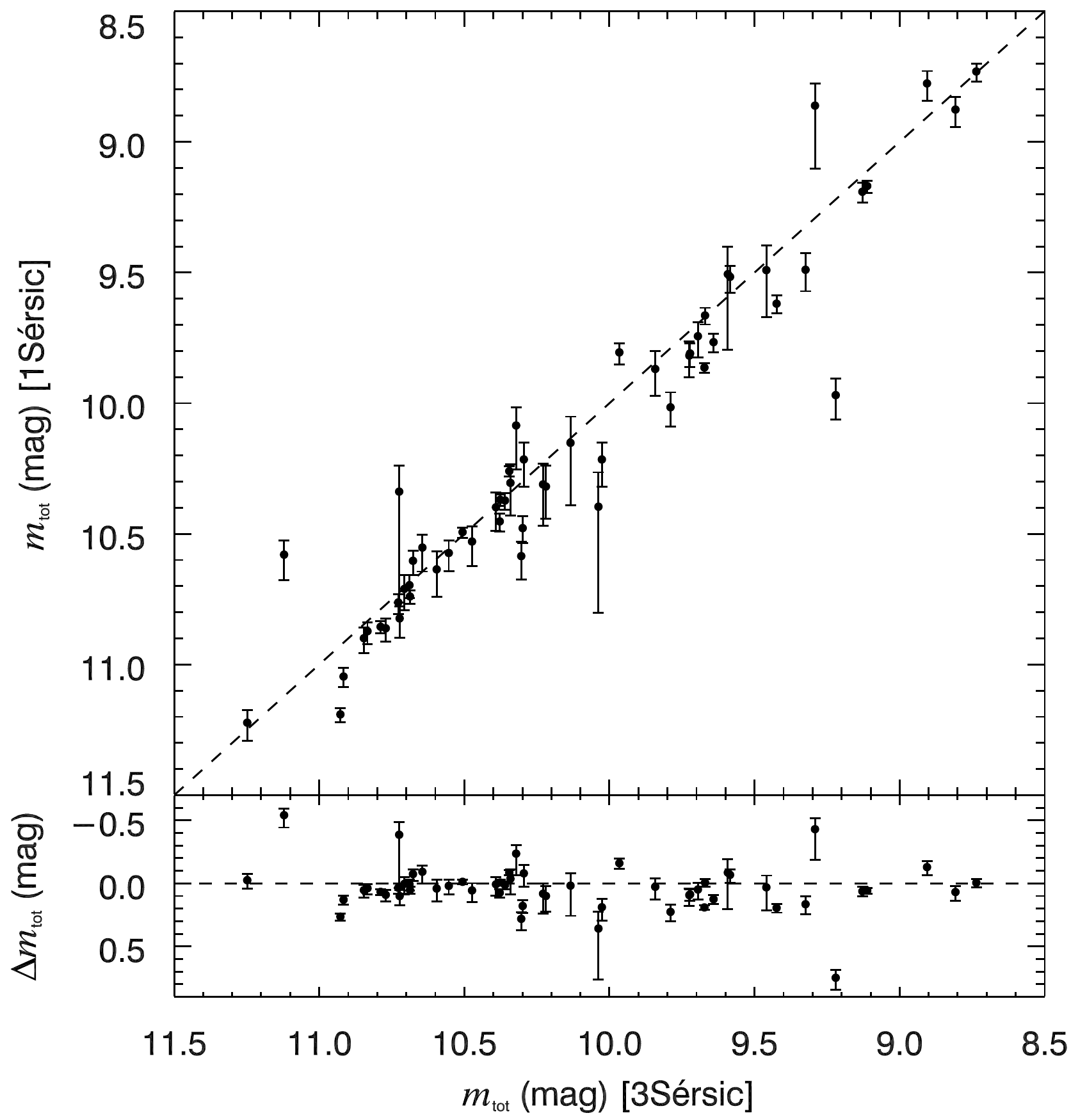}
  \caption{Comparison of total magnitudes of the CGS ellipticals obtained
    through single-\sersic{} fits with those through three-\sersic{} fits. There
    is no significant systematic bias ($\langle \Delta m_{\rm tot} \rangle = 
    0.03 \pm 0.19$ mag).
    \label{fig:comp_mag}}
\end{figure}

Although we choose \sersic{}-based parameters for the ellipticals to facilitate
a consistent comparison with the bulges, there are concerns about these
measurements that need to be addressed. First, the measurements are
model-dependent, as we already know that ellipticals can be better fit by
multiple \sersic{} components \citep{2013ApJ+Huang1}. Second, we do not know the
light distributions beyond the limiting surface brightness of the images, so it
is dangerous to recklessly extrapolate to infinity. The three-component
\sersic{} fits of \citet{2013ApJ+Huang1}, which conform to the two-phase
scenario for the origin of ellipticals \citep{2013ApJ+Huang2}, in principle
should describe more accurately the outermost component of the galaxy and
therefore provide more reasonable extrapolation at large
radii. \citet{2016ApJ+Huang} presented three-component \sersic{} fits of a
subset of the CGS ellipticals in the $R$ band. Figure~\ref{fig:comp_mag} shows
good agreement between the total magnitudes derived from our single-\sersic{}
fits and their three-\sersic{} fits.  Systematic bias is negligible (0.03\,mag),
and the rms scatter is 0.19\,mag.  This comparison assures us that the total
light of the ellipticals obtained from single-\sersic{} fits is reasonable, at
least when compared with state-of-art parametric measurements.

We only have to deal with the extreme outliers. There are four galaxies
(NGC~1172, 2865, 4936, and 4976) that exhibit offsets larger than 2 times the
rms. We remeasure their $\mu_{e}$, $\langle\mu_{e}\rangle$, and $r_{e}$ using
the curve-of-growth method, but using total magnitudes obtained from the
three-\sersic{} fits as input instead of the total magnitudes obtained from the
non-parametric method.  However, this method failed in the case of NGC~4936, for
its $r_{e}$ lies beyond the image boundary, so we apply no further corrections
for it. We also correct for NGC~1439, because it appears as an outlier in the
Kormendy relation.  Correcting the structural parameters of these four
galaxies has little effect on the scaling relation.

\begin{figure*}
  \epsscale{1.12}
  \plotone{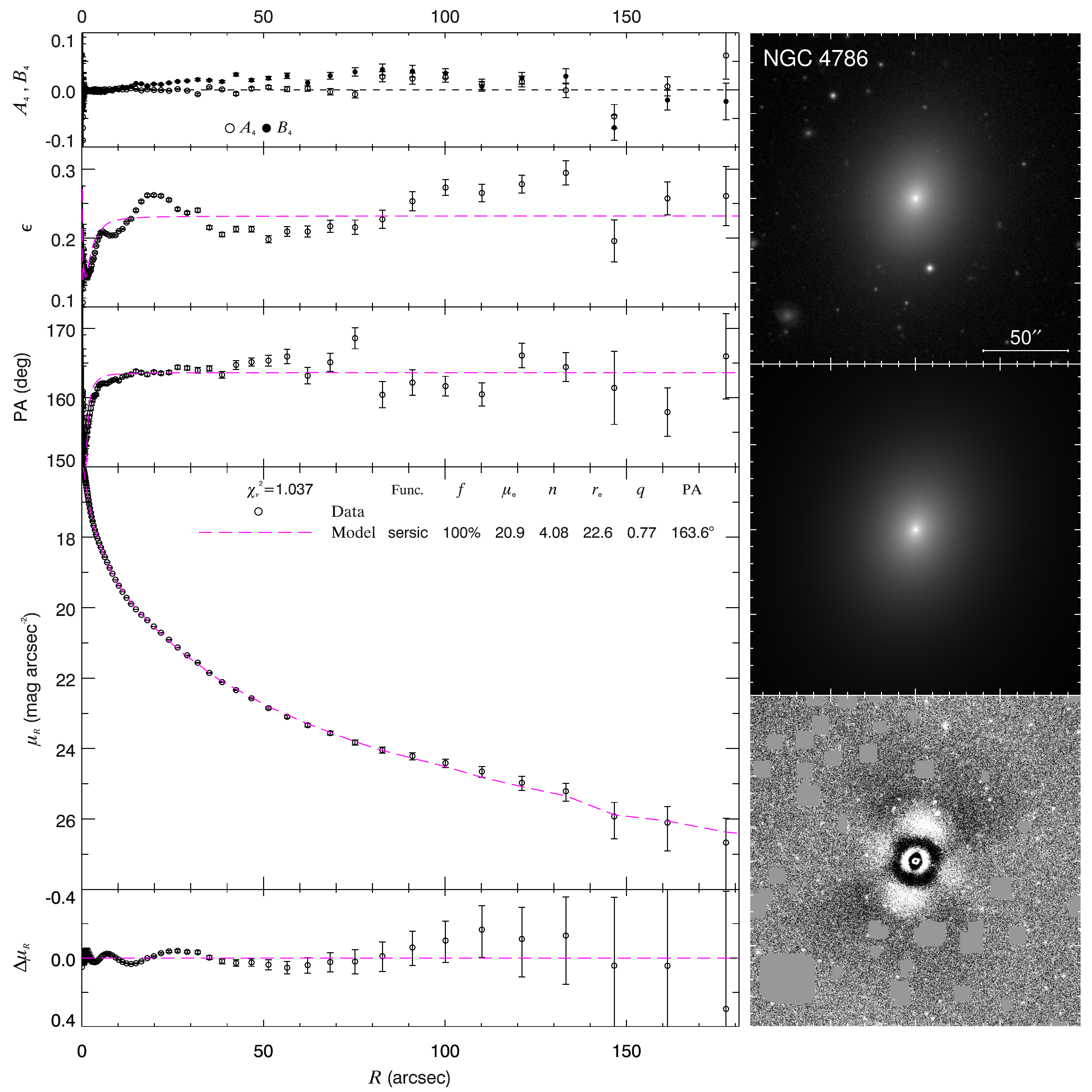}
  \caption{Best-fit model of NGC~4786. The left panels display the isophotal
    analysis of the 2D image fitting. From top to bottom, the panels show the
    radial profiles of the fourth harmonic deviations from an ellipse ($A_{4}$
    and $B_{4}$), ellipticity ($\epsilon$), position angle (PA), $R$-band
    surface brightness ($\mu_{R}$), and fitting residuals
    ($\bigtriangleup\mu_{R}$). Profiles of the data and model are encoded
    consistently with different symbols, line styles, and colors, as explained
    in the legends. The text to the right of the legends gives detailed
    information on each component; from left to right, the columns describe the
    radial profile functions (\sersic{}), the light fractions, the effective
    surface brightness $\mu_{e}$, the \sersic{} index $n$, the effective radius
    $r_{e}$, the axis ratio $q$, and the PA. Note that the
    surface brightness profile of the model is generated by fixing the geometric
    parameters to those of the data surface brightness profile. The right panels
    display, from top to bottom, the grayscale $R$-band image, the best-fit
    model image, and the residual image. The images are shown using the same
    logarithmic stretch for the data and model image, and histogram equalization
    stretch for the residual image. All images are cropped to have the same size
    of $1.5D_{25}$, with $D_{25}$ the isophotal galaxy diameter at
    $\mu_{B}=25$~mag~arcsec$^{-2}$, and are centered on the galaxy centroid,
    with north up and east to the left. \\(The complete figure set for 83
    elliptical galaxies is available in the online journal.)
    \label{fig:NGC4786}}
\end{figure*}

The best-fit models for the ellipticals are displayed in
Figure~\ref{fig:NGC4786}, and their structural parameters are listed in
Table~\ref{tab:ell_param}.


\figsetstart
\figsetnum{19}
\figsettitle{Best-fit models of CGS ellipticals.}

\figsetgrpstart
\figsetgrpnum{19.59}
\figsetgrptitle{NGC 4786}
\figsetplot{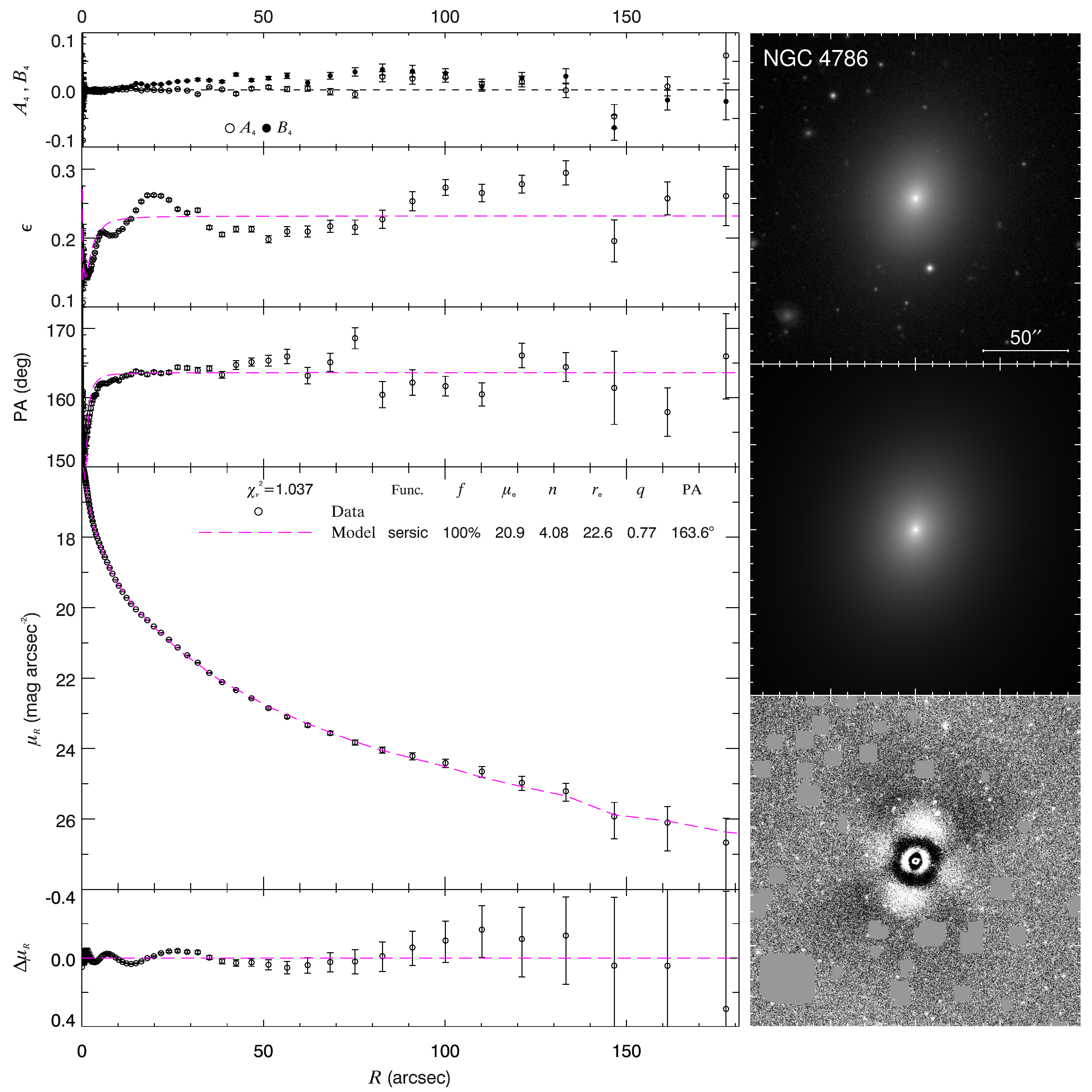}
\figsetgrpnote{Best-fit model of NGC 4786. The left panels display the isophotal analysis of the 2D image fitting. From top to bottom, the panels show the radial profiles of the fourth harmonic deviations from an ellipse ($A_{4}$ and $B_{4}$), ellipticity ($\epsilon$), position angle (PA), $R$-band surface brightness ($\mu_{R}$), and fitting residuals ($\bigtriangleup\mu_{R}$). The right panels display, from top to bottom, the grayscale $R$-band data image, the best-fit model image, and the residual images. The legends and explanatory text follow the same convention as in the static version of this figure set.}
\figsetgrpend

\figsetend

\clearpage
\startlongtable
\begin{deluxetable}{lCCCC}
  \tablecaption{Best-fit Parameters of the CGS Ellipticals \label{tab:ell_param}}
  
  \tablehead{\colhead{Name} & \colhead{$\langle\mu_{e}\rangle$} & \colhead{$n$} &
    \colhead{$r_{e}$} & \colhead{$\epsilon$} \\ \colhead{} &
    \colhead{(mag~arcsec$^{-2}$)} & \colhead{} & \colhead{(kpc)} & \colhead{}
    \\ \colhead{(1)} & \colhead{(2)} & \colhead{(3)} & \colhead{(4)} &
    \colhead{(5)}}
  
  \startdata
  ESO 185--G054 & $20.76\pm 0.20$ & $4.94\pm 0.26$ & $ 18.71\pm  2.46$ & $0.269\pm 0.000$ \\
  IC 1459      & $19.31\pm 0.09$ & $5.19\pm 0.12$ & $  6.61\pm  0.34$ & $0.261\pm 0.000$ \\
  IC 1633      & $21.56\pm 0.25$ & $6.69\pm 0.34$ & $ 35.50\pm  5.58$ & $0.167\pm 0.000$ \\
  IC 2311      & $19.15\pm 0.19$ & $5.18\pm 0.29$ & $  2.90\pm  0.34$ & $0.043\pm 0.000$ \\
  IC 2597      & $20.35\pm 0.12$ & $4.97\pm 0.17$ & $  7.04\pm  0.54$ & $0.284\pm 0.000$ \\
  IC 3370      & $19.54\pm 0.08$ & $3.92\pm 0.15$ & $  4.53\pm  0.23$ & $0.164\pm 0.000$ \\
  IC 3896      & $19.97\pm 0.29$ & $5.49\pm 0.42$ & $  7.03\pm  1.31$ & $0.208\pm 0.002$ \\
  IC 4296      & $20.28\pm 0.19$ & $4.97\pm 0.23$ & $ 13.37\pm  1.56$ & $0.084\pm 0.001$ \\
  IC 4742      & $20.05\pm 0.29$ & $3.89\pm 0.51$ & $  8.28\pm  1.71$ & $0.192\pm 0.002$ \\
  IC 4765      & $22.53\pm 0.59$ & $6.05\pm 0.66$ & $ 52.04\pm 21.53$ & $0.362\pm 0.001$ \\
  IC 4797      & $18.71\pm 0.09$ & $4.52\pm 0.16$ & $  3.37\pm  0.20$ & $0.399\pm 0.001$ \\
  IC 4889      & $19.02\pm 0.06$ & $4.07\pm 0.09$ & $  3.56\pm  0.13$ & $0.336\pm 0.000$ \\
  IC 5328      & $19.62\pm 0.12$ & $4.26\pm 0.19$ & $  4.93\pm  0.36$ & $0.288\pm 0.002$ \\
  NGC 596      & $19.59\pm 0.06$ & $5.14\pm 0.10$ & $  3.30\pm  0.12$ & $0.126\pm 0.001$ \\
  NGC 636      & $20.22\pm 0.23$ & $6.42\pm 0.40$ & $  4.63\pm  0.63$ & $0.119\pm 0.000$ \\
  NGC 720      & $18.85\pm 0.07$ & $3.22\pm 0.09$ & $  4.36\pm  0.20$ & $0.420\pm 0.000$ \\
  NGC 1052     & $18.88\pm 0.06$ & $3.97\pm 0.10$ & $  2.96\pm  0.10$ & $0.295\pm 0.000$ \\
  NGC 1172     & $21.05\pm 0.86$ & \nodata       & $  7.03\pm  3.41$ & $0.207\pm 0.002$ \\
  NGC 1199     & $19.79\pm 0.13$ & $4.61\pm 0.20$ & $  4.84\pm  0.37$ & $0.249\pm 0.001$ \\
  NGC 1209     & $18.92\pm 0.05$ & $4.58\pm 0.10$ & $  3.85\pm  0.13$ & $0.524\pm 0.000$ \\
  NGC 1339     & $18.87\pm 0.08$ & $4.58\pm 0.16$ & $  1.78\pm  0.09$ & $0.284\pm 0.000$ \\
  NGC 1340     & $19.13\pm 0.07$ & $3.59\pm 0.12$ & $  3.51\pm  0.17$ & $0.354\pm 0.000$ \\
  NGC 1374     & $19.48\pm 0.05$ & $4.62\pm 0.09$ & $  2.66\pm  0.09$ & $0.097\pm 0.000$ \\
  NGC 1379     & $19.56\pm 0.05$ & $3.12\pm 0.08$ & $  2.47\pm  0.08$ & $0.019\pm 0.000$ \\
  NGC 1395     & $19.65\pm 0.09$ & $4.33\pm 0.12$ & $  5.89\pm  0.33$ & $0.163\pm 0.001$ \\
  NGC 1399     & $19.64\pm 0.10$ & $4.58\pm 0.12$ & $  5.93\pm  0.35$ & $0.093\pm 0.001$ \\
  NGC 1404     & $18.06\pm 0.11$ & $3.31\pm 0.18$ & $  2.14\pm  0.15$ & $0.136\pm 0.001$ \\
  NGC 1407     & $20.17\pm 0.14$ & $4.42\pm 0.17$ & $  9.32\pm  0.85$ & $0.046\pm 0.000$ \\
  NGC 1426     & $19.87\pm 0.18$ & $5.09\pm 0.33$ & $  3.72\pm  0.42$ & $0.350\pm 0.000$ \\
  NGC 1427     & $20.11\pm 0.11$ & $5.45\pm 0.18$ & $  4.84\pm  0.35$ & $0.299\pm 0.000$ \\
  NGC 1439     & $21.02\pm 0.41$ & \nodata       & $  6.95\pm  1.58$ & $0.090\pm 0.000$ \\
  NGC 1453     & $20.07\pm 0.20$ & $5.63\pm 0.29$ & $ 10.81\pm  1.35$ & $0.163\pm 0.001$ \\
  NGC 1521     & $20.32\pm 0.18$ & $5.25\pm 0.26$ & $ 10.91\pm  1.22$ & $0.300\pm 0.001$ \\
  NGC 1549     & $19.64\pm 0.10$ & $4.96\pm 0.14$ & $  4.64\pm  0.29$ & $0.103\pm 0.000$ \\
  NGC 1600     & $20.32\pm 0.20$ & $3.75\pm 0.27$ & $ 12.39\pm  1.72$ & $0.329\pm 0.002$ \\
  NGC 1700     & $19.11\pm 0.17$ & $5.33\pm 0.28$ & $  5.31\pm  0.55$ & $0.287\pm 0.000$ \\
  NGC 2305     & $19.70\pm 0.14$ & $4.82\pm 0.19$ & $  4.98\pm  0.43$ & $0.251\pm 0.000$ \\
  NGC 2325     & $20.18\pm 0.11$ & $2.95\pm 0.12$ & $  5.53\pm  0.41$ & $0.364\pm 0.000$ \\
  NGC 2434     & $19.74\pm 0.18$ & $4.71\pm 0.27$ & $  4.21\pm  0.48$ & $0.089\pm 0.001$ \\
  NGC 2663     & $20.05\pm 0.29$ & $4.73\pm 0.35$ & $ 13.22\pm  2.57$ & $0.332\pm 0.001$ \\
  NGC 2865     & $18.44\pm 0.17$ & \nodata       & $  2.28\pm  0.27$ & $0.250\pm 0.000$ \\
  NGC 2986     & $20.44\pm 0.11$ & $5.57\pm 0.17$ & $ 10.70\pm  0.76$ & $0.151\pm 0.000$ \\
  NGC 3078     & $19.23\pm 0.17$ & $5.00\pm 0.27$ & $  4.93\pm  0.51$ & $0.241\pm 0.000$ \\
  NGC 3087     & $19.64\pm 0.16$ & $5.33\pm 0.24$ & $  5.95\pm  0.59$ & $0.149\pm 0.000$ \\
  NGC 3091     & $20.24\pm 0.19$ & $5.24\pm 0.27$ & $ 10.81\pm  1.32$ & $0.286\pm 0.001$ \\
  NGC 3136     & $20.68\pm 0.43$ & $6.73\pm 0.58$ & $ 12.43\pm  3.52$ & $0.264\pm 0.001$ \\
  NGC 3250     & $19.26\pm 0.12$ & $3.85\pm 0.17$ & $  6.30\pm  0.47$ & $0.256\pm 0.001$ \\
  NGC 3258     & $20.16\pm 0.20$ & $4.80\pm 0.26$ & $  7.74\pm  0.97$ & $0.127\pm 0.001$ \\
  NGC 3268     & $21.22\pm 0.50$ & $5.51\pm 0.61$ & $ 14.45\pm  4.92$ & $0.203\pm 0.000$ \\
  NGC 3557     & $19.54\pm 0.10$ & $4.58\pm 0.13$ & $  8.11\pm  0.49$ & $0.242\pm 0.000$ \\
  NGC 3585     & $19.98\pm 0.17$ & $6.24\pm 0.25$ & $  8.22\pm  0.86$ & $0.445\pm 0.001$ \\
  NGC 3923     & $20.20\pm 0.14$ & $4.71\pm 0.17$ & $ 10.69\pm  0.97$ & $0.343\pm 0.001$ \\
  NGC 3962     & $19.96\pm 0.11$ & $4.91\pm 0.16$ & $  7.49\pm  0.54$ & $0.202\pm 0.000$ \\
  NGC 4696     & $20.72\pm 0.12$ & $2.99\pm 0.13$ & $ 16.65\pm  1.46$ & $0.219\pm 0.000$ \\
  NGC 4709     & $21.32\pm 0.38$ & $5.70\pm 0.47$ & $ 15.42\pm  3.80$ & $0.155\pm 0.000$ \\
  NGC 4742     & $18.06\pm 0.08$ & $7.99\pm 0.17$ & $  1.26\pm  0.06$ & $0.400\pm 0.000$ \\
  NGC 4760     & $20.73\pm 0.19$ & $4.42\pm 0.25$ & $ 12.68\pm  1.61$ & $0.155\pm 0.001$ \\
  NGC 4767     & $19.23\pm 0.10$ & $4.17\pm 0.16$ & $  5.03\pm  0.33$ & $0.423\pm 0.001$ \\
  NGC 4786     & $19.37\pm 0.10$ & $4.08\pm 0.16$ & $  6.87\pm  0.44$ & $0.232\pm 0.000$ \\
  NGC 4936     & $20.21\pm 0.17$ & $4.14\pm 0.22$ & $  8.13\pm  0.94$ & $0.170\pm 0.001$ \\
  NGC 4976     & $19.02\pm 0.37$ & \nodata       & $  3.25\pm  0.82$ & $0.320\pm 0.001$ \\
  NGC 5011     & $20.17\pm 0.15$ & $5.06\pm 0.22$ & $  7.23\pm  0.68$ & $0.158\pm 0.000$ \\
  NGC 5018     & $18.52\pm 0.06$ & $4.39\pm 0.10$ & $  4.79\pm  0.18$ & $0.297\pm 0.000$ \\
  NGC 5044     & $20.27\pm 0.12$ & $3.12\pm 0.14$ & $  8.23\pm  0.69$ & $0.060\pm 0.000$ \\
  NGC 5061     & $19.50\pm 0.15$ & $6.48\pm 0.25$ & $  5.70\pm  0.53$ & $0.099\pm 0.001$ \\
  NGC 5077     & $19.14\pm 0.14$ & $4.56\pm 0.21$ & $  5.45\pm  0.47$ & $0.302\pm 0.001$ \\
  NGC 5328     & $19.80\pm 0.14$ & $5.23\pm 0.21$ & $  7.94\pm  0.66$ & $0.313\pm 0.000$ \\
  NGC 5419     & $20.63\pm 0.18$ & $4.55\pm 0.23$ & $ 15.78\pm  1.95$ & $0.210\pm 0.000$ \\
  NGC 5791     & $19.21\pm 0.09$ & $3.98\pm 0.15$ & $  4.43\pm  0.26$ & $0.425\pm 0.000$ \\
  NGC 5796     & $19.55\pm 0.14$ & $5.06\pm 0.22$ & $  5.23\pm  0.45$ & $0.131\pm 0.000$ \\
  NGC 5812     & $19.32\pm 0.10$ & $5.24\pm 0.17$ & $  3.63\pm  0.23$ & $0.044\pm 0.000$ \\
  NGC 5898     & $19.27\pm 0.20$ & $4.59\pm 0.28$ & $  3.88\pm  0.49$ & $0.019\pm 0.001$ \\
  NGC 5903     & $20.48\pm 0.24$ & $4.87\pm 0.31$ & $  9.05\pm  1.39$ & $0.232\pm 0.000$ \\
  NGC 6851     & $18.89\pm 0.07$ & $4.20\pm 0.14$ & $  2.74\pm  0.13$ & $0.266\pm 0.000$ \\
  NGC 6868     & $20.04\pm 0.18$ & $4.94\pm 0.23$ & $  8.08\pm  0.90$ & $0.186\pm 0.000$ \\
  NGC 6876     & $20.19\pm 0.08$ & $3.05\pm 0.10$ & $  9.05\pm  0.52$ & $0.136\pm 0.000$ \\
  NGC 6909     & $20.05\pm 0.14$ & $4.37\pm 0.25$ & $  5.33\pm  0.49$ & $0.442\pm 0.000$ \\
  NGC 6958     & $19.18\pm 0.07$ & $5.17\pm 0.11$ & $  3.55\pm  0.16$ & $0.144\pm 0.000$ \\
  NGC 7029     & $19.41\pm 0.07$ & $5.23\pm 0.11$ & $  4.75\pm  0.19$ & $0.364\pm 0.000$ \\
  NGC 7145     & $20.42\pm 0.15$ & $4.94\pm 0.21$ & $  4.41\pm  0.42$ & $0.036\pm 0.000$ \\
  NGC 7196     & $19.28\pm 0.13$ & $4.58\pm 0.21$ & $  5.41\pm  0.42$ & $0.217\pm 0.000$ \\
  NGC 7507     & $18.80\pm 0.09$ & $4.99\pm 0.13$ & $  4.11\pm  0.22$ & $0.038\pm 0.000$ \\
  NGC 7796     & $20.03\pm 0.13$ & $4.91\pm 0.19$ & $  8.13\pm  0.66$ & $0.153\pm 0.000$ 
  \enddata

  \tablecomments{Col.~(1): Galaxy name. Cols.~(2)--(5): Average effective
    surface brightness, \sersic{} index, effective radius, and
    ellipticity. The parameters are measured in the $R$ band and have been
      corrected for Galactic extinction.}
\end{deluxetable} 

\bibliographystyle{aasjournal}
\bibliography{myref}

\begin{thebibliography}{}
\expandafter\ifx\csname natexlab\endcsname\relax\def\natexlab#1{#1}\fi
\providecommand{\url}[1]{\href{#1}{#1}}

\bibitem[{{Aguerri} {et~al.}(2001){Aguerri}, {Balcells}, \&
  {Peletier}}]{2001A&A+Aguerri}
{Aguerri}, J.~A.~L., {Balcells}, M., \& {Peletier}, R.~F. 2001, \aap, 367, 428

\bibitem[{{Allen} {et~al.}(2006){Allen}, {Driver}, {Graham}, {Cameron},
  {Liske}, \& {de Propris}}]{2006MNRAS+Allen}
{Allen}, P.~D., {Driver}, S.~P., {Graham}, A.~W., {et~al.} 2006, \mnras, 371, 2

\bibitem[{{Athanassoula}(2005)}]{2005MNRAS+Athanassoula}
{Athanassoula}, E. 2005, \mnras, 358, 1477

\bibitem[{{Athanassoula} {et~al.}(2013){Athanassoula}, {Machado}, \&
  {Rodionov}}]{2013MNRAS+Athanassoula}
{Athanassoula}, E., {Machado}, R.~E.~G., \& {Rodionov}, S.~A. 2013, \mnras,
  429, 1949

\bibitem[{{Balcells} {et~al.}(2003){Balcells}, {Graham},
  {Dom{\'{\i}}nguez-Palmero}, \& {Peletier}}]{2003ApJ+Balcells}
{Balcells}, M., {Graham}, A.~W., {Dom{\'{\i}}nguez-Palmero}, L., \& {Peletier},
  R.~F. 2003, \apjl, 582, L79

\bibitem[{{Bekki}(1999)}]{1999ApJ+Bekki}
{Bekki}, K. 1999, \apjl, 510, L15

\bibitem[{{Bekki} \& {Couch}(2011)}]{2011MNRAS+Bekki}
{Bekki}, K., \& {Couch}, W.~J. 2011, \mnras, 415, 1783

\bibitem[{{Bender}(1987)}]{1987MitAG+Bender}
{Bender}, R. 1987, MitAG, 70, 226

\bibitem[{{Bender}(1988)}]{1988A&A+Bender}
{Bender}, R. 1988, \aap, 193, L7

\bibitem[{{Bender} {et~al.}(1992){Bender}, {Burstein}, \&
  {Faber}}]{1992ApJ+Bender}
{Bender}, R., {Burstein}, D., \& {Faber}, S.~M. 1992, \apj, 399, 462

\bibitem[{{Bertin} {et~al.}(2002){Bertin}, {Ciotti}, \& {Del
  Principe}}]{2002A&A+Bertin}
{Bertin}, G., {Ciotti}, L., \& {Del Principe}, M. 2002, \aap, 386, 149

\bibitem[{{Binggeli}(1994)}]{1994ESOC+Binggeli}
{Binggeli}, B. 1994, in ESO/OHP Workshop on Dwarf Galaxies, ed. G.~{Meylan} \&
  P.~{Prugniel} (Garching: ESO), 13

\bibitem[{{Binney} \& {Tremaine}(2008)}]{2008gady+Binney}
{Binney}, J., \& {Tremaine}, S. 2008, {Galactic Dynamics: Second Edition}
  (Princeton, NJ: Princeton Univ. Press)

\bibitem[{{Block} {et~al.}(2002){Block}, {Bournaud}, {Combes}, {Puerari}, \&
  {Buta}}]{2002A&A+Block}
{Block}, D.~L., {Bournaud}, F., {Combes}, F., {Puerari}, I., \& {Buta}, R.
  2002, \aap, 394, L35

\bibitem[{{Borriello} {et~al.}(2003){Borriello}, {Salucci}, \&
  {Danese}}]{2003MNRAS+Borriello}
{Borriello}, A., {Salucci}, P., \& {Danese}, L. 2003, \mnras, 341, 1109

\bibitem[{{Bournaud}(2016)}]{2016ASSL+Bournaud}
{Bournaud}, F. 2016, in Galactic Bulges, ed. E.~{Laurikainen}, R.~{Peletier},
  \& D.~{Gadotti} (New York: Springer), 355

\bibitem[{{Bournaud} \& {Combes}(2002)}]{2002A&A+Bournaud}
{Bournaud}, F., \& {Combes}, F. 2002, \aap, 392, 83

\bibitem[{{Bournaud} {et~al.}(2005){Bournaud}, {Combes}, \&
  {Semelin}}]{2005MNRAS+Bournaud}
{Bournaud}, F., {Combes}, F., \& {Semelin}, B. 2005, \mnras, 364, L18

\bibitem[{{Bournaud} {et~al.}(2007){Bournaud}, {Elmegreen}, \&
  {Elmegreen}}]{2007ApJ+Bournaud}
{Bournaud}, F., {Elmegreen}, B.~G., \& {Elmegreen}, D.~M. 2007, \apj, 670, 237

\bibitem[{{Bournaud} {et~al.}(2009){Bournaud}, {Elmegreen}, \&
  {Martig}}]{2009ApJ+Bournaud}
{Bournaud}, F., {Elmegreen}, B.~G., \& {Martig}, M. 2009, \apjl, 707, L1

\bibitem[{{Brooks} \& {Christensen}(2016)}]{2016ASSL+Brooks}
{Brooks}, A., \& {Christensen}, C. 2016, in Galactic Bulges, ed.
  E.~{Laurikainen}, R.~{Peletier}, \& D.~{Gadotti} (New York: Springer), 317

\bibitem[{{Busarello} {et~al.}(1997){Busarello}, {Capaccioli}, {Capozziello},
  {Longo}, \& {Puddu}}]{1997A&A+Busarello}
{Busarello}, G., {Capaccioli}, M., {Capozziello}, S., {Longo}, G., \& {Puddu},
  E. 1997, \aap, 320, 415

\bibitem[{{Cappellari} {et~al.}(2006){Cappellari}, {Bacon}, {Bureau}, {Damen},
  {Davies}, {de Zeeuw}, {Emsellem}, {Falc{\'o}n-Barroso}, {Krajnovi{\'c}},
  {Kuntschner}, {McDermid}, {Peletier}, {Sarzi}, {van den Bosch}, \& {van de
  Ven}}]{2006MNRAS+Cappellari}
{Cappellari}, M., {Bacon}, R., {Bureau}, M., {et~al.} 2006, \mnras, 366, 1126

\bibitem[{{Cappellari} {et~al.}(2012){Cappellari}, {McDermid}, {Alatalo},
  {Blitz}, {Bois}, {Bournaud}, {Bureau}, {Crocker}, {Davies}, {Davis}, {de
  Zeeuw}, {Duc}, {Emsellem}, {Khochfar}, {Krajnovi{\'c}}, {Kuntschner},
  {Lablanche}, {Morganti}, {Naab}, {Oosterloo}, {Sarzi}, {Scott}, {Serra},
  {Weijmans}, \& {Young}}]{2012Natur+Cappellari}
{Cappellari}, M., {McDermid}, R.~M., {Alatalo}, K., {et~al.} 2012, \nat, 484,
  485

\bibitem[{{Cappellari} {et~al.}(2013{\natexlab{a}}){Cappellari}, {Scott},
  {Alatalo}, {Blitz}, {Bois}, {Bournaud}, {Bureau}, {Crocker}, {Davies},
  {Davis}, {de Zeeuw}, {Duc}, {Emsellem}, {Khochfar}, {Krajnovi{\'c}},
  {Kuntschner}, {McDermid}, {Morganti}, {Naab}, {Oosterloo}, {Sarzi}, {Serra},
  {Weijmans}, \& {Young}}]{2013MNRAS+Cappellari1}
{Cappellari}, M., {Scott}, N., {Alatalo}, K., {et~al.} 2013{\natexlab{a}},
  \mnras, 432, 1709

\bibitem[{{Cappellari} {et~al.}(2013{\natexlab{b}}){Cappellari}, {McDermid},
  {Alatalo}, {Blitz}, {Bois}, {Bournaud}, {Bureau}, {Crocker}, {Davies},
  {Davis}, {de Zeeuw}, {Duc}, {Emsellem}, {Khochfar}, {Krajnovi{\'c}},
  {Kuntschner}, {Morganti}, {Naab}, {Oosterloo}, {Sarzi}, {Scott}, {Serra},
  {Weijmans}, \& {Young}}]{2013MNRAS+Cappellari2}
{Cappellari}, M., {McDermid}, R.~M., {Alatalo}, K., {et~al.}
  2013{\natexlab{b}}, \mnras, 432, 1862

\bibitem[{{Carollo}(1999)}]{1999ApJ+Carollo}
{Carollo}, C.~M. 1999, \apj, 523, 566

\bibitem[{{Carollo} {et~al.}(1997){Carollo}, {Stiavelli}, {de Zeeuw}, \&
  {Mack}}]{1997AJ+Carollo}
{Carollo}, C.~M., {Stiavelli}, M., {de Zeeuw}, P.~T., \& {Mack}, J. 1997, \aj,
  114, 2366

\bibitem[{{Carollo} {et~al.}(2002){Carollo}, {Stiavelli}, {Seigar}, {de Zeeuw},
  \& {Dejonghe}}]{2002AJ+Carollo}
{Carollo}, C.~M., {Stiavelli}, M., {Seigar}, M., {de Zeeuw}, P.~T., \&
  {Dejonghe}, H. 2002, \aj, 123, 159

\bibitem[{{Chown} {et~al.}(2019){Chown}, {Li}, {Athanassoula}, {Li}, {Wilson},
  {Lin}, {Mo}, {Parker}, \& {Xiao}}]{2019MNRAS+Chown}
{Chown}, R., {Li}, C., {Athanassoula}, E., {et~al.} 2019, \mnras, 484, 5192

\bibitem[{{Ciotti} {et~al.}(1996){Ciotti}, {Lanzoni}, \&
  {Renzini}}]{1996MNRAS+Ciotti}
{Ciotti}, L., {Lanzoni}, B., \& {Renzini}, A. 1996, \mnras, 282, 1

\bibitem[{Cleveland \& Devlin(1988)}]{1988J.AM.STAT.ASSOC+Cleveland}
Cleveland, W.~S., \& Devlin, S.~J. 1988, Journal of the American Statistical
  Association, 83, 596

\bibitem[{{Combes}(1996)}]{1996ASPC+Combes}
{Combes}, F. 1996, in ASP Conf. Ser. 91, IAU Coll. 157: Barred Galaxies, ed.
  R.~{Buta}, D.~A. {Crocker}, \& B.~G. {Elmegreen} (San Francisco, CA: ASP),
  286

\bibitem[{{Combes} {et~al.}(1990){Combes}, {Debbasch}, {Friedli}, \&
  {Pfenniger}}]{1990A&A+Combes}
{Combes}, F., {Debbasch}, F., {Friedli}, D., \& {Pfenniger}, D. 1990, \aap,
  233, 82

\bibitem[{{Combes} \& {Sanders}(1981)}]{1981A&A+Combes}
{Combes}, F., \& {Sanders}, R.~H. 1981, \aap, 96, 164

\bibitem[{{Costantin} {et~al.}(2018){Costantin}, {Corsini}, {M{\'e}ndez-Abreu},
  {Morelli}, {Dalla Bont{\`a}}, \& {Pizzella}}]{2018MNRAS+Costantin}
{Costantin}, L., {Corsini}, E.~M., {M{\'e}ndez-Abreu}, J., {et~al.} 2018,
  \mnras, 481, 3623

\bibitem[{{Costantin} {et~al.}(2017){Costantin}, {M{\'e}ndez-Abreu}, {Corsini},
  {Morelli}, {Aguerri}, {Dalla Bont{\`a}}, \& {Pizzella}}]{2017A&A+Costantin}
{Costantin}, L., {M{\'e}ndez-Abreu}, J., {Corsini}, E.~M., {et~al.} 2017, \aap,
  601, A84

\bibitem[{{Davies} {et~al.}(1983){Davies}, {Efstathiou}, {Fall}, {Illingworth},
  \& {Schechter}}]{1983ApJ+Davies}
{Davies}, R.~L., {Efstathiou}, G., {Fall}, S.~M., {Illingworth}, G., \&
  {Schechter}, P.~L. 1983, \apj, 266, 41

\bibitem[{{de Jong}(1996)}]{1996A&A+de_Jong3}
{de Jong}, R.~S. 1996, \aap, 313, 45

\bibitem[{{de Lorenzo-C{\'a}ceres} {et~al.}(2019){de Lorenzo-C{\'a}ceres},
  {M{\'e}ndez-Abreu}, {Thorne}, \& {Costantin}}]{2019MNRAS+de_Lorenzo-Caceres}
{de Lorenzo-C{\'a}ceres}, A., {M{\'e}ndez-Abreu}, J., {Thorne}, B., \&
  {Costantin}, L. 2019, \mnras, 484, 665

\bibitem[{{de Vaucouleurs}(1948)}]{1948AnAp+de_Vaucouleurs}
{de Vaucouleurs}, G. 1948, AnAp, 11, 247

\bibitem[{{Djorgovski}(1987)}]{1987IAUS+Djorgovski}
{Djorgovski}, S. 1987, in IAU Symp. 127, Structure and Dynamics of Elliptical
  Galaxies, ed. P.~T. {de Zeeuw} (Dordrecht: Reidel), 79

\bibitem[{{Djorgovski} \& {Davis}(1987)}]{1987ApJ+Djorgovski}
{Djorgovski}, S., \& {Davis}, M. 1987, \apj, 313, 59

\bibitem[{{Djorgovski} {et~al.}(1988){Djorgovski}, {de Carvalho}, \&
  {Han}}]{1988ASPC+Djorgovski}
{Djorgovski}, S., {de Carvalho}, R., \& {Han}, M.-S. 1988, in ASP Conf. Ser. 4,
  The Extragalactic Distance Scale, ed. S.~{van den Bergh} \& C.~J. {Pritchet}
  (San Francisco, CA: ASP), 329

\bibitem[{{D'Onofrio} {et~al.}(2017){D'Onofrio}, {Cariddi}, {Chiosi}, {Chiosi},
  \& {Marziani}}]{2017ApJ+DOnofrio}
{D'Onofrio}, M., {Cariddi}, S., {Chiosi}, C., {Chiosi}, E., \& {Marziani}, P.
  2017, \apj, 838, 163

\bibitem[{{D'Onofrio} {et~al.}(2013){D'Onofrio}, {Fasano}, {Moretti},
  {Marziani}, {Bindoni}, {Fritz}, {Varela}, {Bettoni}, {Cava}, {Poggianti},
  {Gullieuszik}, {Kj{\ae}rgaard}, {Moles}, {Vulcani}, {Omizzolo}, {Couch}, \&
  {Dressler}}]{2013MNRAS+DOnofrio}
{D'Onofrio}, M., {Fasano}, G., {Moretti}, A., {et~al.} 2013, \mnras, 435, 45

\bibitem[{{Dressler} {et~al.}(1987){Dressler}, {Lynden-Bell}, {Burstein},
  {Davies}, {Faber}, {Terlevich}, \& {Wegner}}]{1987ApJ+Dressler}
{Dressler}, A., {Lynden-Bell}, D., {Burstein}, D., {et~al.} 1987, \apj, 313, 42

\bibitem[{{Eliche-Moral} {et~al.}(2006){Eliche-Moral}, {Balcells}, {Aguerri},
  \& {Gonz{\'a}lez-Garc{\'{\i}}a}}]{2006A&A+Eliche-Moral}
{Eliche-Moral}, M.~C., {Balcells}, M., {Aguerri}, J.~A.~L., \&
  {Gonz{\'a}lez-Garc{\'{\i}}a}, A.~C. 2006, \aap, 457, 91

\bibitem[{{Eliche-Moral} {et~al.}(2018){Eliche-Moral},
  {Rodr{\'{\i}}guez-P{\'e}rez}, {Borlaff}, {Querejeta}, \&
  {Tapia}}]{2018A&A+Eliche-Moral}
{Eliche-Moral}, M.~C., {Rodr{\'{\i}}guez-P{\'e}rez}, C., {Borlaff}, A.,
  {Querejeta}, M., \& {Tapia}, T. 2018, \aap, 617, A113

\bibitem[{{Elmegreen} {et~al.}(2008){Elmegreen}, {Bournaud}, \&
  {Elmegreen}}]{2008ApJ+Elmegreen}
{Elmegreen}, B.~G., {Bournaud}, F., \& {Elmegreen}, D.~M. 2008, \apj, 688, 67

\bibitem[{{Emsellem} {et~al.}(2011){Emsellem}, {Cappellari}, {Krajnovi{\'c}},
  {Alatalo}, {Blitz}, {Bois}, {Bournaud}, {Bureau}, {Davies}, {Davis}, {de
  Zeeuw}, {Khochfar}, {Kuntschner}, {Lablanche}, {McDermid}, {Morganti},
  {Naab}, {Oosterloo}, {Sarzi}, {Scott}, {Serra}, {van de Ven}, {Weijmans}, \&
  {Young}}]{2011MNRAS+Emsellem}
{Emsellem}, E., {Cappellari}, M., {Krajnovi{\'c}}, D., {et~al.} 2011, \mnras,
  414, 888

\bibitem[{{Erwin} {et~al.}(2015){Erwin}, {Saglia}, {Fabricius}, {Thomas},
  {Nowak}, {Rusli}, {Bender}, {Vega Beltr{\'a}n}, \&
  {Beckman}}]{2015MNRAS+Erwin}
{Erwin}, P., {Saglia}, R.~P., {Fabricius}, M., {et~al.} 2015, \mnras, 446, 4039

\bibitem[{{Faber}(1977)}]{1977egsp.conf+Faber}
{Faber}, S.~M. 1977, in Evolution of Galaxies and Stellar Populations, ed.
  B.~M. {Tinsley} \& R.~B. {Larson} (New Haven, CT: Yale Univ. Obs.), 157

\bibitem[{{Faber} {et~al.}(1987){Faber}, {Dressler}, {Davies}, {Burstein}, \&
  {Lynden-Bell}}]{1987nngp+Faber}
{Faber}, S.~M., {Dressler}, A., {Davies}, R.~L., {Burstein}, D., \&
  {Lynden-Bell}, D. 1987, in Nearly Normal Galaxies. From the Planck Time to
  the Present, ed. S.~M. {Faber} (New York: Springer), 175

\bibitem[{{Faber} \& {Jackson}(1976)}]{1976ApJ+Faber2}
{Faber}, S.~M., \& {Jackson}, R.~E. 1976, \apj, 204, 668

\bibitem[{{Faber} {et~al.}(1997){Faber}, {Tremaine}, {Ajhar}, {Byun},
  {Dressler}, {Gebhardt}, {Grillmair}, {Kormendy}, {Lauer}, \&
  {Richstone}}]{1997AJ+Faber}
{Faber}, S.~M., {Tremaine}, S., {Ajhar}, E.~A., {et~al.} 1997, \aj, 114, 1771

\bibitem[{{Fisher} \& {Drory}(2008)}]{2008AJ+Fisher}
{Fisher}, D.~B., \& {Drory}, N. 2008, \aj, 136, 773

\bibitem[{{Fisher} \& {Drory}(2010)}]{2010ApJ+Fisher}
{Fisher}, D.~B., \& {Drory}, N. 2010, \apj, 716, 942

\bibitem[{{Fisher} \& {Drory}(2016)}]{2016ASSL+Fisher}
{Fisher}, D.~B., \& {Drory}, N. 2016, in Galactic Bulges, ed. E.~{Laurikainen},
  R.~{Peletier}, \& D.~{Gadotti} (New York: Springer), 41

\bibitem[{{Fisher} {et~al.}(2009){Fisher}, {Drory}, \&
  {Fabricius}}]{2009ApJ+Fisher}
{Fisher}, D.~B., {Drory}, N., \& {Fabricius}, M.~H. 2009, \apj, 697, 630

\bibitem[{{Fragkoudi} {et~al.}(2016){Fragkoudi}, {Athanassoula}, \&
  {Bosma}}]{2016MNRAS+Fragkoudi}
{Fragkoudi}, F., {Athanassoula}, E., \& {Bosma}, A. 2016, \mnras, 462, L41

\bibitem[{{Fukugita} {et~al.}(1995){Fukugita}, {Shimasaku}, \&
  {Ichikawa}}]{1995+Fukugita}
{Fukugita}, M., {Shimasaku}, K., \& {Ichikawa}, T. 1995, \pasp, 107, 945

\bibitem[{{Gadotti}(2008)}]{2008MNRAS+Gadotti}
{Gadotti}, D.~A. 2008, \mnras, 384, 420

\bibitem[{{Gadotti}(2009)}]{2009MNRAS+Gadotti}
{Gadotti}, D.~A. 2009, \mnras, 393, 1531

\bibitem[{{Gallagher} {et~al.}(1982){Gallagher}, {Goad}, \&
  {Mould}}]{1982ApJ+Gallagher}
{Gallagher}, J.~S., {Goad}, J.~W., \& {Mould}, J. 1982, \apj, 263, 101

\bibitem[{{Gao} \& {Ho}(2017)}]{2017ApJ+Gao}
{Gao}, H., \& {Ho}, L.~C. 2017, \apj, 845, 114

\bibitem[{{Gao} {et~al.}(2018){Gao}, {Ho}, {Barth}, \& {Li}}]{2018ApJ+Gao}
{Gao}, H., {Ho}, L.~C., {Barth}, A.~J., \& {Li}, Z.-Y. 2018, \apj, 862, 100

\bibitem[{{Gao} {et~al.}(2019){Gao}, {Ho}, {Barth}, \& {Li}}]{2019ApJS+Gao}
{Gao}, H., {Ho}, L.~C., {Barth}, A.~J., \& {Li}, Z.-Y. 2019, \apjs, 244, 34

\bibitem[{{Genzel} {et~al.}(2008){Genzel}, {Burkert}, {Bouch{\'e}}, {Cresci},
  {F{\"o}rster Schreiber}, {Shapley}, {Shapiro}, {Tacconi}, {Buschkamp},
  {Cimatti}, {Daddi}, {Davies}, {Eisenhauer}, {Erb}, {Genel}, {Gerhard},
  {Hicks}, {Lutz}, {Naab}, {Ott}, {Rabien}, {Renzini}, {Steidel}, {Sternberg},
  \& {Lilly}}]{2008ApJ+Genzel}
{Genzel}, R., {Burkert}, A., {Bouch{\'e}}, N., {et~al.} 2008, \apj, 687, 59

\bibitem[{{Gott}(1977)}]{1977ARA&A+Gott}
{Gott}, III, J.~R. 1977, \araa, 15, 235

\bibitem[{{Grosb{\o}l} {et~al.}(2004){Grosb{\o}l}, {Patsis}, \&
  {Pompei}}]{2004A&A+Grosbol}
{Grosb{\o}l}, P., {Patsis}, P.~A., \& {Pompei}, E. 2004, \aap, 423, 849

\bibitem[{{Ho} {et~al.}(2011){Ho}, {Li}, {Barth}, {Seigar}, \&
  {Peng}}]{2011ApJS+Ho}
{Ho}, L.~C., {Li}, Z.-Y., {Barth}, A.~J., {Seigar}, M.~S., \& {Peng}, C.~Y.
  2011, \apjs, 197, 21

\bibitem[{{Hopkins} {et~al.}(2009{\natexlab{a}}){Hopkins}, {Cox}, {Younger}, \&
  {Hernquist}}]{2009ApJ+Hopkins}
{Hopkins}, P.~F., {Cox}, T.~J., {Younger}, J.~D., \& {Hernquist}, L.
  2009{\natexlab{a}}, \apj, 691, 1168

\bibitem[{{Hopkins} {et~al.}(2009{\natexlab{b}}){Hopkins}, {Somerville}, {Cox},
  {Hernquist}, {Jogee}, {Kere{\v s}}, {Ma}, {Robertson}, \&
  {Stewart}}]{2009MNRAS+Hopkins}
{Hopkins}, P.~F., {Somerville}, R.~S., {Cox}, T.~J., {et~al.}
  2009{\natexlab{b}}, \mnras, 397, 802

\bibitem[{{Hopkins} {et~al.}(2010){Hopkins}, {Bundy}, {Croton}, {Hernquist},
  {Keres}, {Khochfar}, {Stewart}, {Wetzel}, \& {Younger}}]{2010ApJ+Hopkins}
{Hopkins}, P.~F., {Bundy}, K., {Croton}, D., {et~al.} 2010, \apj, 715, 202

\bibitem[{{Huang} {et~al.}(2013{\natexlab{a}}){Huang}, {Ho}, {Peng}, {Li}, \&
  {Barth}}]{2013ApJ+Huang1}
{Huang}, S., {Ho}, L.~C., {Peng}, C.~Y., {Li}, Z.-Y., \& {Barth}, A.~J.
  2013{\natexlab{a}}, \apj, 766, 47

\bibitem[{{Huang} {et~al.}(2013{\natexlab{b}}){Huang}, {Ho}, {Peng}, {Li}, \&
  {Barth}}]{2013ApJ+Huang2}
{Huang}, S., {Ho}, L.~C., {Peng}, C.~Y., {Li}, Z.-Y., \& {Barth}, A.~J.
  2013{\natexlab{b}}, \apjl, 768, L28

\bibitem[{{Huang} {et~al.}(2016){Huang}, {Ho}, {Peng}, {Li}, \&
  {Barth}}]{2016ApJ+Huang}
{Huang}, S., {Ho}, L.~C., {Peng}, C.~Y., {Li}, Z.-Y., \& {Barth}, A.~J. 2016,
  \apj, 821, 114

\bibitem[{{Illingworth}(1977)}]{1977ApJ+Illingworth}
{Illingworth}, G. 1977, \apjl, 218, L43

\bibitem[{{Immeli} {et~al.}(2004){Immeli}, {Samland}, {Gerhard}, \&
  {Westera}}]{2004A&A+Immeli}
{Immeli}, A., {Samland}, M., {Gerhard}, O., \& {Westera}, P. 2004, \aap, 413,
  547

\bibitem[{{Inoue} \& {Saitoh}(2012)}]{2012MNRAS+Inoue}
{Inoue}, S., \& {Saitoh}, T.~R. 2012, \mnras, 422, 1902

\bibitem[{{Izquierdo-Villalba} {et~al.}(2019){Izquierdo-Villalba}, {Bonoli},
  {Spinoso}, {Rosas-Guevara}, {Henriques}, \&
  {Hern{\'a}ndez-Monteagudo}}]{2019MNRAS+Izquierdo-Villalba}
{Izquierdo-Villalba}, D., {Bonoli}, S., {Spinoso}, D., {et~al.} 2019, \mnras,
  488, 609

\bibitem[{{Kaviraj}(2014)}]{2014MNRAS+Kaviraj}
{Kaviraj}, S. 2014, \mnras, 440, 2944

\bibitem[{{Kent}(1985)}]{1985ApJS+Kent}
{Kent}, S.~M. 1985, \apjs, 59, 115

\bibitem[{{Kim} \& {Ho}(2019)}]{2019ApJ+Kim}
{Kim}, M., \& {Ho}, L.~C. 2019, \apj, 876, 35

\bibitem[{{Kim} {et~al.}(2017){Kim}, {Ho}, {Peng}, {Barth}, \&
  {Im}}]{2017ApJS+Kim}
{Kim}, M., {Ho}, L.~C., {Peng}, C.~Y., {Barth}, A.~J., \& {Im}, M. 2017, \apjs,
  232, 21

\bibitem[{{Kodaira} {et~al.}(1986){Kodaira}, {Watanabe}, \&
  {Okamura}}]{1986ApJS+Kodaira}
{Kodaira}, K., {Watanabe}, M., \& {Okamura}, S. 1986, \apjs, 62, 703

\bibitem[{Kormendy(1977)}]{1977ApJ+Kormendy2}
Kormendy, J. 1977, \apj, 218, 333

\bibitem[{{Kormendy}(1981)}]{1981seng.proc+Kormendy}
{Kormendy}, J. 1981, in The Structure and Evolution of Normal Galaxies, ed.
  S.~M. {Fall} \& D.~{Lynden-Bell} (Cambridge: Cambridge Univ. Press), 85

\bibitem[{{Kormendy}(1982{\natexlab{a}})}]{1982ApJ+Kormendy2}
{Kormendy}, J. 1982{\natexlab{a}}, \apj, 257, 75

\bibitem[{{Kormendy}(1982{\natexlab{b}})}]{1982SAAS+Kormendy}
{Kormendy}, J. 1982{\natexlab{b}}, in Twelfth Advanced Course of the Swiss
  Society of Astronomy and Astrophysics, Morphology and Dynamics of Galaxies,
  ed. L.~{Martinet} \& M.~{Mayor} (Sauverny: Geneva Obs.), 113

\bibitem[{{Kormendy}(1985)}]{1985ApJ+Kormendy}
{Kormendy}, J. 1985, \apj, 295, 73

\bibitem[{{Kormendy}(1987)}]{1987nngp+Kormendy}
{Kormendy}, J. 1987, in Nearly Normal Galaxies. From the Planck Time to the
  Present, ed. S.~M. {Faber} (New York: Springer), 163

\bibitem[{{Kormendy}(1993)}]{1993IAUS+Kormendy}
{Kormendy}, J. 1993, in IAU Symp. 153, Galactic Bulges, ed. H.~{Dejonghe} \&
  H.~J. {Habing} (Dordrecht: Kluwer), 209

\bibitem[{{Kormendy}(2013)}]{2013seg+Kormendy}
{Kormendy}, J. 2013, in Secular Evolution of Galaxies, ed.
  J.~{Falc{\'o}n-Barroso} \& J.~H. {Knapen} (Cambridge: Cambridge Univ. Press),
  1

\bibitem[{{Kormendy}(2016)}]{2016ASSL+Kormendy}
{Kormendy}, J. 2016, in Galactic Bulges, ed. E.~{Laurikainen}, R.~{Peletier},
  \& D.~{Gadotti} (New York: Springer), 431

\bibitem[{{Kormendy} \& {Barentine}(2010)}]{2010ApJ+Kormendy1}
{Kormendy}, J., \& {Barentine}, J.~C. 2010, \apjl, 715, L176

\bibitem[{{Kormendy} \& {Djorgovski}(1989)}]{1989ARA&A+Kormendy}
{Kormendy}, J., \& {Djorgovski}, S. 1989, \araa, 27, 235

\bibitem[{{Kormendy} {et~al.}(2010){Kormendy}, {Drory}, {Bender}, \&
  {Cornell}}]{2010ApJ+Kormendy2}
{Kormendy}, J., {Drory}, N., {Bender}, R., \& {Cornell}, M.~E. 2010, \apj, 723,
  54

\bibitem[{{Kormendy} \& {Fisher}(2008)}]{2008ASPC+Kormendy}
{Kormendy}, J., \& {Fisher}, D.~B. 2008, in ASP Conf. Ser. 396, Formation and
  Evolution of Galaxy Disks, ed. J.~G. {Funes} \& E.~M. {Corsini} (San
  Francisco, CA: ASP), 297

\bibitem[{{Kormendy} {et~al.}(2009){Kormendy}, {Fisher}, {Cornell}, \&
  {Bender}}]{2009ApJS+Kormendy}
{Kormendy}, J., {Fisher}, D.~B., {Cornell}, M.~E., \& {Bender}, R. 2009, \apjs,
  182, 216

\bibitem[{{Kormendy} \& {Ho}(2013)}]{2013ARA&A+Kormendy}
{Kormendy}, J., \& {Ho}, L.~C. 2013, \araa, 51, 511

\bibitem[{{Kormendy} \& {Illingworth}(1982)}]{1982ApJ+Kormendy1}
{Kormendy}, J., \& {Illingworth}, G. 1982, \apj, 256, 460

\bibitem[{{Kormendy} \& {Illingworth}(1983)}]{1983ApJ+Kormendy}
{Kormendy}, J., \& {Illingworth}, G. 1983, \apj, 265, 632

\bibitem[{{Kormendy} \& {Kennicutt}(2004)}]{2004ARA&A+Kormendy}
{Kormendy}, J., \& {Kennicutt}, Jr., R.~C. 2004, \araa, 42, 603

\bibitem[{{Kormendy} {et~al.}(1996){Kormendy}, {Byun}, {Ajhar}, {Lauer},
  {Dressler}, {Faber}, {Grillmair}, {Gebhardt}, {Richstone}, \&
  {Tremaine}}]{1996IAUS+Kormendy}
{Kormendy}, J., {Byun}, Y., {Ajhar}, E.~A., {et~al.} 1996, in IAU Symp. 171,
  New Light on Galaxy Evolution, ed. R.~{Bender} \& R.~L. {Davies} (Dordrecht:
  Kluwer), 105

\bibitem[{{Kroupa}(2001)}]{2001MNRAS+Kroupa}
{Kroupa}, P. 2001, \mnras, 322, 231

\bibitem[{{L{\"a}sker} {et~al.}(2016){L{\"a}sker}, {Greene}, {Seth}, {van de
  Ven}, {Braatz}, {Henkel}, \& {Lo}}]{2016ApJ+Lasker}
{L{\"a}sker}, R., {Greene}, J.~E., {Seth}, A., {et~al.} 2016, \apj, 825, 3

\bibitem[{{Lauer}(1985)}]{1985ApJ+Lauer}
{Lauer}, T.~R. 1985, \apj, 292, 104

\bibitem[{{Lauer} {et~al.}(1995){Lauer}, {Ajhar}, {Byun}, {Dressler}, {Faber},
  {Grillmair}, {Kormendy}, {Richstone}, \& {Tremaine}}]{1995AJ+Lauer}
{Lauer}, T.~R., {Ajhar}, E.~A., {Byun}, Y.-I., {et~al.} 1995, \aj, 110, 2622

\bibitem[{{Lauer} {et~al.}(2005){Lauer}, {Faber}, {Gebhardt}, {Richstone},
  {Tremaine}, {Ajhar}, {Aller}, {Bender}, {Dressler}, {Filippenko}, {Green},
  {Grillmair}, {Ho}, {Kormendy}, {Magorrian}, {Pinkney}, \&
  {Siopis}}]{2005AJ+Lauer}
{Lauer}, T.~R., {Faber}, S.~M., {Gebhardt}, K., {et~al.} 2005, \aj, 129, 2138

\bibitem[{{Lauer} {et~al.}(2007){Lauer}, {Gebhardt}, {Faber}, {Richstone},
  {Tremaine}, {Kormendy}, {Aller}, {Bender}, {Dressler}, {Filippenko}, {Green},
  \& {Ho}}]{2007ApJ+Lauer}
{Lauer}, T.~R., {Gebhardt}, K., {Faber}, S.~M., {et~al.} 2007, \apj, 664, 226

\bibitem[{{Laurikainen} {et~al.}(2007){Laurikainen}, {Salo}, {Buta}, \&
  {Knapen}}]{2007MNRAS+Laurikainen}
{Laurikainen}, E., {Salo}, H., {Buta}, R., \& {Knapen}, J.~H. 2007, \mnras,
  381, 401

\bibitem[{{Laurikainen} {et~al.}(2010){Laurikainen}, {Salo}, {Buta}, {Knapen},
  \& {Comer{\'o}n}}]{2010MNRAS+Laurikainen}
{Laurikainen}, E., {Salo}, H., {Buta}, R., {Knapen}, J.~H., \& {Comer{\'o}n},
  S. 2010, \mnras, 405, 1089

\bibitem[{{Li} {et~al.}(2017){Li}, {Ho}, \& {Barth}}]{2017ApJ+Li}
{Li}, Z.-Y., {Ho}, L.~C., \& {Barth}, A.~J. 2017, \apj, 845, 87

\bibitem[{{Li} {et~al.}(2011){Li}, {Ho}, {Barth}, \& {Peng}}]{2011ApJS+Li}
{Li}, Z.-Y., {Ho}, L.~C., {Barth}, A.~J., \& {Peng}, C.~Y. 2011, \apjs, 197, 22

\bibitem[{{Lin} {et~al.}(2017){Lin}, {Li}, {He}, {Xiao}, \&
  {Wang}}]{2017ApJ+Lin}
{Lin}, L., {Li}, C., {He}, Y., {Xiao}, T., \& {Wang}, E. 2017, \apj, 838, 105

\bibitem[{{Lynden-Bell} {et~al.}(1988){Lynden-Bell}, {Faber}, {Burstein},
  {Davies}, {Dressler}, {Terlevich}, \& {Wegner}}]{1988ApJ+Lynden-Bell}
{Lynden-Bell}, D., {Faber}, S.~M., {Burstein}, D., {et~al.} 1988, \apj, 326, 19

\bibitem[{{M{\'e}ndez-Abreu} {et~al.}(2008){M{\'e}ndez-Abreu}, {Aguerri},
  {Corsini}, \& {Simonneau}}]{2008A&A+Mendez-Abreu}
{M{\'e}ndez-Abreu}, J., {Aguerri}, J.~A.~L., {Corsini}, E.~M., \& {Simonneau},
  E. 2008, \aap, 478, 353

\bibitem[{{M{\'e}ndez-Abreu} {et~al.}(2010){M{\'e}ndez-Abreu}, {Simonneau},
  {Aguerri}, \& {Corsini}}]{2010A&A+Mendez-Abreu}
{M{\'e}ndez-Abreu}, J., {Simonneau}, E., {Aguerri}, J.~A.~L., \& {Corsini},
  E.~M. 2010, \aap, 521, A71

\bibitem[{{M{\'e}ndez-Abreu} {et~al.}(2017){M{\'e}ndez-Abreu}, {Ruiz-Lara},
  {S{\'a}nchez-Menguiano}, {de Lorenzo-C{\'a}ceres}, {Costantin},
  {Catal{\'a}n-Torrecilla}, {Florido}, {Aguerri}, {Bland-Hawthorn}, {Corsini},
  {Dettmar}, {Galbany}, {Garc{\'{\i}}a-Benito}, {Marino}, {M{\'a}rquez},
  {Ortega-Minakata}, {Papaderos}, {S{\'a}nchez}, {S{\'a}nchez-Blazquez},
  {Spekkens}, {van de Ven}, {Wild}, \& {Ziegler}}]{2017A&A+Mendez-Abreu}
{M{\'e}ndez-Abreu}, J., {Ruiz-Lara}, T., {S{\'a}nchez-Menguiano}, L., {et~al.}
  2017, \aap, 598, A32

\bibitem[{{M{\'e}ndez-Abreu} {et~al.}(2018){M{\'e}ndez-Abreu}, {Aguerri},
  {Falc{\'o}n-Barroso}, {Ruiz-Lara}, {S{\'a}nchez-Menguiano}, {de
  Lorenzo-C{\'a}ceres}, {Costantin}, {Catal{\'a}n-Torrecilla}, {Zhu},
  {S{\'a}nchez-Blazquez}, {Florido}, {Corsini}, {Wild}, {Lyubenova}, {van de
  Ven}, {S{\'a}nchez}, {Bland-Hawthorn}, {Galbany}, {Garc{\'{\i}}a-Benito},
  {Garc{\'{\i}}a-Lorenzo}, {Gonz{\'a}lez Delgado}, {L{\'o}pez-S{\'a}nchez},
  {Marino}, {M{\'a}rquez}, {Ziegler}, \& {CALIFA
  Collaboration}}]{2018MNRAS+Mendez-Abreu}
{M{\'e}ndez-Abreu}, J., {Aguerri}, J.~A.~L., {Falc{\'o}n-Barroso}, J., {et~al.}
  2018, \mnras, 474, 1307

\bibitem[{{Moore} {et~al.}(1996){Moore}, {Katz}, {Lake}, {Dressler}, \&
  {Oemler}}]{1996Natur+Moore}
{Moore}, B., {Katz}, N., {Lake}, G., {Dressler}, A., \& {Oemler}, A. 1996,
  \nat, 379, 613

\bibitem[{{Moore} {et~al.}(1998){Moore}, {Lake}, \& {Katz}}]{1998ApJ+Moore}
{Moore}, B., {Lake}, G., \& {Katz}, N. 1998, \apj, 495, 139

\bibitem[{{Moore} {et~al.}(1999){Moore}, {Lake}, {Quinn}, \&
  {Stadel}}]{1999MNRAS+Moore}
{Moore}, B., {Lake}, G., {Quinn}, T., \& {Stadel}, J. 1999, \mnras, 304, 465

\bibitem[{{Neumann} {et~al.}(2017){Neumann}, {Wisotzki}, {Choudhury},
  {Gadotti}, {Walcher}, {Bland-Hawthorn}, {Garc{\'{\i}}a-Benito}, {Gonz{\'a}lez
  Delgado}, {Husemann}, {Marino}, {M{\'a}rquez}, {S{\'a}nchez}, {Ziegler}, \&
  {Califa Collaboration}}]{2017A&A+Neumann}
{Neumann}, J., {Wisotzki}, L., {Choudhury}, O.~S., {et~al.} 2017, \aap, 604,
  A30

\bibitem[{{Nieto} {et~al.}(1991){Nieto}, {Bender}, \& {Surma}}]{1991A&A+Nieto}
{Nieto}, J.-L., {Bender}, R., \& {Surma}, P. 1991, \aap, 244, L37

\bibitem[{{Noguchi}(1999)}]{1999ApJ+Noguchi}
{Noguchi}, M. 1999, \apj, 514, 77

\bibitem[{{Oklop{\v c}i{\'c}} {et~al.}(2017){Oklop{\v c}i{\'c}}, {Hopkins},
  {Feldmann}, {Kere{\v s}}, {Faucher-Gigu{\`e}re}, \&
  {Murray}}]{2017MNRAS+Oklopcic}
{Oklop{\v c}i{\'c}}, A., {Hopkins}, P.~F., {Feldmann}, R., {et~al.} 2017,
  \mnras, 465, 952

\bibitem[{{Pahre} {et~al.}(1998){Pahre}, {de Carvalho}, \&
  {Djorgovski}}]{1998AJ+Pahre}
{Pahre}, M.~A., {de Carvalho}, R.~R., \& {Djorgovski}, S.~G. 1998, \aj, 116,
  1606

\bibitem[{{Paturel} {et~al.}(2003){Paturel}, {Petit}, {Prugniel}, {Theureau},
  {Rousseau}, {Brouty}, {Dubois}, \& {Cambr{\'e}sy}}]{2003A&A+Paturel}
{Paturel}, G., {Petit}, C., {Prugniel}, P., {et~al.} 2003, \aap, 412, 45

\bibitem[{{Peng} {et~al.}(2002){Peng}, {Ho}, {Impey}, \& {Rix}}]{2002AJ+Peng}
{Peng}, C.~Y., {Ho}, L.~C., {Impey}, C.~D., \& {Rix}, H.-W. 2002, \aj, 124, 266

\bibitem[{{Peng} {et~al.}(2010){Peng}, {Ho}, {Impey}, \& {Rix}}]{2010AJ+Peng}
{Peng}, C.~Y., {Ho}, L.~C., {Impey}, C.~D., \& {Rix}, H.-W. 2010, \aj, 139,
  2097

\bibitem[{{Pfenniger}(1984)}]{1984A&A+Pfenniger}
{Pfenniger}, D. 1984, \aap, 134, 373

\bibitem[{{Pfenniger}(1985)}]{1985A&A+Pfenniger}
{Pfenniger}, D. 1985, \aap, 150, 112

\bibitem[{{Pfenniger} \& {Norman}(1990)}]{1990ApJ+Pfenniger}
{Pfenniger}, D., \& {Norman}, C. 1990, \apj, 363, 391

\bibitem[{{Poggianti} {et~al.}(2017){Poggianti}, {Jaff{\'e}}, {Moretti},
  {Gullieuszik}, {Radovich}, {Tonnesen}, {Fritz}, {Bettoni}, {Vulcani},
  {Fasano}, {Bellhouse}, {Hau}, \& {Omizzolo}}]{2017Natur+Poggianti}
{Poggianti}, B.~M., {Jaff{\'e}}, Y.~L., {Moretti}, A., {et~al.} 2017, \nat,
  548, 304

\bibitem[{{Renzini}(1999)}]{1999fgb+Renzini}
{Renzini}, A. 1999, in The Formation of Galactic Bulges, ed. C.~M. {Carollo},
  H.~C. {Ferguson}, \& R.~F.~G. {Wyse} (Cambridge: Cambridge Univ. Press), 9

\bibitem[{{Renzini} \& {Ciotti}(1993)}]{1993ApJ+Renzini}
{Renzini}, A., \& {Ciotti}, L. 1993, \apjl, 416, L49

\bibitem[{{Roberts} \& {Haynes}(1994)}]{1994ARA&A+Roberts}
{Roberts}, M.~S., \& {Haynes}, M.~P. 1994, \araa, 32, 115

\bibitem[{{Rodriguez-Gomez} {et~al.}(2017){Rodriguez-Gomez}, {Sales}, {Genel},
  {Pillepich}, {Zjupa}, {Nelson}, {Griffen}, {Torrey}, {Snyder},
  {Vogelsberger}, {Springel}, {Ma}, \& {Hernquist}}]{2017MNRAS+Rodriguez-Gomez}
{Rodriguez-Gomez}, V., {Sales}, L.~V., {Genel}, S., {et~al.} 2017, \mnras, 467,
  3083

\bibitem[{{Sandage} {et~al.}(1970){Sandage}, {Freeman}, \&
  {Stokes}}]{1970ApJ+Sandage}
{Sandage}, A., {Freeman}, K.~C., \& {Stokes}, N.~R. 1970, \apj, 160, 831

\bibitem[{{Sauvaget} {et~al.}(2018){Sauvaget}, {Hammer}, {Puech}, {Yang},
  {Flores}, \& {Rodrigues}}]{2018MNRAS+Sauvaget}
{Sauvaget}, T., {Hammer}, F., {Puech}, M., {et~al.} 2018, \mnras, 473, 2521

\bibitem[{{Sellwood}(2014)}]{2014RvMP+Sellwood}
{Sellwood}, J.~A. 2014, RvMP, 86, 1

\bibitem[{{Sellwood} \& {Wilkinson}(1993)}]{1993RPPh+Sellwood}
{Sellwood}, J.~A., \& {Wilkinson}, A. 1993, RPPh, 56, 173

\bibitem[{{S{\'e}rsic}(1968)}]{1968adga+Sersic}
{S{\'e}rsic}, J.~L. 1968, {Atlas de Galaxias Australes} (C{\'o}rdoba: Obs.
  Astron., Univ. Nac. C{\'o}rdoba)

\bibitem[{{Shen} \& {Sellwood}(2004)}]{2004ApJ+Shen}
{Shen}, J., \& {Sellwood}, J.~A. 2004, \apj, 604, 614

\bibitem[{{Simien} \& {de Vaucouleurs}(1986)}]{1986ApJ+Simien}
{Simien}, F., \& {de Vaucouleurs}, G. 1986, \apj, 302, 564

\bibitem[{{Tabor} {et~al.}(2017){Tabor}, {Merrifield}, {Arag{\'o}n-Salamanca},
  {Cappellari}, {Bamford}, \& {Johnston}}]{2017MNRAS+Tabor}
{Tabor}, M., {Merrifield}, M., {Arag{\'o}n-Salamanca}, A., {et~al.} 2017,
  \mnras, 466, 2024

\bibitem[{{Tonini} {et~al.}(2016){Tonini}, {Mutch}, {Croton}, \&
  {Wyithe}}]{2016MNRAS+Tonini}
{Tonini}, C., {Mutch}, S.~J., {Croton}, D.~J., \& {Wyithe}, J.~S.~B. 2016,
  \mnras, 459, 4109

\bibitem[{{Toomre}(1977)}]{1977egsp+Tmoore}
{Toomre}, A. 1977, in Evolution of Galaxies and Stellar Populations, ed. B.~M.
  {Tinsley} \& R.~B. {Larson} (New Haven, CT: Yale Univ. Obs.), 401

\bibitem[{{Treu} {et~al.}(2005){Treu}, {Ellis}, {Liao}, \& {van
  Dokkum}}]{2005ApJ+Treu}
{Treu}, T., {Ellis}, R.~S., {Liao}, T.~X., \& {van Dokkum}, P.~G. 2005, \apjl,
  622, L5

\bibitem[{{Trujillo} {et~al.}(2001){Trujillo}, {Graham}, \&
  {Caon}}]{2001MNRAS+Trujillo}
{Trujillo}, I., {Graham}, A.~W., \& {Caon}, N. 2001, \mnras, 326, 869

\bibitem[{{van den Bergh}(1976)}]{1976ApJ+van_den_Bergh1}
{van den Bergh}, S. 1976, \apj, 203, 764

\bibitem[{{Wang} {et~al.}(2015){Wang}, {Hammer}, {Puech}, {Yang}, \&
  {Flores}}]{2015MNRAS+Wang}
{Wang}, J., {Hammer}, F., {Puech}, M., {Yang}, Y., \& {Flores}, H. 2015,
  \mnras, 452, 3551

\bibitem[{{Wang} {et~al.}(2012){Wang}, {Kauffmann}, {Overzier}, {Tacconi},
  {Kong}, {Saintonge}, {Catinella}, {Schiminovich}, {Moran}, \&
  {Johnson}}]{2012MNRAS+Wang}
{Wang}, J., {Kauffmann}, G., {Overzier}, R., {et~al.} 2012, \mnras, 423, 3486

\bibitem[{{Weinzirl} {et~al.}(2009){Weinzirl}, {Jogee}, {Khochfar}, {Burkert},
  \& {Kormendy}}]{2009ApJ+Weinzirl}
{Weinzirl}, T., {Jogee}, S., {Khochfar}, S., {Burkert}, A., \& {Kormendy}, J.
  2009, \apj, 696, 411

\bibitem[{{Wyse} {et~al.}(1997){Wyse}, {Gilmore}, \& {Franx}}]{1997ARA&A+Wyse}
{Wyse}, R.~F.~G., {Gilmore}, G., \& {Franx}, M. 1997, \araa, 35, 637

\bibitem[{{Zhang} \& {Zaritsky}(2016)}]{2016MNRAS+Zhang}
{Zhang}, H., \& {Zaritsky}, D. 2016, \mnras, 455, 1364

\bibitem[{{Zhao} {et~al.}(2019){Zhao}, {Ho}, {Zhao}, {Shangguan}, \&
  {Kim}}]{2019ApJ+Zhao}
{Zhao}, D., {Ho}, L.~C., {Zhao}, Y., {Shangguan}, J., \& {Kim}, M. 2019, \apj,
  877, 52

\end{thebibliography}
\end{CJK*}

\end{document}